\documentclass{aa}
\usepackage[utf8]{inputenc}
\usepackage[colorlinks]{hyperref}
\usepackage{amsmath}
\usepackage{amssymb}
\usepackage{microtype}
\usepackage[T1]{fontenc}
\usepackage{lmodern}
\usepackage{natbib}
\usepackage{graphicx}
\usepackage{microtype}
\usepackage{color}
\usepackage[dvipsnames]{xcolor}
\usepackage{multirow}
\usepackage{xspace}
\usepackage[normalem]{ulem}
\usepackage[export]{adjustbox}

\bibpunct{(}{)}{;}{a}{}{,} 

\newcommand{\ik}[1] {}

\def\srg{\textit{SRG}}
\def\art{ART-XC}
\def\erosita{eROSITA}

\def\arcdeg{\hbox{$^\circ$}}
\def\arcmin{\hbox{$^\prime$}}
\def\arcsec{\hbox{$^{\prime\prime}$}}

\begin{document}

\title{SRG X-ray orbital observatory. Its telescopes and first scientific results}

\author{R. Sunyaev\inst{\ref{iki},\ref{mpa}}
\and V. Arefiev\inst{\ref{iki}}
\and V. Babyshkin\inst{\ref{npol}}
\and A. Bogomolov\inst{\ref{iki}}
\and K. Borisov\inst{\ref{sscr}}
\and M. Buntov\inst{\ref{iki}}
\and H. Brunner\inst{\ref{mpe}}
\and R.~Burenin\inst{\ref{iki}}
\and E. Churazov\inst{\ref{iki},\ref{mpa}}
\and D. Coutinho\inst{\ref{mpe}}
\and J. Eder\inst{\ref{mpe}}
\and N. Eismont\inst{\ref{iki}}
\and M. Freyberg\inst{\ref{mpe}}
\and M. Gilfanov\inst{\ref{iki},\ref{mpa}}
\and P. Gureyev\inst{\ref{npol}}
\and G. Hasinger\inst{\ref{esac}}
\and I. Khabibullin\inst{\ref{iki},\ref{mpa}}
\and V. Kolmykov\inst{\ref{npol}}
\and S. Komovkin\inst{\ref{npol}}
\and R. Krivonos\inst{\ref{iki}}
\and I. Lapshov\inst{\ref{iki}}
\and V. Levin\inst{\ref{iki}} 
\and I.~Lomakin\inst{\ref{npol}}
\and A. Lutovinov\inst{\ref{iki}}
\and P. Medvedev\inst{\ref{iki}}
\and A. Merloni\inst{\ref{mpe}}
\and T. Mernik\inst{\ref{dlr}}
\and E. Mikhailov\inst{\ref{npol}}
\and {V. Molodtsov}\inst{\ref{npol}}
\and P.~Mzhelsky\inst{\ref{npol}}
\and S.~M\"uller\inst{\ref{mpe}}
\and K. Nandra\inst{\ref{mpe}}
\and V. Nazarov\inst{\ref{iki}}
\and M. Pavlinsky\inst{\ref{iki}}
\and A. Poghodin\inst{\ref{npol}}
\and P. Predehl\inst{\ref{mpe}}
\and J. Robrade\inst{\ref{unih}}
\and S.~Sazonov\inst{\ref{iki}}
\and H.~Scheuerle\inst{\ref{dlr}}
\and A. Shirshakov\inst{\ref{npol}}
\and A. Tkachenko\inst{\ref{iki}}
\and V. Voron\inst{\ref{sscr}}
}

\institute{Space Research Institute (IKI), Russian Academy of Sciences, Profsoyuznaya ul. 84/32, Moscow, 117997 Russia\label{iki}
\and Max-Planck-Institut f\"{u}r Astrophysik (MPA), Karl-Schwarzschild-Str. 1, D-85741 Garching, Germany\label{mpa}
\and Lavochkin Association, 24 Leningradskaya ul., Khimki 141400, Moscow Region, Russia\label{npol}
\and State Space Corporation Roscosmos, Schepkina ulitsa 42, Moscow, 107996, Russia\label{sscr}
\and Max-Planck-Institut f\"{u}r extraterrestrische Physik (MPE), Giessenbachstr. D-85748 Garching, Germany\label{mpe}
\and ESAC Camino bajo de Casillo S/N, Villanueva de la Canada, Madrid, Spain\label{esac}
\and Deutsches Zentrum f\"{ur}r Luft- und Raumfahrt, K\"{o}nigswinterer Str. 522-524, D-53227 Bonn, Germany\label{dlr}
\and Universit\"at Hamburg, Hamburger Sternwarte, Gojenbergsweg 112, D-21029 Hamburg\label{unih}
}


\abstract{
The orbital observatory \textit{Spectrum-Roentgen-Gamma} (\srg), equipped with the grazing-incidence X-ray telescopes Mikhail Pavlinsky ART-XC and eROSITA, was launched by Roscosmos to the Lagrange L2 point of the Sun--Earth system on July 13, 2019. The launch was carried out from the Baikonur Cosmodrome by a Proton-M rocket with a DM-03 upper stage. The German telescope eROSITA was installed on \srg\ under an agreement between Roskosmos and the DLR, the German Aerospace Agency. In December 2019, \srg\ started to perform its main scientific task: scanning the celestial sphere to obtain X-ray maps of the entire sky in several energy ranges {(from 0.2} to 8\,keV with eROSITA, and from 4 to 30\,keV with ART-XC). By mid-June 2021, the third six-month all-sky survey had been completed. Over a period of four years, it is planned to obtain eight independent maps of the entire sky in each of the energy ranges. The sum of these maps will provide high sensitivity and reveal more than three million quasars and over one hundred thousand massive galaxy clusters and galaxy groups. The availability of eight sky maps will enable monitoring of long-term variability (every six months) of a huge number of extragalactic and Galactic X-ray sources, including hundreds of thousands of stars with hot coronae. In addition, the rotation of the satellite around the axis directed toward the Sun with a period of four hours enables tracking the faster variability of bright X-ray sources during one day every half year. The chosen strategy of scanning the sky leads to the formation of deep survey zones near both ecliptic poles. The paper presents sky maps obtained by the telescopes on board \srg\ during the first survey of the entire sky and a number of results of deep observations performed during the flight to the L2 point in the frame of the performance verification program, demonstrating the capabilities of the observatory in imaging, spectroscopy, and timing of X-ray sources. It is planned that in December 2023, the observatory will for at least two years switch to observations of the most interesting sources in the sky in triaxial orientation mode and deep scanning of selected celestial fields with an area of up to 150\,square degrees. These modes of operation were tested during the performance verification phase. Every day, data from the \srg\ observatory are dumped onto the largest antennas of the Russian Deep Space Network in Bear Lakes and near Ussuriysk. 
}

\authorrunning{R. Sunyaev et al.}

\maketitle

\section{Introduction}
\label{s:intro}

\begin{figure}
\centering
\includegraphics[width=0.98\columnwidth,clip]{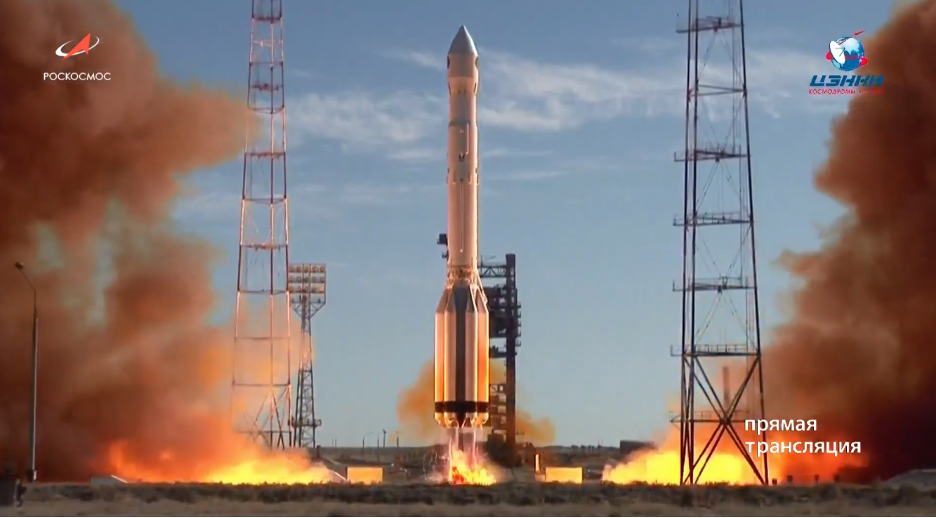}
\caption{Baikonur launch site (Kazakhstan): Proton rocket and the DM-03 upper stage with the \srg\ spacecraft.}
\label{fig:launch}
\end{figure}

\begin{figure}
\centering
\includegraphics[width=0.98\columnwidth,clip]{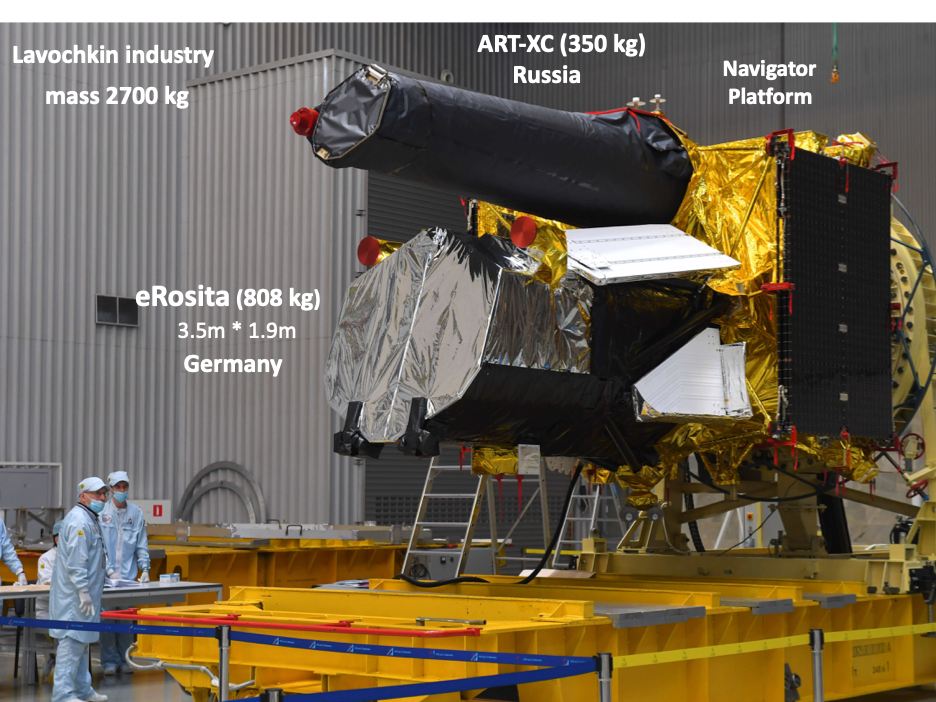}
\caption{\srg\ orbital observatory with the folded solar panels in NPO Lavochkin's assembly hall before shipment to Baikonur.}
\label{fig:srg_npol}
\end{figure}

\begin{figure}
\centering
\includegraphics[width=0.98\columnwidth,clip]{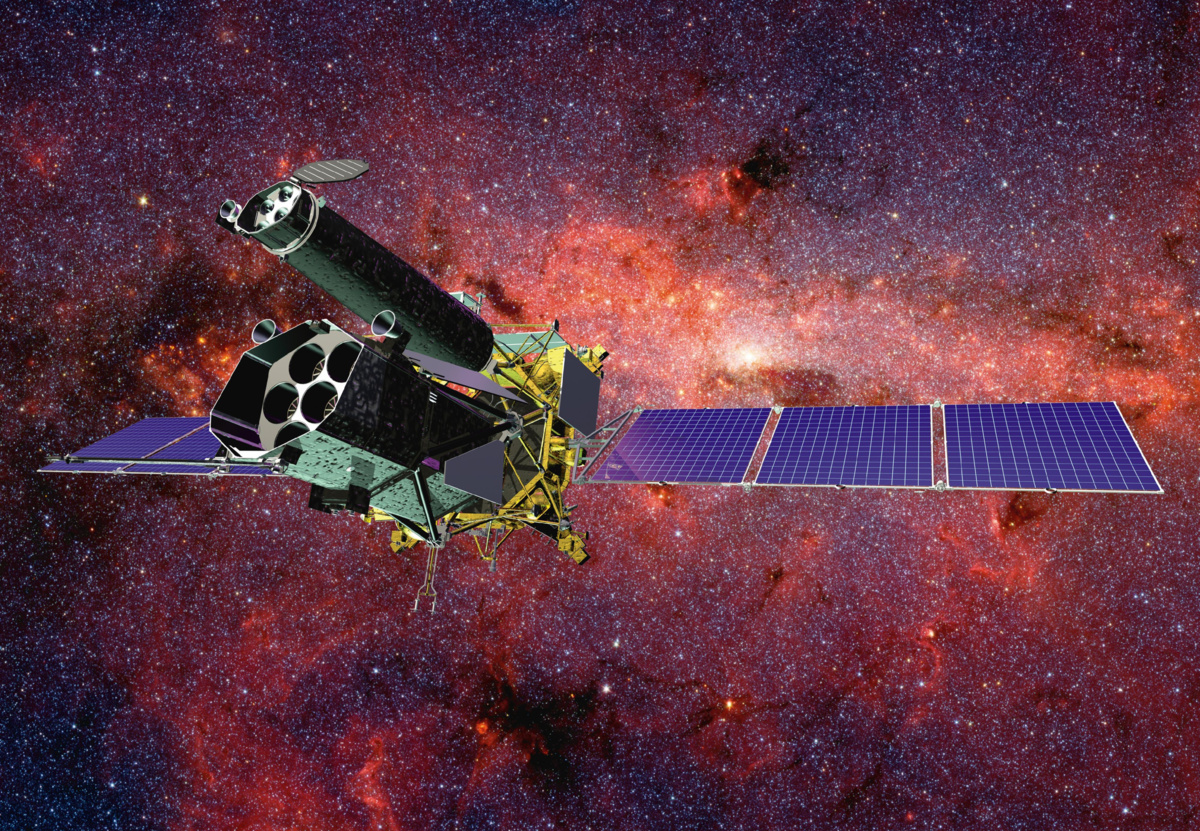}
\caption{\srg\ observatory in flight (artist's impression). Each X-ray telescope consists of seven independent mirror modules.}\label{fig:flight}
\end{figure}

The \textit{Spectrum-Roentgen-Gamma} (\srg) orbital observatory\footnote{\url{http://srg.cosmos.ru}} was launched into a halo orbit around the Lagrange L2 point of the Sun--Earth system on July 13, 2019, from the Baikonur Cosmodrome by a Russian Proton rocket with a DM-03 booster (Fig.~\ref{fig:launch}). The Navigator platform (total mass 2712\,kg), developed by NPO Lavochkin (NPOL) in Khimki near Moscow, carries a scientific payload (total mass 1170\,kg) consisting of two X-ray telescopes with grazing-incidence optics (Figs.~\ref{fig:srg_npol} and \ref{fig:flight}): extended Roentgen Survey with an Imaging Telescope Array (\erosita, \citealt{Predehl_2020}), developed in the Max Planck Institute for extraterrestrial Physics (MPE), Germany, and the Mikhail Pavlinsky ART-XC (Astronomical Roentgen Telescope -- X-ray Concentrator), developed in Russia \citep{pavlinsky21}. This scheme is implemented in accordance with a memorandum signed in 2007 between the Russian Space Agency (at that time) Roscosmos and the German Aerospace Agency, Deutsches Zentrum f\"ur Luft- und Raumfahrt (DLR). The \erosita\ telescope is sensitive to X-rays in the energy range from {200\,eV} to 8\,keV, and ART-XC is sensitive at 4--30\,keV. 

The Navigator platform has been developed as a universal medium-class platform for scientific and meteorological missions to be launched into various orbits. Since January 2011, the Navigator platform has been used in the three {\it Elekro-L} meteorological satellite missions, as well as in the scientific {\it Spektr-R} mission ({\it RadioAstron}, \citealt{Kardashev2013}), which was launched in 2011 and operated until 2018.

\begin{figure}
\centering
\includegraphics[width=0.98\columnwidth,clip]{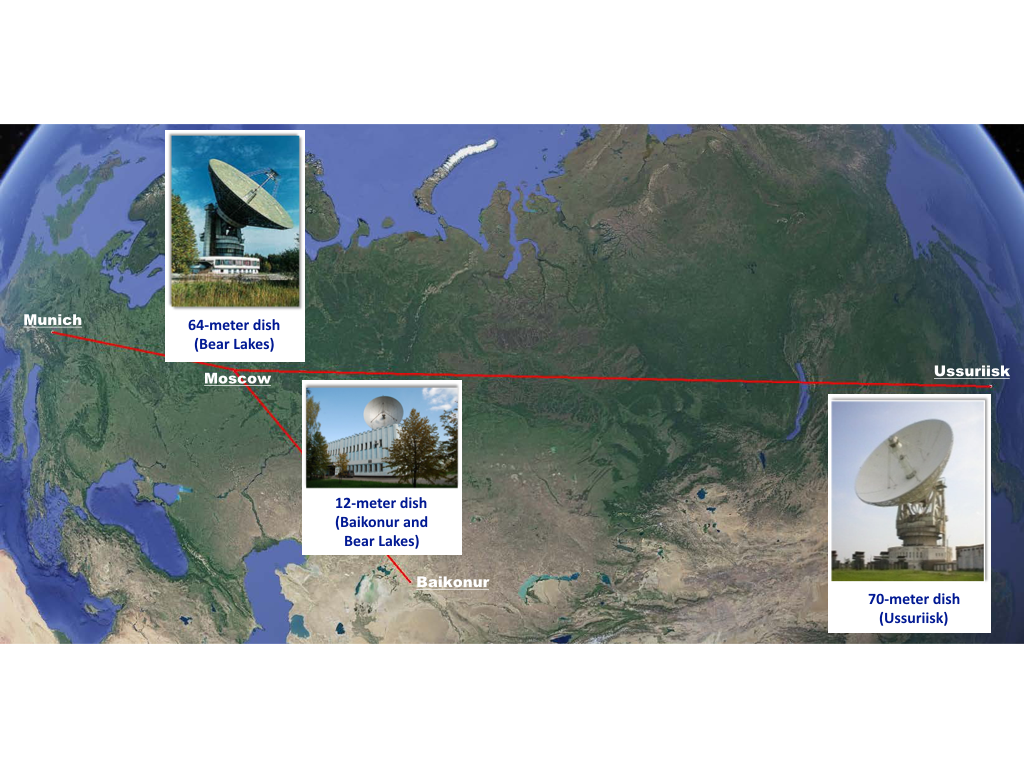}

\caption{{Russian Deep Space Network for the \srg~ mission.}
}

\label{fig:map_with_antennae}
\end{figure}

The \srg\ spacecraft is operated by NPOL's Control Center, while the responsibility for operation of the X-ray telescopes lies with the MPE in Garching near Munich (\erosita) and the Space Research Institute (IKI) of the Russian Academy of Sciences in Moscow (\art). { The downlink and uplink of spacecraft telemetry and commanding data is performed by the Centers of Deep Space Communications in Bear Lakes (64 m diameter antenna) near Moscow and Ussuriysk (70 m diameter antenna) in the Russian Far East as well as Baikonur (12 m antenna) placed in Kazakhstan Fig.~\ref{fig:map_with_antennae}.}


The daily command uplink and scientific data downlink take about one hour for \art\ and from 2.5 to 4 hours for \erosita\ (including health checks of the telescopes' detectors and subsystems). The uplink of commands to the spacecraft and telescopes is performed using the antennas mentioned above or additional 12m diameter antennas in Baikonur and Bear Lakes. 

The \srg\ orbit allows observations to be conducted around the clock with just short breaks for uploading and implementation of operating commands for the Navigator platform and the telescopes. The Navigator platform and both telescopes have onboard mass memory, allowing them to store the data accumulated over several days and dump them to the receiving stations during daily communication sessions. During downlink, the telescopes can (in most cases) continue observations and accumulation of data. 

The \erosita\ and \art\ telescopes each have seven independent mirror systems and seven independent positionally sensitive detectors in their focal planes (see Fig.~\ref{fig:artxc} for \art\ and Fig.~\ref{fig:erosita} for \erosita), six of which are coaligned on a hexagon surrounding an identical central mirror system and detector. To focus X-rays, both telescopes use grazing incidence optics \citep{Wolter1952a,Wolter1952b}. Figure~\ref{fig:wolter} demonstrates the Wolter I optical scheme of each of the seven \art\ telescope modules, consisting of 28 pairs of coaxial parabolic and hyperbolic mirror shells, and a detector at the focus of each.

During the 100-day flight of \srg\ to the L2 point, {a thorough check-up of subsystems of the Navigator platform and both  telescopes as well as calibrations of the telescopes} were carried out. An extensive program of observations of known point and diffuse X-ray sources with various spectral and timing characteristics was carried out during the flight to L2 to {calibrate the alignment of the optical axes of the 14 mirror systems of the telescopes (and their alignment with the axes of the satellite, of the optical star trackers, and of the \erosita\ and \art\ telescopes)} and to calibrate the imaging and spectral characteristics of the detectors as well as parameters relevant for timing studies of X-ray sources. 

The main operation regime of the \srg\ orbital observatory is to scan the entire sky in X-rays with the goal of constructing all-sky X-ray maps in several energy bands. A detailed catalog of compact and diffuse X-ray sources in our Galaxy and the extragalactic Universe is created thereby. 

It was decided several years prior to launch that the first four years of the \srg\ mission after the satellite's arrival at the operational orbit near L2 will be devoted to an all-sky survey through eight consecutive scans of the entire sky. This will allow the observatory to monitor the time variability of numerous X-ray sources: its telescopes will for each sufficiently bright source {provide} a light curve consisting of eight data points separated by half a year. The total number of these bright sources may reach many tens of thousands at energies between 300\,eV and 2\,keV and several hundreds at energies above 4\,keV. Among the most interesting X-ray transients expected to be found by \srg\ in large numbers (for the first time) are tidal disruptions of stars by supermassive and intermediate-mass black holes \citep{Khabibullin_2014,Malyali2019,Jonker2020}.

Furthermore, the telescopes of the \srg\ observatory can trace the ``fast'' variability of X-ray sources during the all-sky survey because each of the eight data points in the long-term light curves will in fact consist of six measurements separated by four hours. This capability should be useful, in particular, for observations of afterglows of gamma-ray bursts (GRBs, \citealt{Khabibullin_2012,Ghirlanda2015,Ascenzi2020}), even if a GRB is directed away from the Sun or obscured by Earth from the view of near-Earth spacecraft (e.g., the {\it Fermi} observatory). 

It is planned that after four years of continuous scanning of the sky and the creation of eight independent X-ray maps of the entire sky, the \srg\ observatory will switch for {at least two years} to a mode of detailed observation of the most interesting X-ray sources in triaxial stabilization. At this stage, long-term scanning of extended X-ray sources and selected regions in the sky with an area of up to 150\,square degrees will also be possible. Table~\ref{tab:srg} summarizes the main parameters of the \srg\ observatory and its telescopes. Further parameters of the \art\ and \erosita\ telescopes are provided in Table~\ref{tab:art_keypar} in \S\ref{s:art} and Tables~\ref{tab:erosita} and \ref{tab:erosita_sens} in \S\ref{s:erosita}, respectively.

\begin{table}
\caption{Overview parameters of the \srg\ observatory and its telescopes.}
\label{tab:srg}
\begin{tabular}{lcc}
\hline
\multicolumn{3}{c}{Observatory} \\
\hline
Orbit & \multicolumn{2}{c}{Halo orbit around L2 point} \\
& \multicolumn{2}{c}{of the Sun--Earth system} \\
Total mass & \multicolumn{2}{c}{2712\,kg} \\
Power consumption & \multicolumn{2}{c}{1700\,W} \\
Telemetry rate & \multicolumn{2}{c}{512\,Kbit\,s$^{-1}$} \\
{Pointing accuracy} & \multicolumn{2}{c}{$4\arcsec$} \\
Observation modes: & \multicolumn{2}{c}{a) all-sky survey} \\
& \multicolumn{2}{c}{b) scanning (up to 150\,sq. deg)} \\
& \multicolumn{2}{c}{c) triaxial pointing} \\
Planned mission duration & \multicolumn{2}{c}{4 years all-sky survey and} \\
& \multicolumn{2}{c}{$>2$ years pointed/scanning obs.} \\
\hline
Telescope & eROSITA & ART-XC \\
\hline
Energy range & 0.2--8\,keV & 4--30\,keV \\
Energy resolution & $\sim 80\,{\rm eV}$ at 1.5\,keV & 9\% at {13.9\,keV} \\
FoV (FWZI) & $\diameter=62\arcmin$ & $\diameter=36\arcmin$ \\
Angular resolution & $30$\arcsec\,HEW & 53\arcsec\,FWHM \\
(during survey) & & \\
Pixel angular size & 9.6\arcsec\ & 45\arcsec\ \\
Sensitivity & $<\mu{\rm Crab}$ & {$\sim 100\mu{\rm Crab}$} \\
(4-year all-sky survey) & & \\
\hline
\end{tabular}

Notes: FWZI -- full width at zero intensity, HEW -- half-energy width, FWHM -- full width at half maximum.

\end{table}

The paper is organized as follows: in Section 2 we present the orbit and operation regimes of the \srg\ observatory. In Section 3 we demonstrate the sensitivities and capabilities of the two X-ray telescopes of the \srg\ and discuss the heterogeneity of the X-ray maps obtained during the first two scans of the entire sky. In Section 4 we describe the scientific goals and first results of the mission. In Sections 5, 6, and 7 we describe the key properties of the Navigator platform and of the \art\ and \erosita\ telescopes. In Section 8 we briefly describe the history of the \srg\ project in Russia and the responsibilities of the German and Russian consortia of scientists in the processing and scientific analysis of the \srg/\erosita\ data coming from two complementary hemispheres of the sky. We finally draw our conclusions.

\section{Orbit and operation regimes of the SRG observatory} 
\label{s:orbit}

\begin{figure}
\centering
\includegraphics[width=0.95\columnwidth,clip]{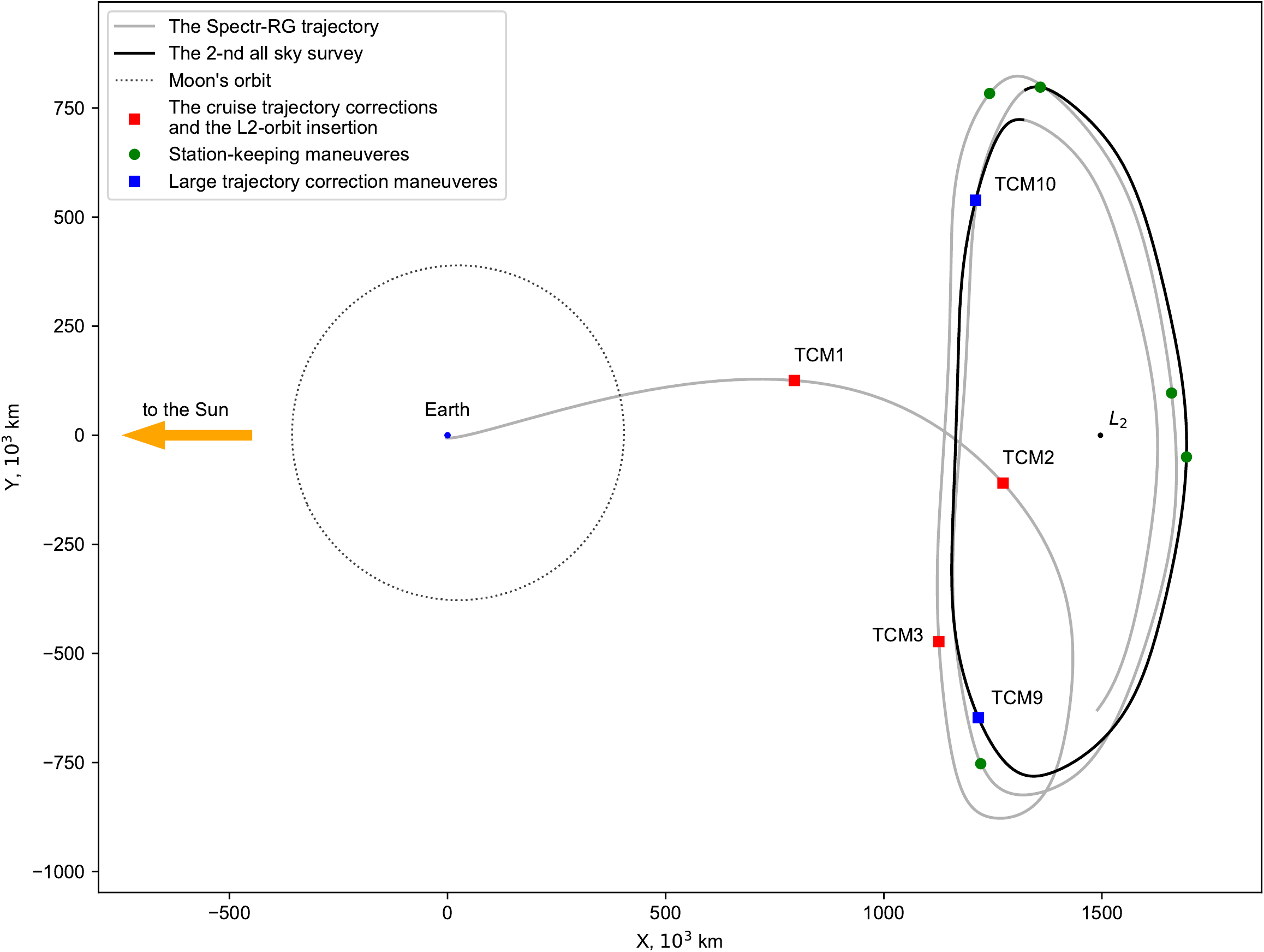}
\includegraphics[width=0.95\columnwidth,clip]{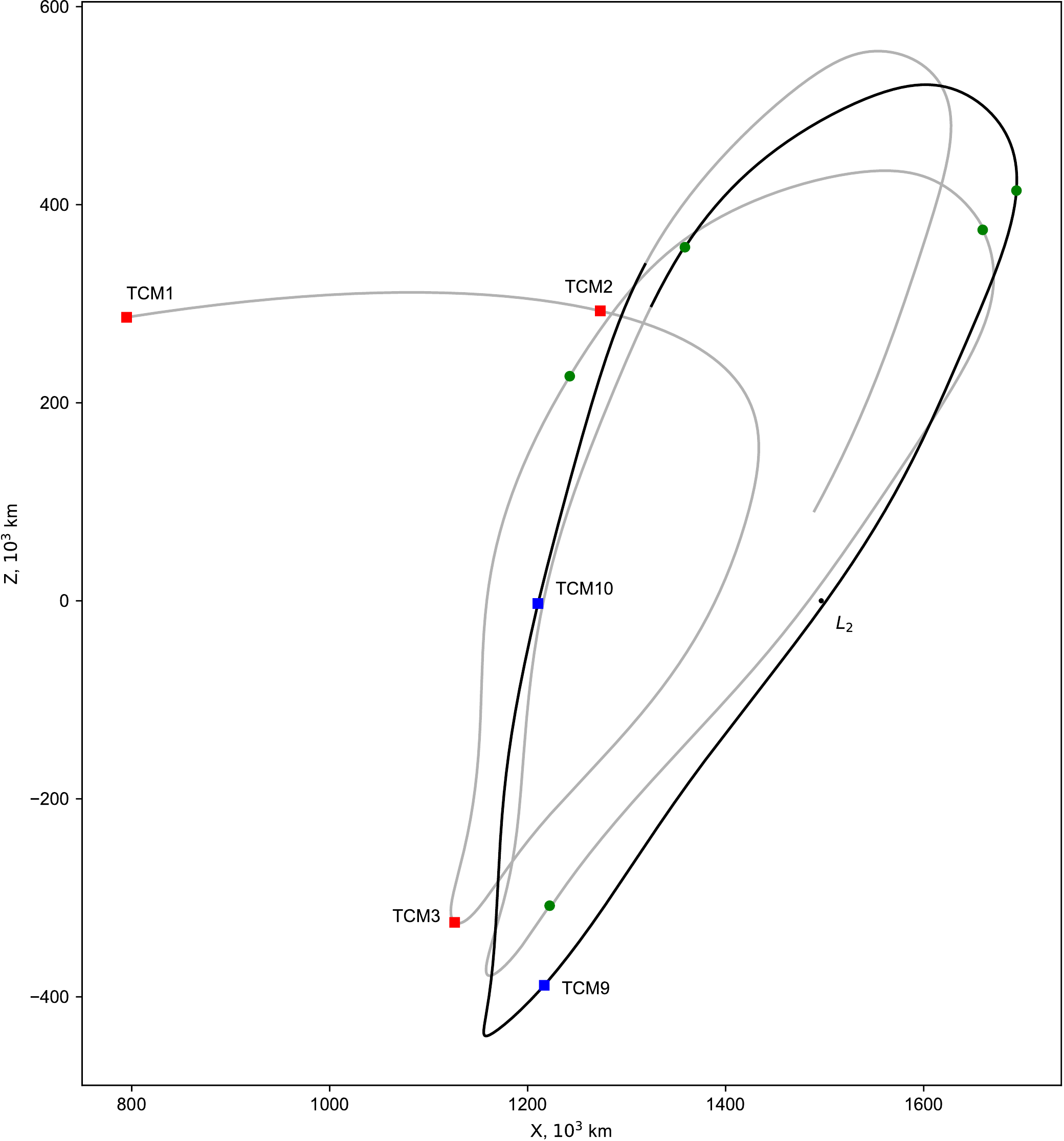}
\includegraphics[width=0.95\columnwidth,clip]{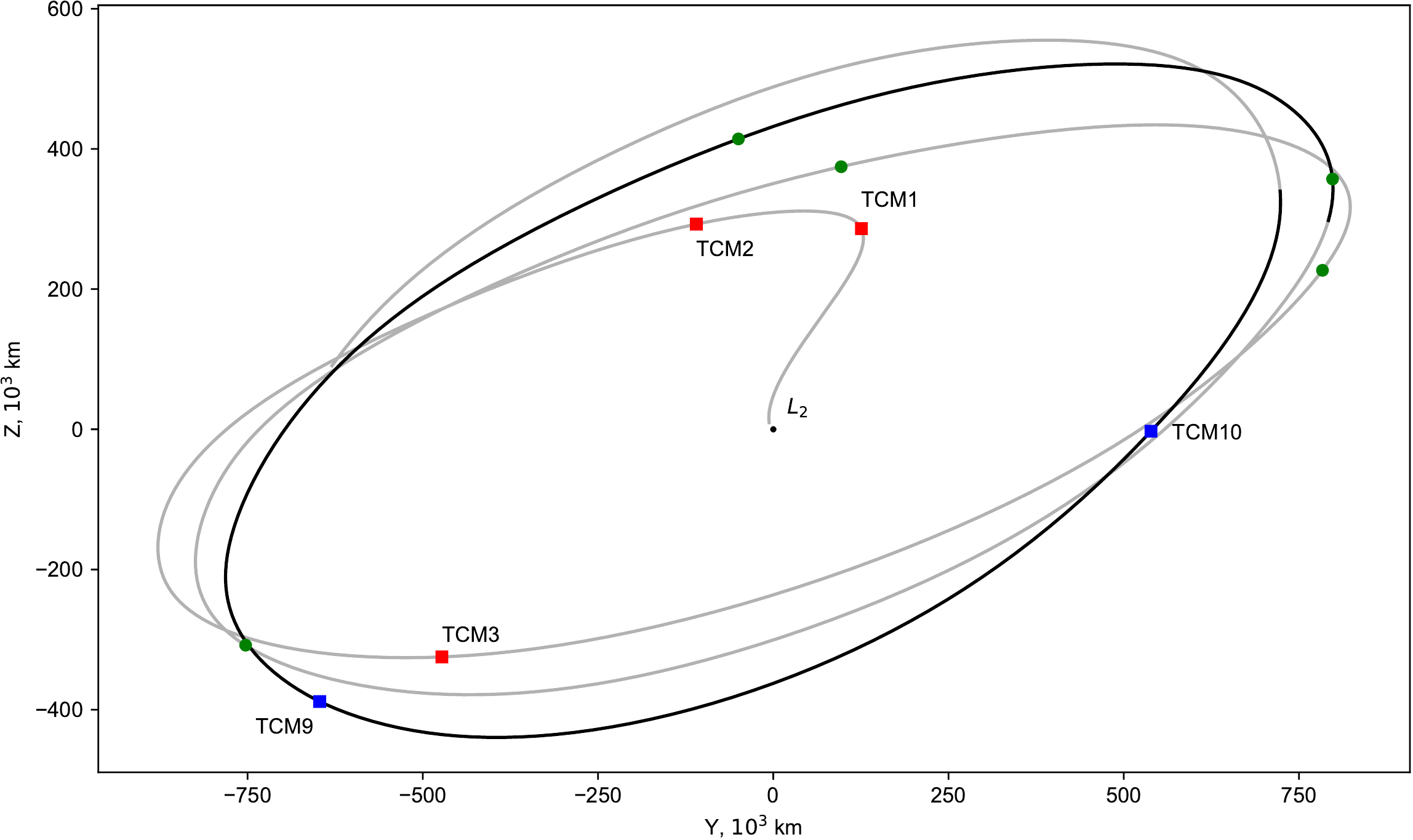}
\caption{Three projections of the SRG orbit onto the ecliptic planes and the scheme of the flight to the L2 point. The dotted line shows the lunar orbit. The points TCM1 and TCM2 denote major trajectory correction maneuvers of the orbit (see §5.4). The point TCM3 denotes the insertion into the quasi-stationary orbit around L2. The green and blue points indicate subsequent minor corrections of the orbit.  Courtesy of the Keldysh Institute of Applied Mathematics of the Russian Academy of Sciences.}
\label{fig:orbit}
\end{figure}

Figure~\ref{fig:orbit} shows the projection of the \srg\ orbit onto the ecliptic plane and the scheme of the flight to the L2 point \citep{Eismont_2020}. Orbital corrections that were implemented before and after entry into the halo orbit around L2 are also indicated. 

\begin{figure}
\centering
\includegraphics[width=0.98\columnwidth,clip]{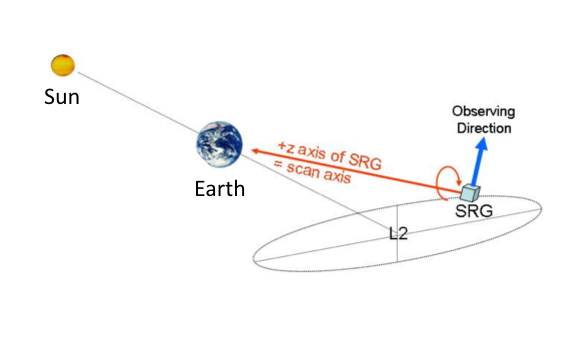}
\caption{Rotation of \srg\ with a period of four hours around the axis pointed at the Sun. This causes scans of large narrow circles in the sky. The survey plane moves slowly, following the Sun at approximately 1 degree per day.}
\label{fig:rotation}
\end{figure}

During the all-sky survey, the satellite rotates around an axis close to the direction toward the Sun with a period of four hours. The rotation axis {gradually} shifts by approximately one degree per day following the motion of the Sun (see Fig.~\ref{fig:rotation}). As a result, the \erosita\ telescope (with its 1 degree field of view, FoV) observes each point source in the sky six times for 30--40\,s over a day, but typically only once every six months. The full FoV of \art\ is 36 {arcmin}, so that a given celestial source is {typically exposed four} times per day for $\sim 20$\,s, also every six months. The variability of the sources in the vicinity of the ecliptic poles could be monitored much longer. The \srg\ observatory observes the entire sky and builds its full map every six months (Fig.~\ref{fig:early}). 

\begin{figure}
\centering
\includegraphics[width=0.98\columnwidth,clip]{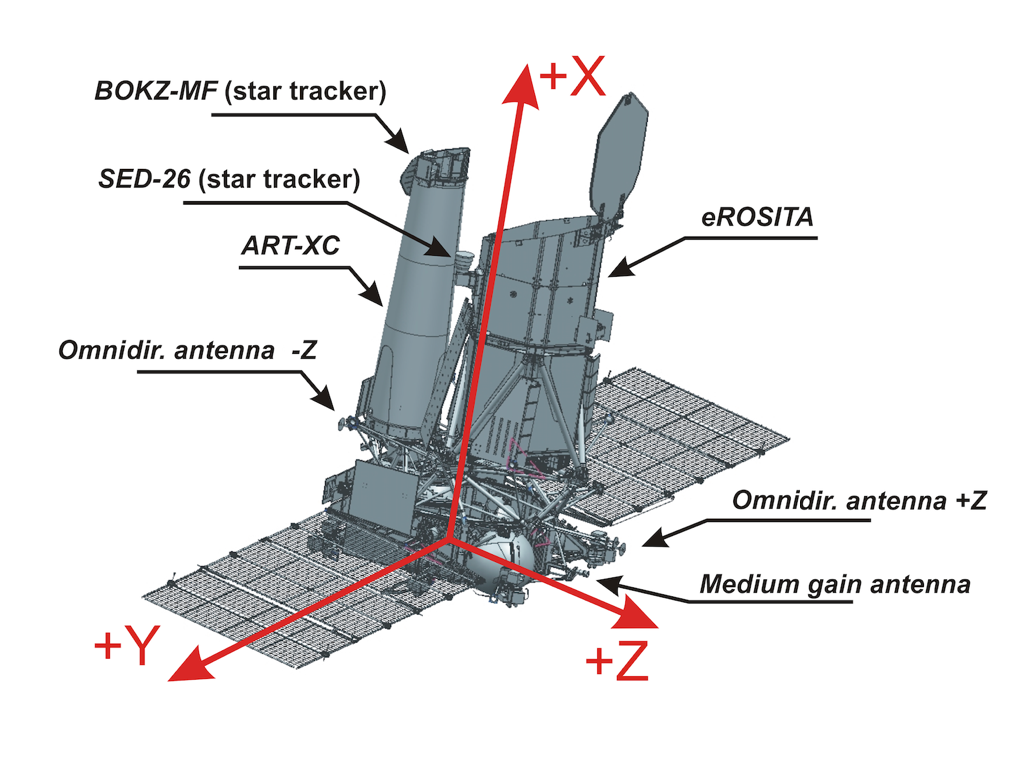}
\caption{\srg\ spacecraft. Positions of the key star trackers, two omnidirectional radio antennas and a medium-gain antenna for transmitting scientific data.}
\label{fig:srg_spacecraft}
\end{figure}

Star trackers are used for the precise orientation of the \srg\ observatory. Both telescopes use star tracker data to refine their pointing accuracy. Fig.~\ref{fig:srg_spacecraft} shows the positions of the main star trackers that are installed on the two telescopes of the observatory, as well as the position of the radio antennas on the Navigator platform.

\subsection{Inhomogeneity of the sky coverage during the survey} 

\begin{figure}
\centering
\includegraphics[width=0.98\columnwidth,clip]{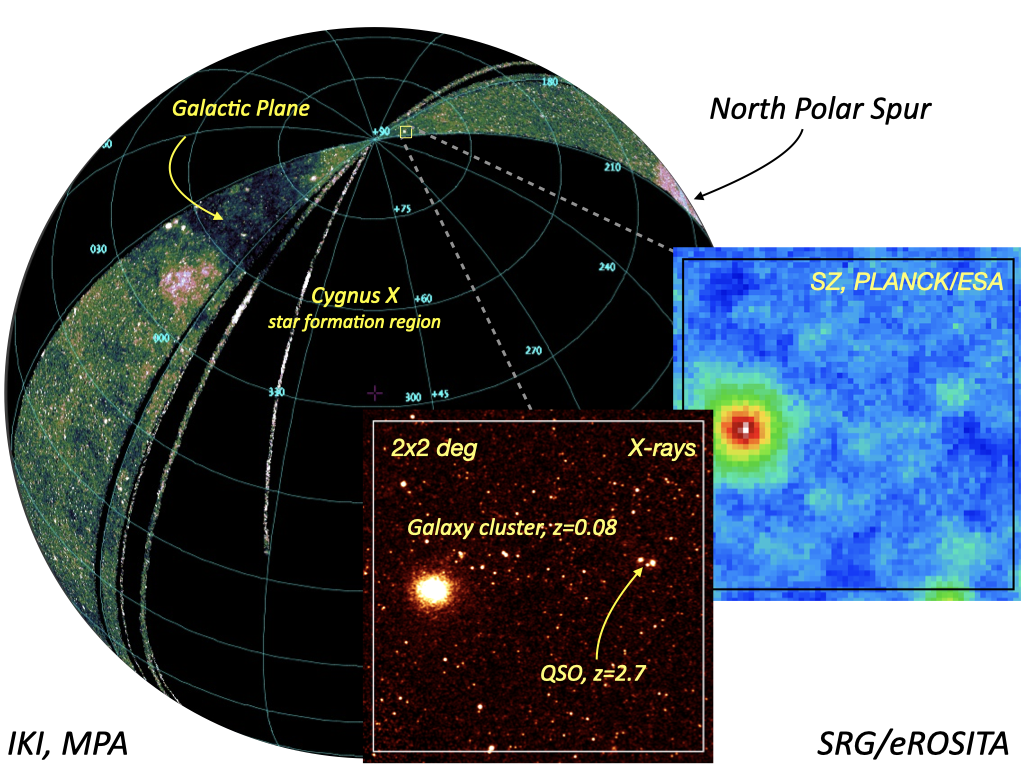}
\caption{Beginning of the \erosita\ all-sky survey. Initial test scans are visible, and a dark stripe shows when there were no observations. Individual scans cross at the north and south ecliptic poles. The inset shows a $2^\circ\times 2^\circ$ region within the deep field near the north ecliptic pole, containing the cluster of galaxies A2255 and a bright quasar.}
\label{fig:early}
\end{figure}

The adopted strategy of the \srg\ survey leads to the appearance of deep fields around the north and south ecliptic poles in the sky map (see Fig.~\ref{fig:early}), where the large circles of all individual scans cross. As a result, the exposure time depends on the ecliptic latitude $\theta$ as $1/\cos\theta$, that is, the exposure is shortest at the ecliptic equator and longest at the ecliptic poles.

\begin{figure*}
\centering
\includegraphics[width=0.95\textwidth,clip]{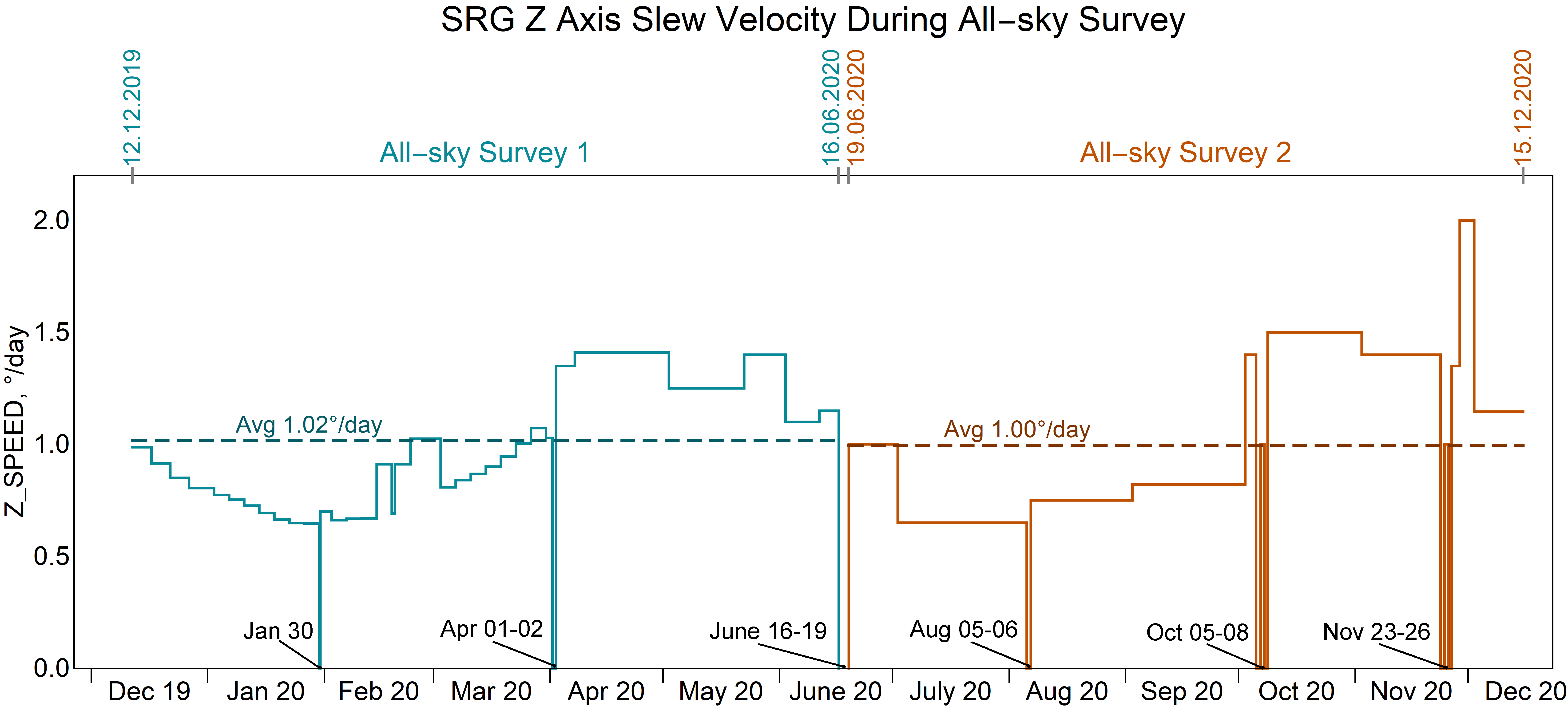}
\caption{Rotation speed of the Z-axis of the \srg\ spacecraft in the ecliptic plane during the first two sky surveys. The spacecraft spins around the Z-axis with a period of four hours, thus performing the all-sky survey. Due to limitations on orientation of the spacecraft, the angle between the Z-axis direction and the Sun and Earth must not exceed 13 and 24 degrees, respectively. This condition defines the speed with which the Z-axis has to follow the Sun and Earth at any point in the halo orbit around L2. Variation in the speed of the Z-axis causes variations in the exposure of the survey. Higher speed leads to lower exposures, and lower speed allows higher exposures of the corresponding sky regions. The vertical bars mark the dates of the gaps in the all-sky survey that are associated with orbit corrections and calibrations.} 
\label{fig:srg_scan_speed}
\end{figure*}

\begin{figure}
\centering
\includegraphics[width=0.98\columnwidth,clip]{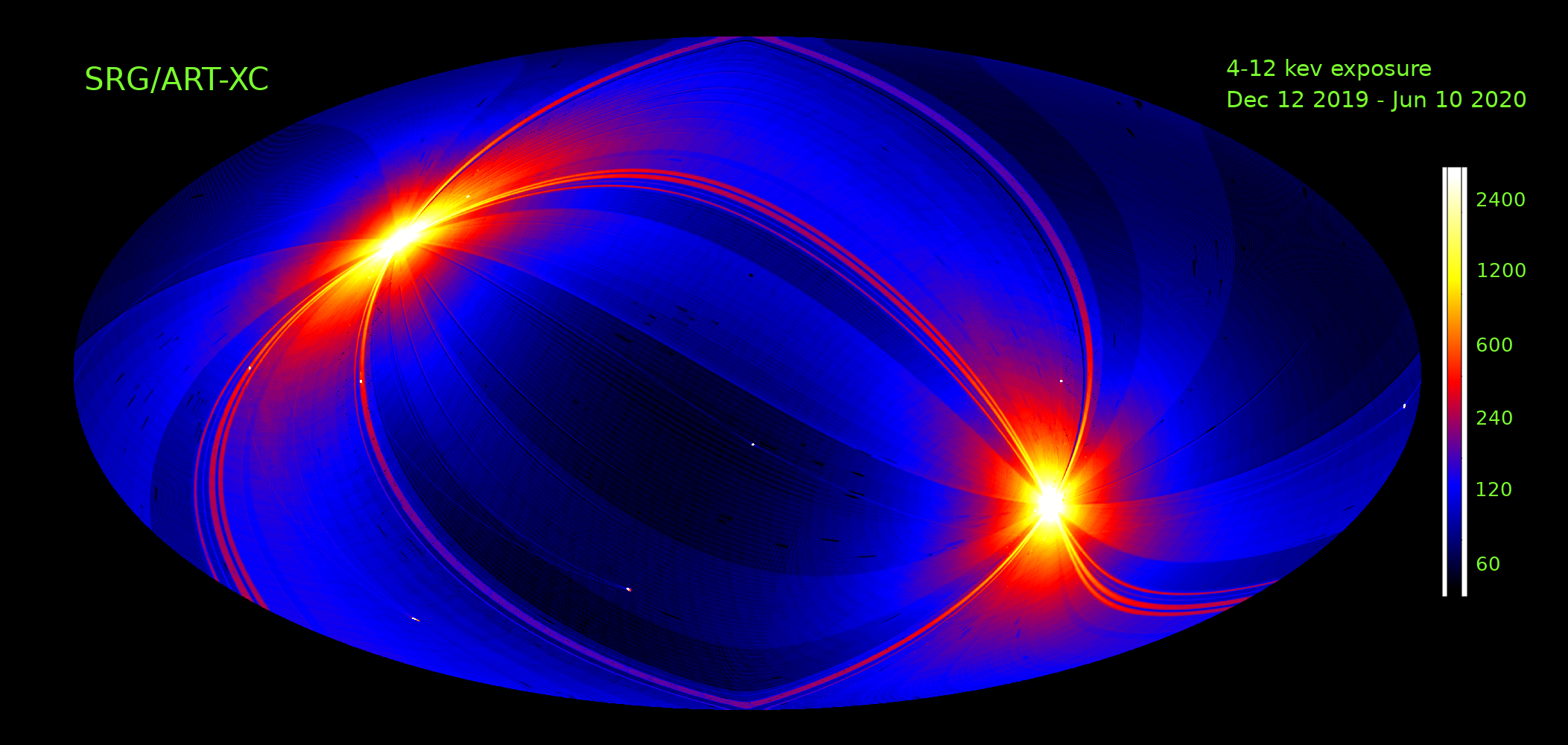}
\caption{Exposure map (in Galactic coordinates) of the first all-sky survey by the \art\ telescope. Exposure time is given in seconds (see the color scale shown on the right). The white spots in the map correspond to the ecliptic poles. Exposure times for \erosita\ are approximately three times longer than for \art,\ in accordance with the instruments' FoV (36\arcmin\ and 1.03\arcdeg, respectively).} 
\label{fig:exposure}
\end{figure}

\begin{figure}
\centering
\includegraphics[width=0.98\columnwidth,clip]{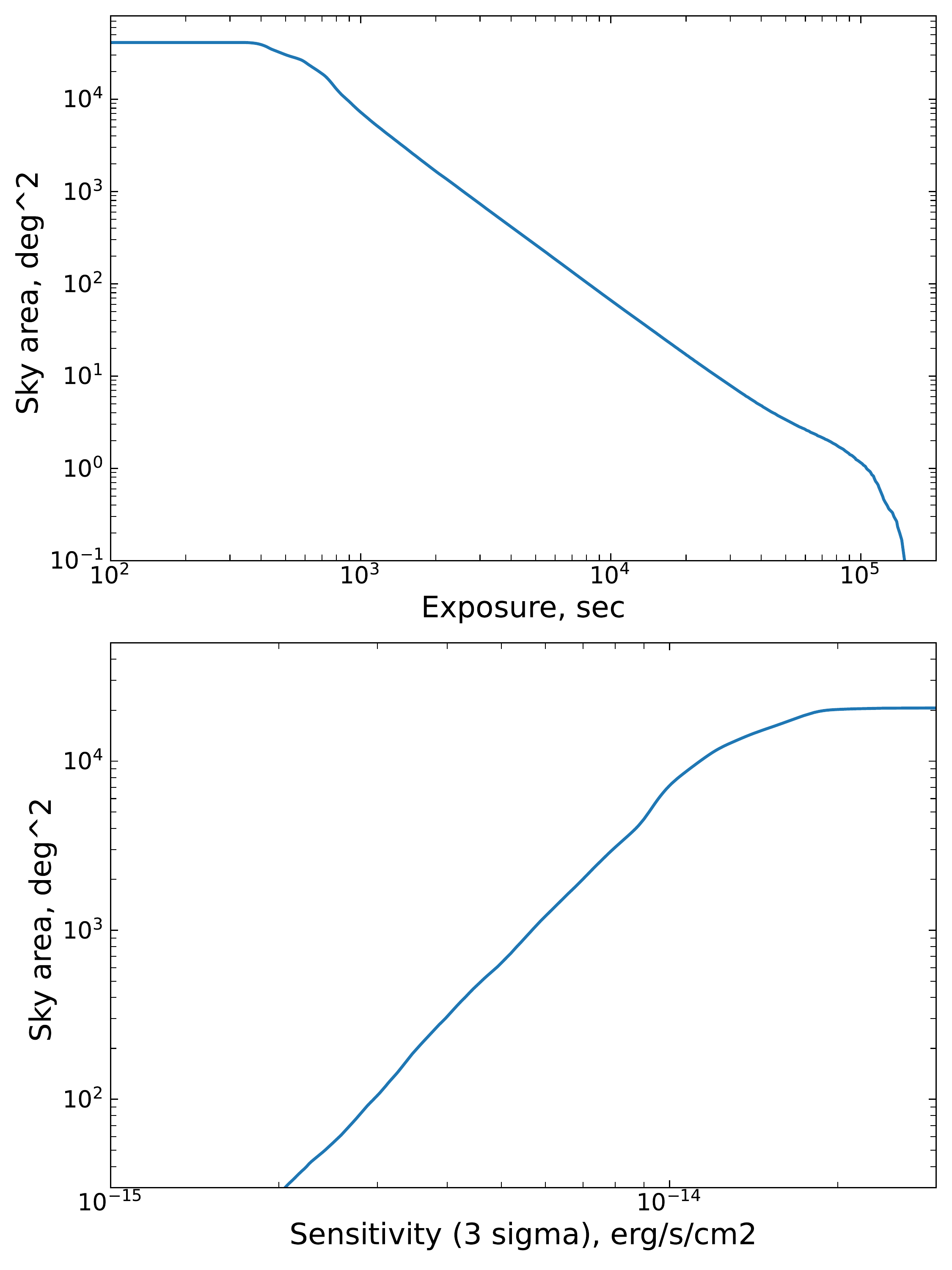}
\caption{Distribution of \erosita\ exposure (upper panel) and sensitivity in the 0.3--2.2\,keV energy band (lower panel) after the first three all-sky surveys. The solid curve in the upper panel shows the area of the sky in which the \erosita\ exposure exceeded the given value. The solid curve in the lower panel shows the area of the sky in the hemisphere analyzed by the Russian consortium in which the $3\sigma$ sensitivity is  better than the given value.} 
\label{fig:erosita_exposure}
\end{figure}

During the first two all-sky surveys, additional inhomogeneity in the sky exposure results from the \srg\ elongated (in the ecliptic plane) halo orbit around L2. The distance to the Sun varies significantly over six months (see Fig.~\ref{fig:orbit}). The satellite moves ahead of Earth for three months and lags Earth for the next three months, which causes a significant variation in drifting rate of the scanning plane, from 0.7\,deg/day {at the} closest to the Earth segment of the orbit to 1.6\,deg/day at its farthest segment (see Fig.~\ref{fig:srg_scan_speed}). Figure~\ref{fig:exposure} shows the actual exposure map for the first \srg\ all-sky survey based on \art\ data.

Figure~\ref{fig:erosita_exposure} (upper panel) shows the sky area covered by \erosita\ with a given exposure. After the first three all-sky surveys, half of the sky has been covered with an exposure of at least 600\,s, while a total area of $\sim 3$\,square degrees around the ecliptic poles has been covered with an exposure exceeding 50~ks, so that source confusion is already high in these regions. The lower panel of Fig.~\ref{fig:erosita_exposure} shows the corresponding sensitivity achieved by \erosita\ in the 0.3--2.2\,keV energy band after the first three all-sky surveys.

\subsection{Deep scanning of sky fields with a size up to 150\,square degrees}
\begin{figure}
\centering
\includegraphics[width=0.95\columnwidth,clip]{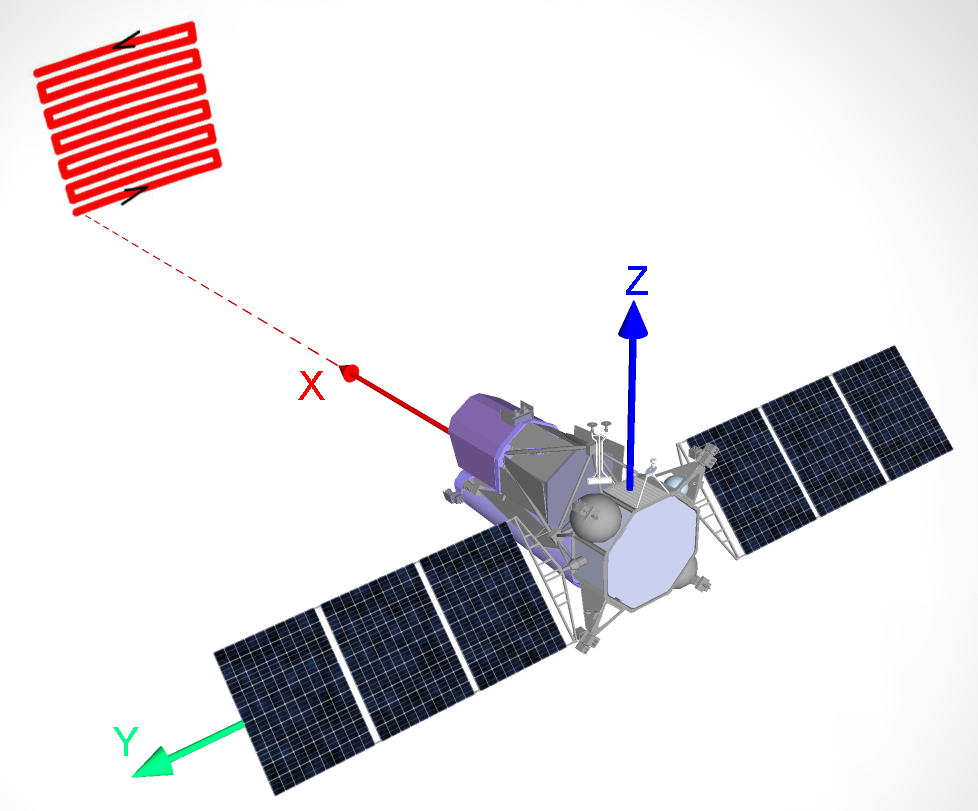}
\caption{Sky visualization of a typical route of the X-axis in scanning mode.} 
\label{fig:scanmode}
\end{figure}
During the final stages of the flight to the L2 point, long observations of a large number of point sources and extended Galactic and extragalactic sources were carried out for calibration (Cal) and performance verification (PV) of the \srg telescopes. In addition, scanning deep-survey observations were performed of a number of extended extragalactic fields, a region of the Galactic X-ray Ridge, and fields in the direction of nearby molecular clouds (to reveal X-ray diffuse emission between them and us). The targets for the PV phase have been chosen by the \art\ science team and the science working groups of the German and Russian \erosita\ consortia (see examples of the maps obtained during these observations in \S\ref{s:survey},\ref{s:deep} below).

To perform deep surveys of extended (up to 150\,square degrees) fields, NPOL and IKI proposed and implemented a novel fruitful method of scanning observations, described in \S\ref{s:spacecraft} below and illustrated in Fig.~\ref{fig:scanmode}. {This method provides a much more uniform coverage of the observed field compared to the standard method, which is based on a grid of pointed observations, and enables obtaining high-quality maps of extended astrophysical objects.}

\subsection{Pointed observations}

The Navigator platform also allows observations of chosen targets to be done with the triaxial orientation of the observatory. This regime was successfully tested during flight calibration and PV observations. The pointing precision is described in \S\ref{s:spacecraft} below. 

\subsection{Instrumental background at L2}
\begin{figure}
\centering
\includegraphics[trim=2cm 5cm 0mm 9cm,width=1\columnwidth]{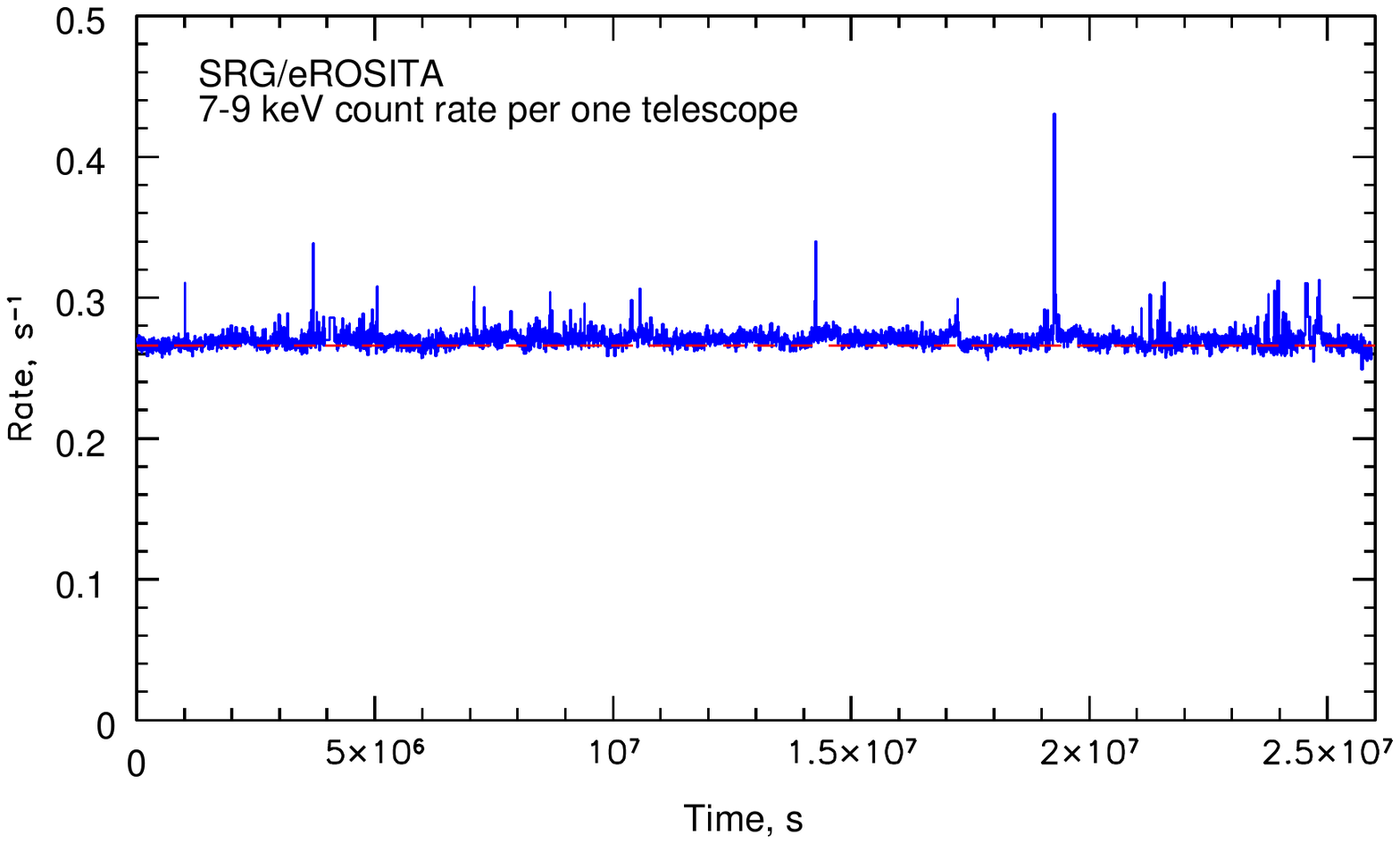}
\caption{\erosita\ count rate (per single detector) in the 7--9\,keV energy band during the first ten months of the all-sky survey in 6000\,s bins. In this energy band, the detector background dominates astrophysical sources. The horizontal dashed red line shows the typical quiescent background, which is globally very stable. Short spikes, whose amplitude rarely exceeds 10-20\%, are plausibly associated with low-energy charged particles, which reach the {CCDs} through the thin entrance filters.} 
\label{fig:bgatl2}
\end{figure}

The particle background recorded by both instruments at L2 has been notably stable during the first 1.5 years of operations, when the solar activity was very modest. This is illustrated in Fig.~\ref{fig:bgatl2}, where the \erosita\ count rate in the 7--9\,keV band is shown. In this energy range, the internal detector background  dominates the astrophysical background. The horizontal dashed red line shows the typical quiescent background. There are rare spikes of counts with a modest amplitude, which are most likely caused by low-energy charged particles reaching the CCD through the thin entrance filter. 

\begin{figure}
\centering
\includegraphics[trim=0cm 5cm 0mm 2cm,width=0.98\columnwidth,clip]{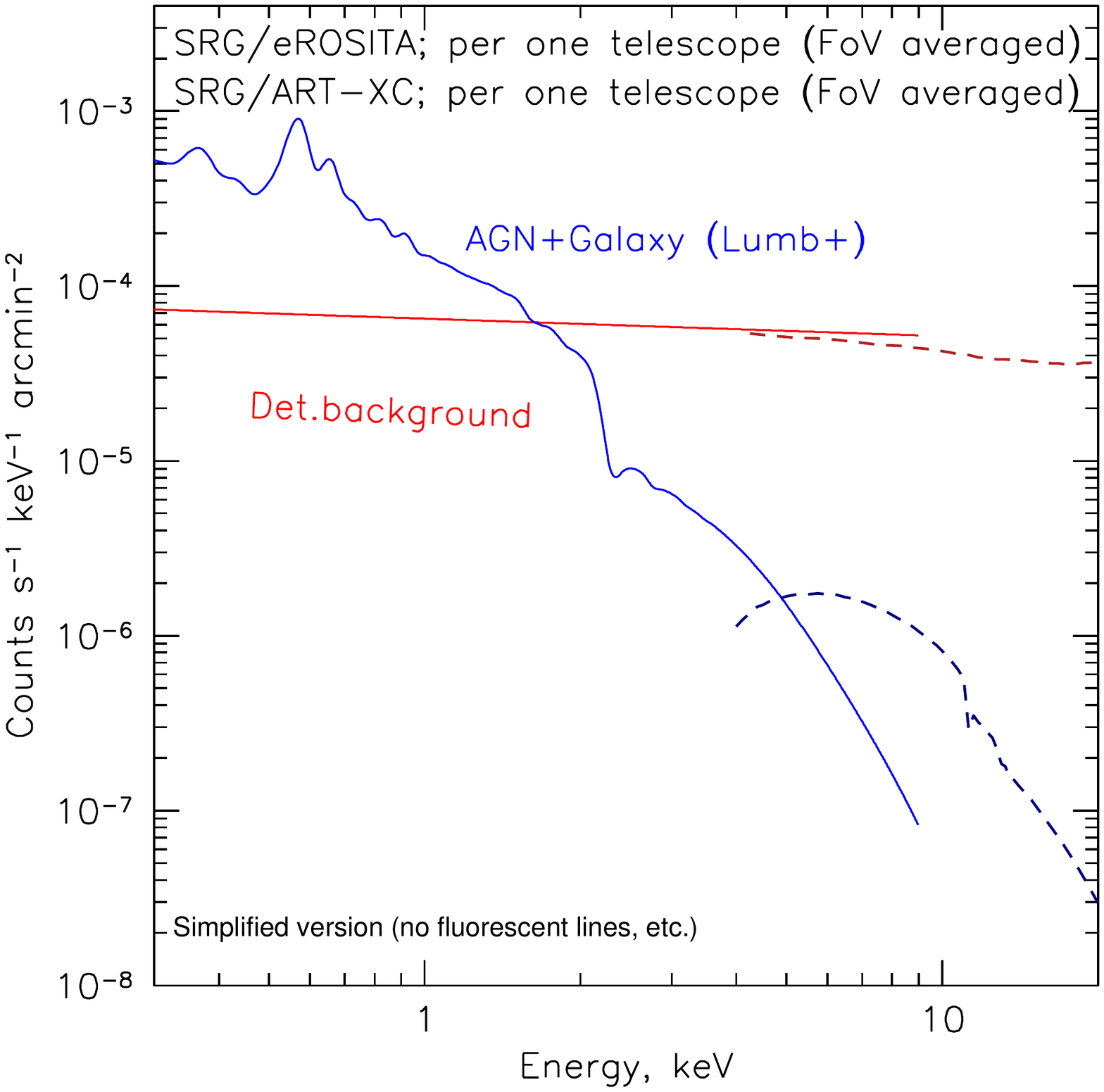}
\caption{Sketch of \art\ (dashed lines) and \erosita\ (solid lines) astrophysical and internal detector backgrounds. 
The blue lines show the estimated sky background using the models of \citet{2002A&A...389...93L} and \citet{1999ApJ...520..124G}, convolved with the telescopes' responses averaged over the FoV. The red lines show the level of the detector internal background (fluorescent lines are not shown). For \erosita, the astrophysical background dominates below $\sim 2$\,keV, while at higher energies, the detector background exceeds the sky background by an order of magnitude.} 
\label{fig:srg_backgrounds}
\end{figure}

The spectrum of the \erosita\ detector background is described in \citet{Predehl_2020}. Here, we show a simplified version of the background model for both instruments (see Fig.~\ref{fig:srg_backgrounds}). In this figure, the blue lines show the (slightly modified) model of sky astrophysical background \citep{1999ApJ...520..124G,2002A&A...389...93L} convolved with the telescopes' responses. For comparison, the red lines illustrate the level of the internal background of the detectors (excluding fluorescent lines). For \erosita, the astrophysical background dominates below $\sim 2$\,keV the astrophysical background, while at higher energies, the detector background is the factor that affects the sensitivity of the telescope. Figure~\ref{fig:srg_backgrounds} also demonstrates that the particle background in both instruments is comparable.

\begin{figure}
\centering
\includegraphics[trim=0cm 0cm 0mm 0cm,width=0.95\columnwidth,clip]{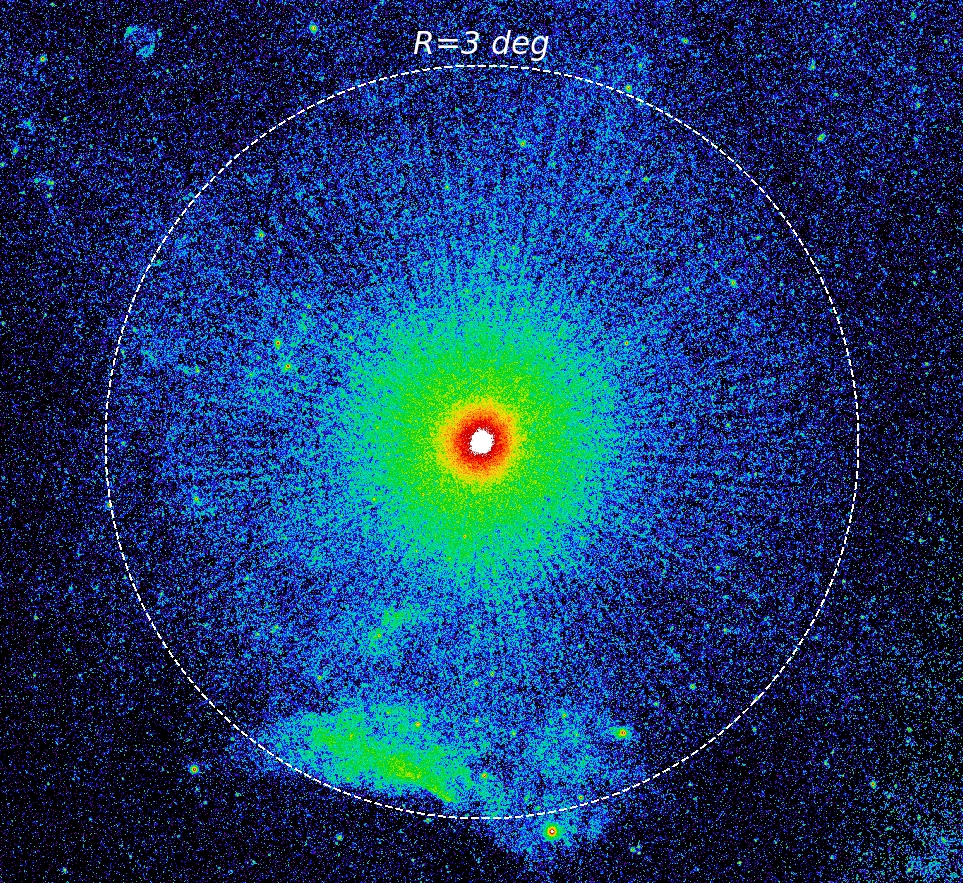}
\caption{\erosita\ image of the region around the extremely bright Galactic black hole binary Cygnus X-1 in the 0.4--4\,keV energy band. The image was saturated in the core to make very faint diffuse structures visible. In particular, the extended halo that is visible up to $\sim 3$ degrees from the source is due to X-ray photons scattered by mirrors only once (stray light). The same effect is largely responsible for a halo around Sco X-1 that is visible in the all-sky map. A special baffle mounted on top of \erosita\ mirrors strongly reduces the magnitude of stray light, so that it is visible only around extremely bright objects. The radial rays that are clearly seen in the image are due to the mirror support structure.}
\label{fig:stray}
\end{figure}

\subsection{Role of stray light}
The \erosita\ telescope optics (Wolter-I scheme) focuses photons that are scattered two times: first by parabolic shells, and then by hyperbolic shells. However, some photons can reach the detector after only one scattering or after two scatterings involving ``wrong scatterings'' , for instance, when they are scattered by the outer surface of a shell. All these photons are collectively called ``stray light''. A special baffle  was introduced to reduce the magnitude of the stray light by an order of magnitude \citep{2014SPIE.9144E..4RF,Predehl_2020}. Nevertheless, a faint and extended halo associated with stray light can be seen when extremely bright sources are observed, such as the X-ray binary Cygnus X-1 (Fig.~\ref{fig:stray}). The role of stray light is more significant in the case of \art\ (see \S\ref{s:art_char} below and \citealt{pavlinsky21} for details).

\section{Sensitivity of the telescopes of the SRG observatory}
\label{s:sens}

\begin{figure}
\centering
\includegraphics[width=0.9\columnwidth,clip]{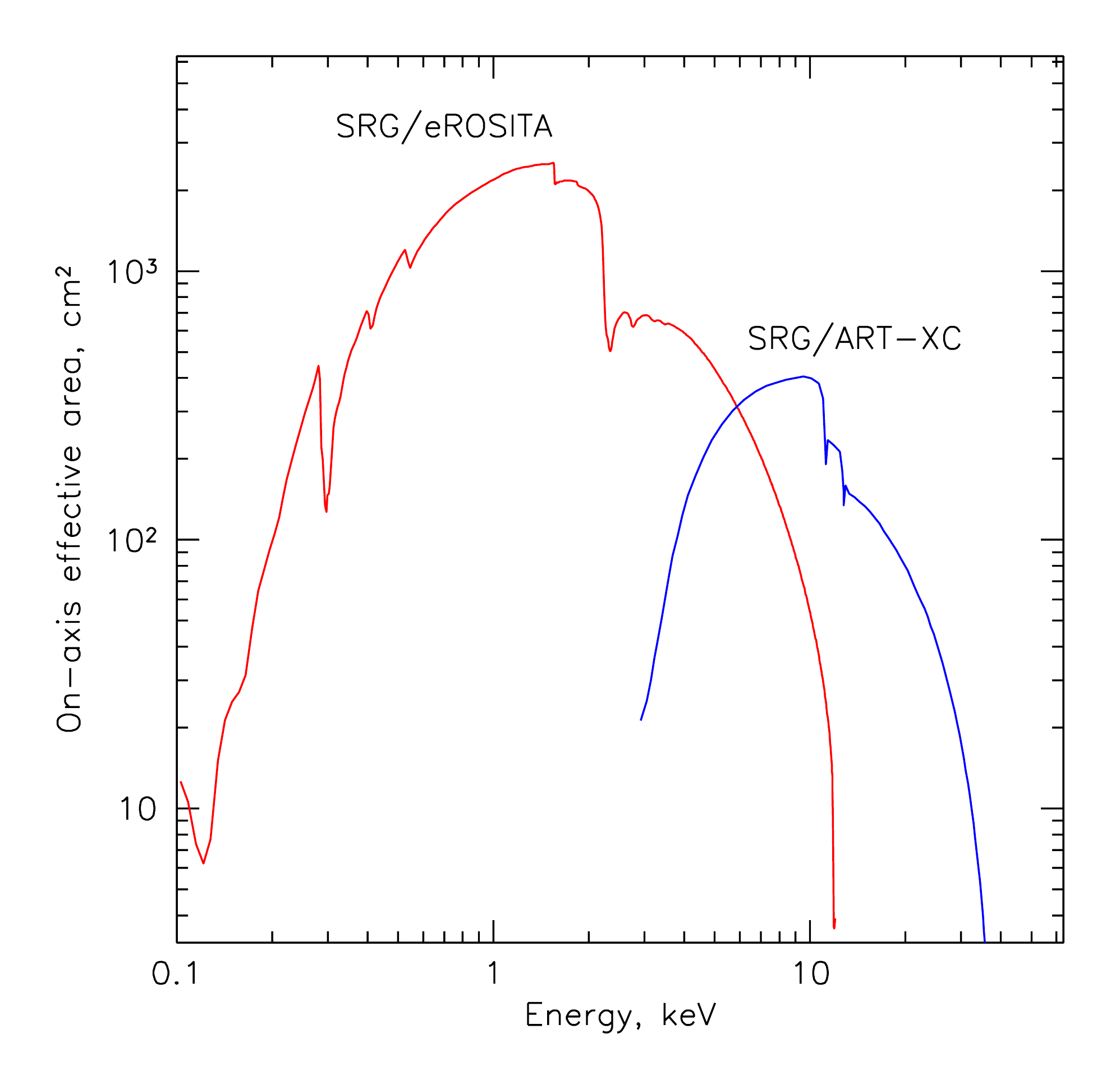}
\caption{On-axis effective area of the {\erosita\ (red) and \art\ (blue)} telescopes of the \srg\ observatory. These curves are based on the pre-flight beam tests and models, which are broadly consistent with the in-flight calibration results.}
\label{fig:effarea}
\end{figure}

Figure~\ref{fig:effarea} compares the on-axis effective areas of the \erosita\ and \art\ telescopes as a function of the energy of registered photons. The \erosita\ and \art\ curves intersect near 5\,keV. The telescopes nicely complement each other, detecting X-ray photons in adjacent energy bands. The results of the first all-sky survey have confirmed the preflight estimates of the effective area of the \srg\ telescopes.

\begin{figure}
\centering
\includegraphics[trim=0cm 5cm 0mm 2cm,width=0.98\columnwidth,clip]{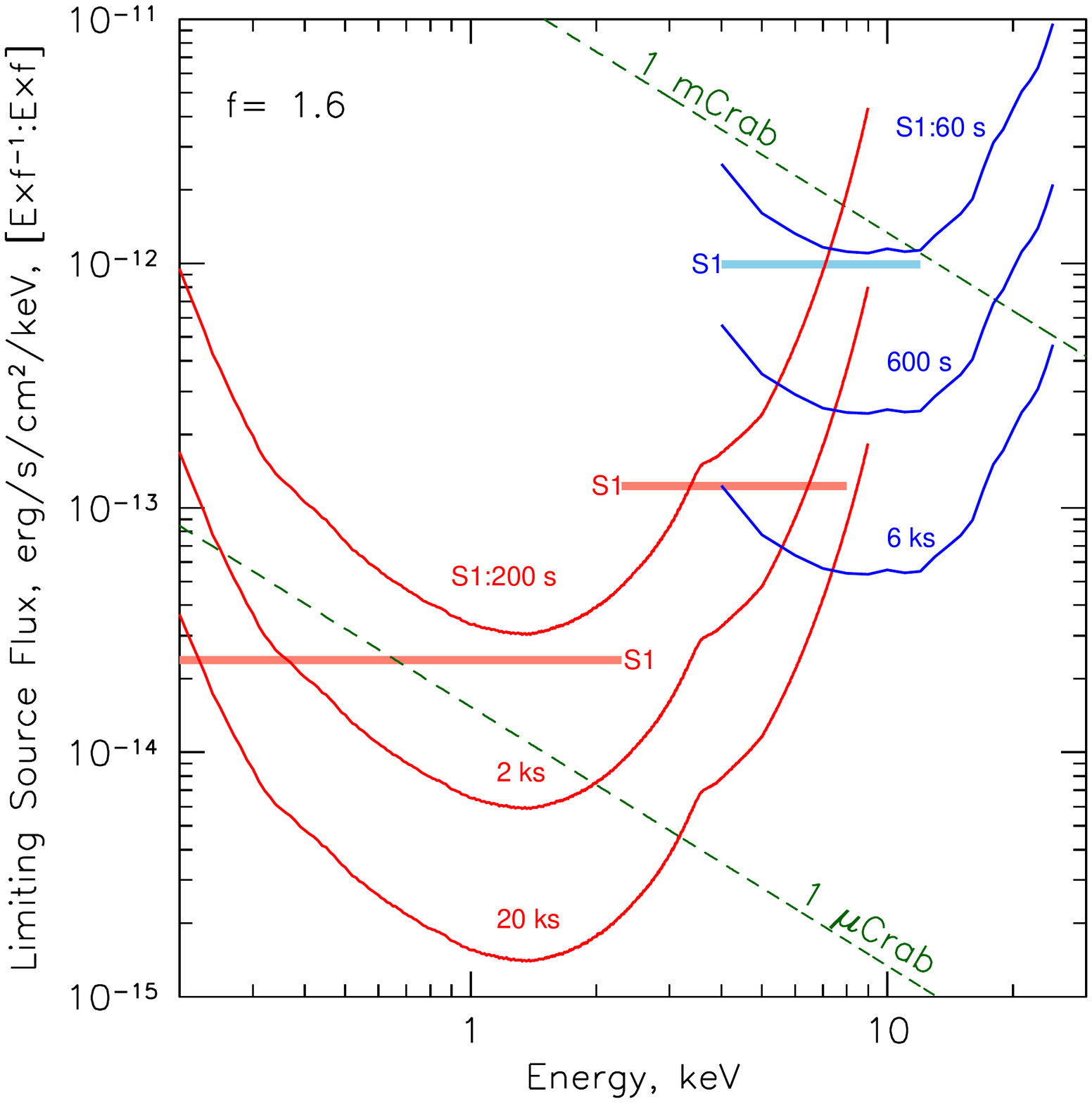}
\caption{Estimated sensitivities of the \art\ (red) and \erosita\ (blue) telescopes for a continuum spectrum in the energy interval $\Delta E \sim E$ as a function of energy. The curves show the sensitivity in units of energy flux density, i.e., ${\rm erg\,s^{-1}\,cm^{-2}\,keV^{-1}}$, evaluated for an interval of energies between $E_1=E/f$ and  $E_2=E\times f$, where $f=1.6$. The top curves, labeled S1, correspond to the sensitivity achieved during the first all-sky survey close to the ecliptic equator, where the exposure time is shortest. The difference in the effective time (200\,s vs 60\,s for \erosita\ and \art,\, respectively) reflects the difference in the FoVs of the two instruments. Also shown are the sensitivity curves for 10 and 100 times deeper exposures. The horizontal bars show the broadband (0.2--2.3\,keV and 2.3--8\,keV for \erosita\ and 4--12\,keV for \art) sensitivities in the first survey, approximately converted into units of flux density. The spectrum of the Crab nebula scaled to a flux of 1\,mCrab and 1\,$\mu$Crab is shown for comparison. 
} 
\label{fig:sensitivity}
\end{figure}

Figure~\ref{fig:sensitivity} shows the sensitivity of the \erosita\ and \art\ telescopes that was achieved during the first half-year survey as a function of energy. The sensitivity has been evaluated {assuming an unabsorbed power-law spectrum with a photon index of 1.8} in the energy interval $\Delta E\sim E$. The in-flight background spectrum (both the intrinsic detector background and unresolved astrophysical sky emission, see \citealt{Freyberg2020} for details) is integrated over this energy interval to predict the image surface brightness. Using this value, we can predict the expected distribution of fluxes (due to Poisson fluctuations of the number of counts) at a given position of an image convolved with the telescope point spread function (PSF). The sensitivity is then estimated as the flux associated with peaks that for a Gaussian distribution would correspond to 5$\,\sigma$ deviations. The corresponding curves are labeled S1. In the survey, the exposure is shortest near the ecliptic equator, where it is 200 and 60\,s for \erosita\ and \art, respectively. The difference in exposure is due to the difference in the solid angles subtended by the two telescopes at any given moment. The horizontal bars show the broadband sensitivities in the first survey, approximately converted into units of flux density. 

\section{Science goals and first results of the SRG observatory}
\label{s:tasks}

The main scientific objective of the \srg\ observatory is the construction of detailed X-ray maps of the sky and catalogs of point sources and extended X-ray sources in different energy bands from 0.3 to 12\,keV. The \art\ and \erosita\ telescopes were developed specifically for performing these tasks. 

It is expected that during the eight all-sky surveys spanning a period of four years, up to four to five million compact X-ray sources will be discovered: about three million active galactic nuclei (AGN), nearly one hundred thousand rich clusters of galaxies, up to a million stars with bright coronae (mostly M dwarfs), and tens of thousands of other Galactic objects, including cooling neutron stars and radiopulsars, accreting neutron stars (in particular, X-ray pulsars and bursters), black holes, and numerous white dwarfs in binary stellar systems. 

Also of interest is the exploration of extended X-ray sources, including supernova remnants (SNRs), pulsar wind nebulae, the hot gas filling most of the interstellar volume near the Galactic plane, rarefied gas in the halo of the Galaxy, and gas fountains in nearby galaxies. In addition, there is interest in the Local Bubble, comets in the Solar System, and in a search for traces of a shock wave at the boundary of the heliosphere, where the solar wind is halted by the interstellar gas surrounding the Solar System. It is also possible to continue the investigation that was begun by {\it ROSAT} \citep{Snowden_1997} of the origin of the Galactic soft X-ray background, which depends on the distribution of molecular and atomic gas and dust in the Galaxy, which absorb the soft X-ray emission. 

As demonstrated by the highly successful {\it ROSAT} spacecraft (all-sky X-ray survey in 1990, \citealt{Voges_1999}), these data are eagerly anticipated and will be widely used by the global astronomical community. Furthermore, the availability of eight all-sky surveys, each lasting six months, should help the \srg\ observatory to discover a huge number of variable compact sources of Galactic and extragalactic origin.  

\subsection{X-ray maps obtained during the first all-sky survey and the CalPV phase}
\label{s:survey}

\begin{figure*}
\centering
\includegraphics[width=0.95\textwidth,clip]{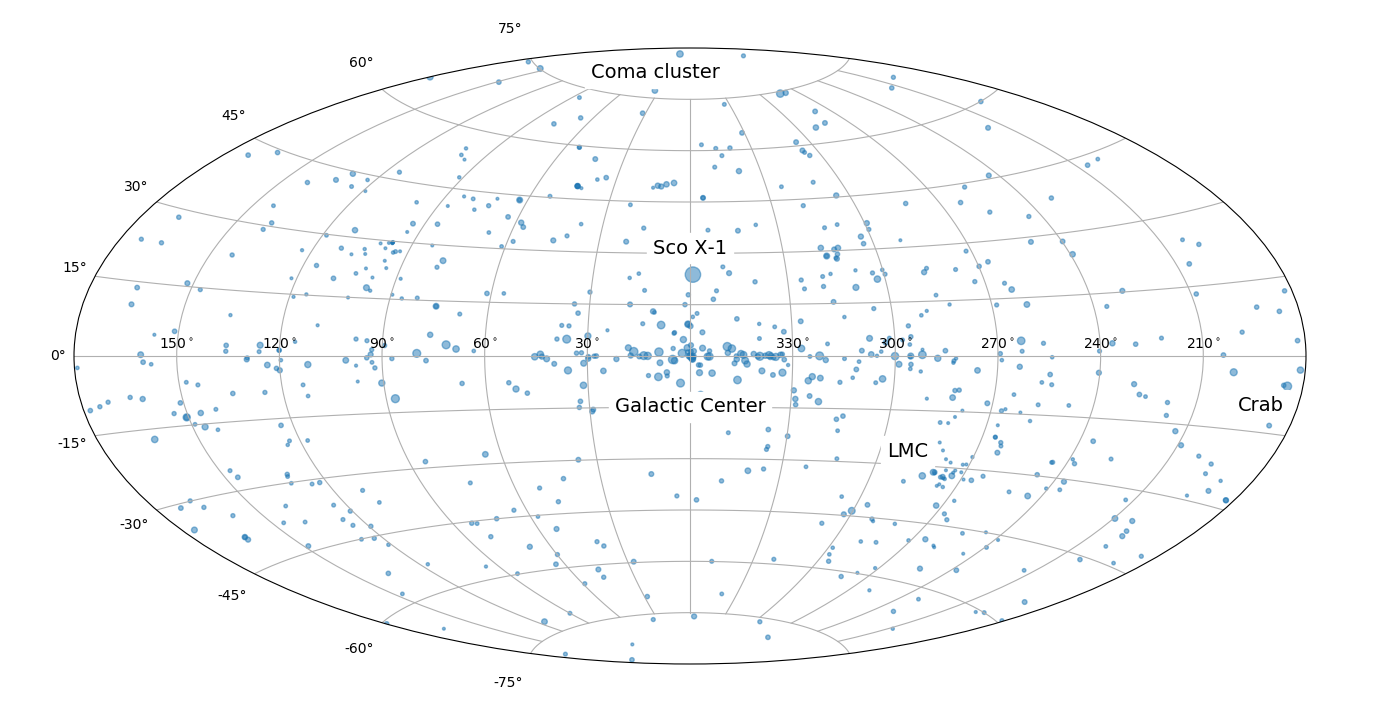}
\caption{Positions (in Galactic coordinates) of the X-ray sources detected by \srg/\art\ during its first all-sky survey (typical exposure per point is just $\sim 60$\,s). Nearly 600 sources have been detected in the 4--12\,keV energy band, $\sim 60$\% of which are Galactic (black holes, neutron stars, white dwarfs, coronally active stars, SNRs, etc.) and $\sim 40$\% are extragalactic (AGN and a few dozen massive clusters of galaxies), as well as a number of newly discovered sources. The symbol size reflects X-ray brightness of the sources.} 
\label{fig:art_sky}
\end{figure*}

\begin{figure}
\centering
\includegraphics[trim=0cm 0cm 0cm 0cm,width=0.48\textwidth,clip]{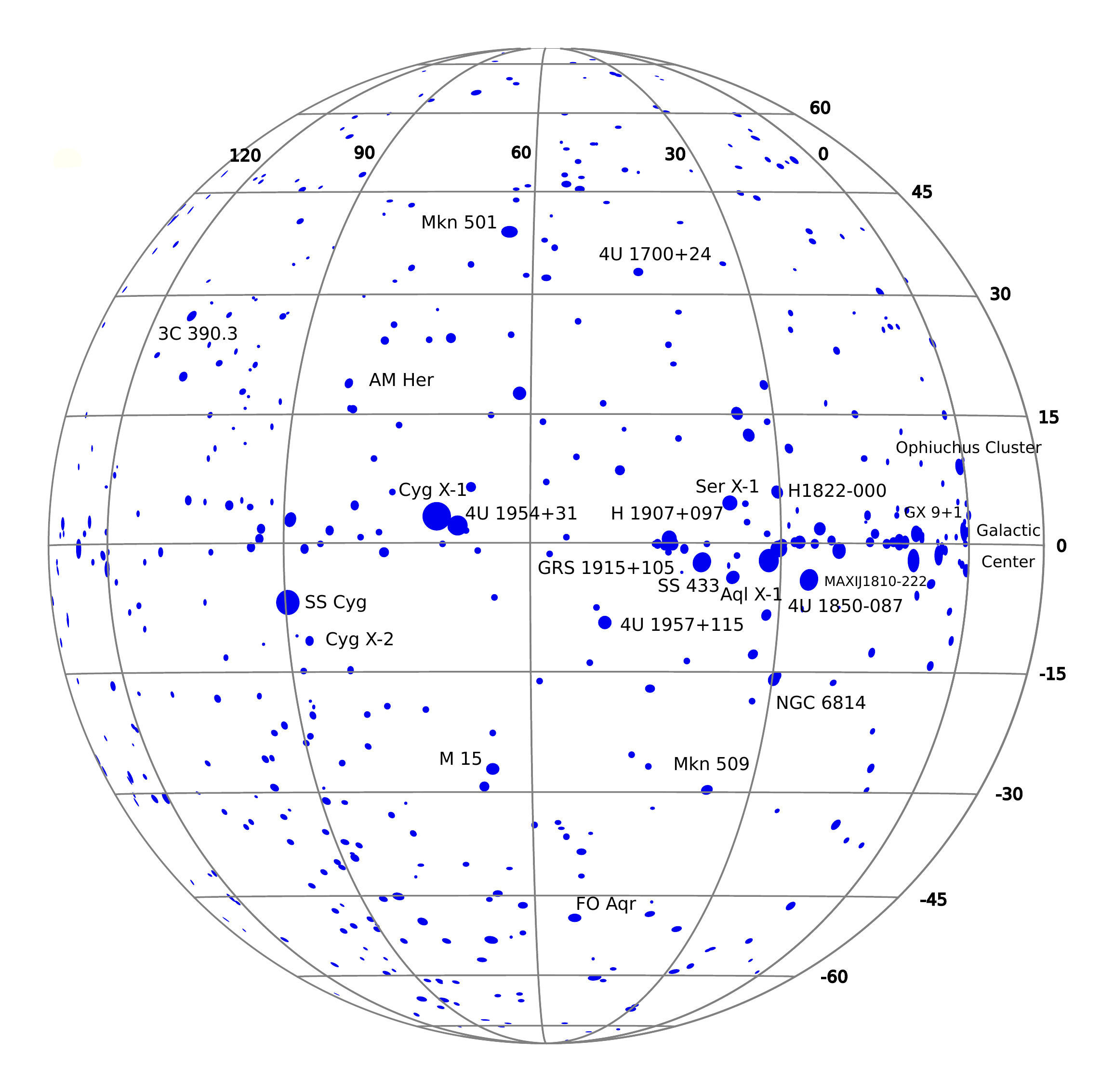}
\caption{Positions (in Galactic coordinates) of X-ray sources detected by \srg/\erosita\ in the 4--8\,keV energy band during the first two all-sky surveys on that half of the sky in which the Russian Consortium is responsible for the data analysis (the Galactic center is on the right side of the image). The image includes $\sim$600 sources, and some of the brightest are labeled.
} 
\label{fig:erosita_48}
\end{figure}

Figure~\ref{fig:art_sky} shows the distribution over the sky of the $\sim 600$ sources detected by \art\ in the 4--12\,keV energy band during its first sky survey (December 12, 2019 -- June 10, 2020). Recently, an updated version of this map has been presented by \cite{Pavlinsky2021}, which is based on the sum of the first two sky surveys (December 2019 -- Dec. 2020). A total of 867 sources (821 point sources and 46 extended sources) have been detected by \art. The 750 sources of known or suspected origin in this catalog include 56\% extragalactic sources (mostly AGN and also 52 rich low-redshift clusters of galaxies), and the rest are Galactic (X-ray binaries, cataclysmic variables (CVs), SNRs, etc.). {From} 114 sources, \art\ has detected X-rays for the first time. Although the majority of them ($\sim 80$) are expected to be spurious given the chosen detection threshold, there are expected to be $\sim 35$ newly discovered astrophysical objects. An ongoing program of optical follow-up observations of these sources has already led to the identification of several new AGN and CVs \citep{zaznobin21,Zaznobin2021b}.

The achieved sensitivity to point sources after the first year of the \art\ all-sky survey varies between $\sim 4\times 10^{-12}$\,erg\,s$^{-1}$\,cm$^{-2}$ near the ecliptic plane and $\sim 8\times 10^{-13}$\,erg\,s$^{-1}$\,cm$^{-2}$ (4--12\,keV) near the ecliptic poles. The typical depth of the \art\ survey is already comparable to that achieved in a similar energy band (4--10\,keV) in the recent {\it MAXI} ({\it Monitor of All-sky X-ray Image}) all-sky survey \citep{Kawamuro_2018} and is just slightly poorer than the sensitivity of the {\it XMM-Newton} Slew Survey in the 2--12\,keV band \citep{Saxton_2008}. However, the \art\ survey greatly improves on the former in terms of angular resolution and provides full and regular sky coverage in contrast to the latter. The \art\ survey sensitivity will be growing during the mission as a result of increasing exposure and probably also due to a decreasing flux of Galactic cosmic rays {as the next solar maximum approaches}. As many as $\sim 5000$ sources can be found by \art\ by the end of the four-year all-sky survey.

As noted above, the \erosita\ telescope is noticeably inferior in sensitivity to the \art\ telescope at energies above 5--6\,keV, but between 4 and 5\,keV, \erosita\ is more sensitive. As shown in Fig.~\ref{fig:erosita_48}, during the first two surveys of the sky, \erosita\ detected over 600 X-ray sources in one hemisphere in the range from 4 to 8\,keV. This number is comparable to the number of sources detected by \art\ over the entire sky during one survey in the 4--12\,keV range.

\begin{figure*}
\centering
\includegraphics[width=1.9\columnwidth,clip]{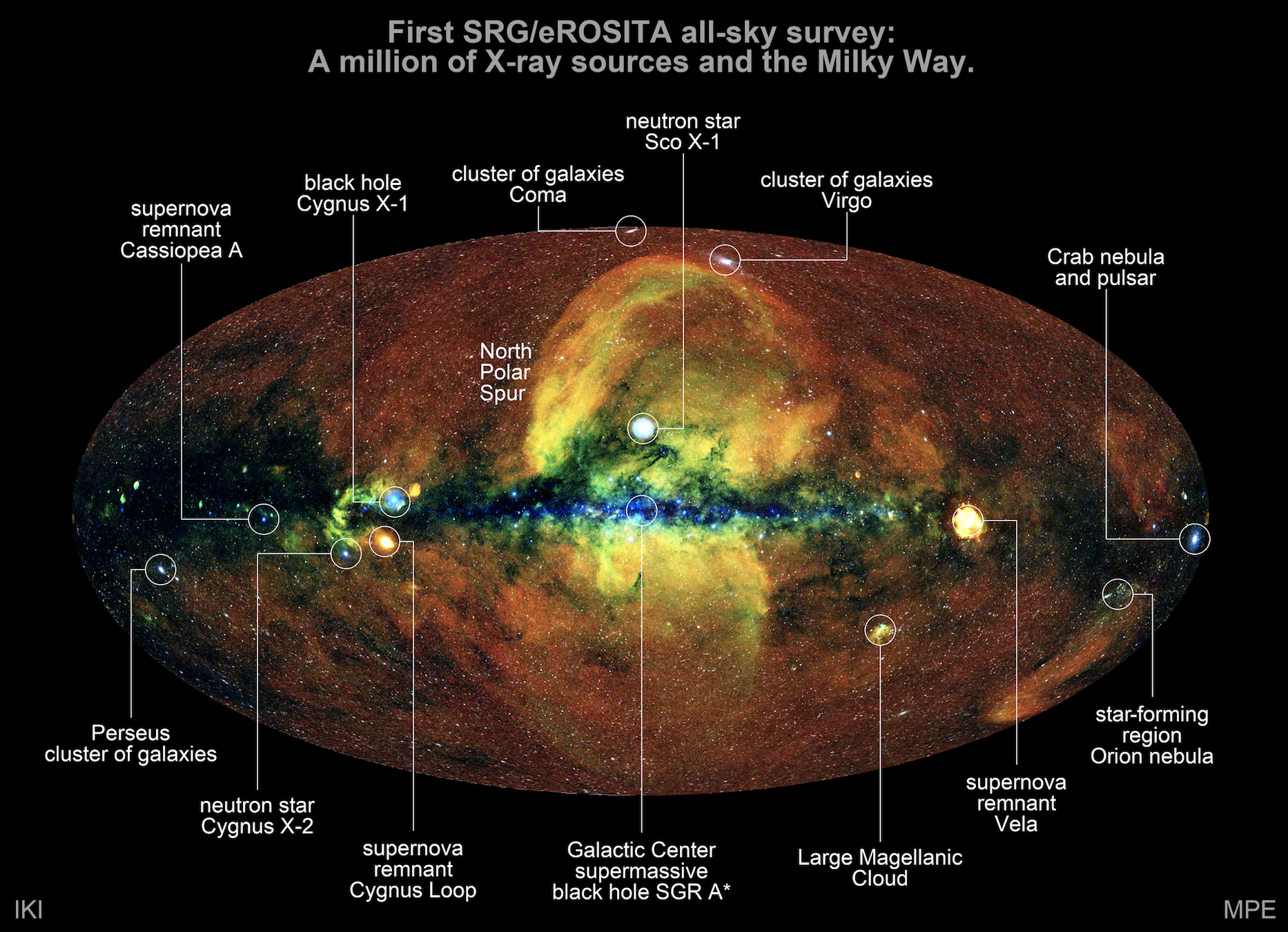}
\caption{Annotated version of the \srg/\erosita\ first all-sky image. Several prominent X-ray features are marked, ranging from distant galaxy clusters (Coma, Virgo, Fornax, and Perseus) to extended sources such as SNRs and nebulae to bright point sources, e.g., Sco X-1 (the first known extrasolar X-ray source). The map of the whole sky is constructed by the two scientific consortia of \srg/\erosita\ in Germany and Russia. Each consortium created their images for one half of the sky. The map is an RGB map, where photons of different energies are shown in different colors: from 300 to 600\,eV in red, from 600\,eV to 1.0\,keV in green, and from 1 to 2.3\,keV in blue. The colors on this map, obtained from about 400~million photons registered by \erosita\ during six months of the survey, allow one to immediately judge the temperature of radiating gas, ranging from 3 to more than 10~million\,K.  At the very center of the map, the supermassive black hole Sgr A* is located with a mass of about 4 million solar masses. It is a rather weak X-ray source and is nearly invisible on this map. In the middle plane of the picture lies the disk of the Milky Way. It looks dark because the molecular gas and dust in the plane of the Galaxy absorb X-rays. The blue dots located in this region reveal a large number of bright and powerful X-ray sources in the Milky Way: X-ray pulsars, accreting black holes, neutron stars, white dwarfs in binary stellar systems, and remnants of supernova explosions.} 
\label{fig:erosita_skylabels}
\end{figure*}

\begin{figure}
\centering
\includegraphics[width=0.98\columnwidth,clip]{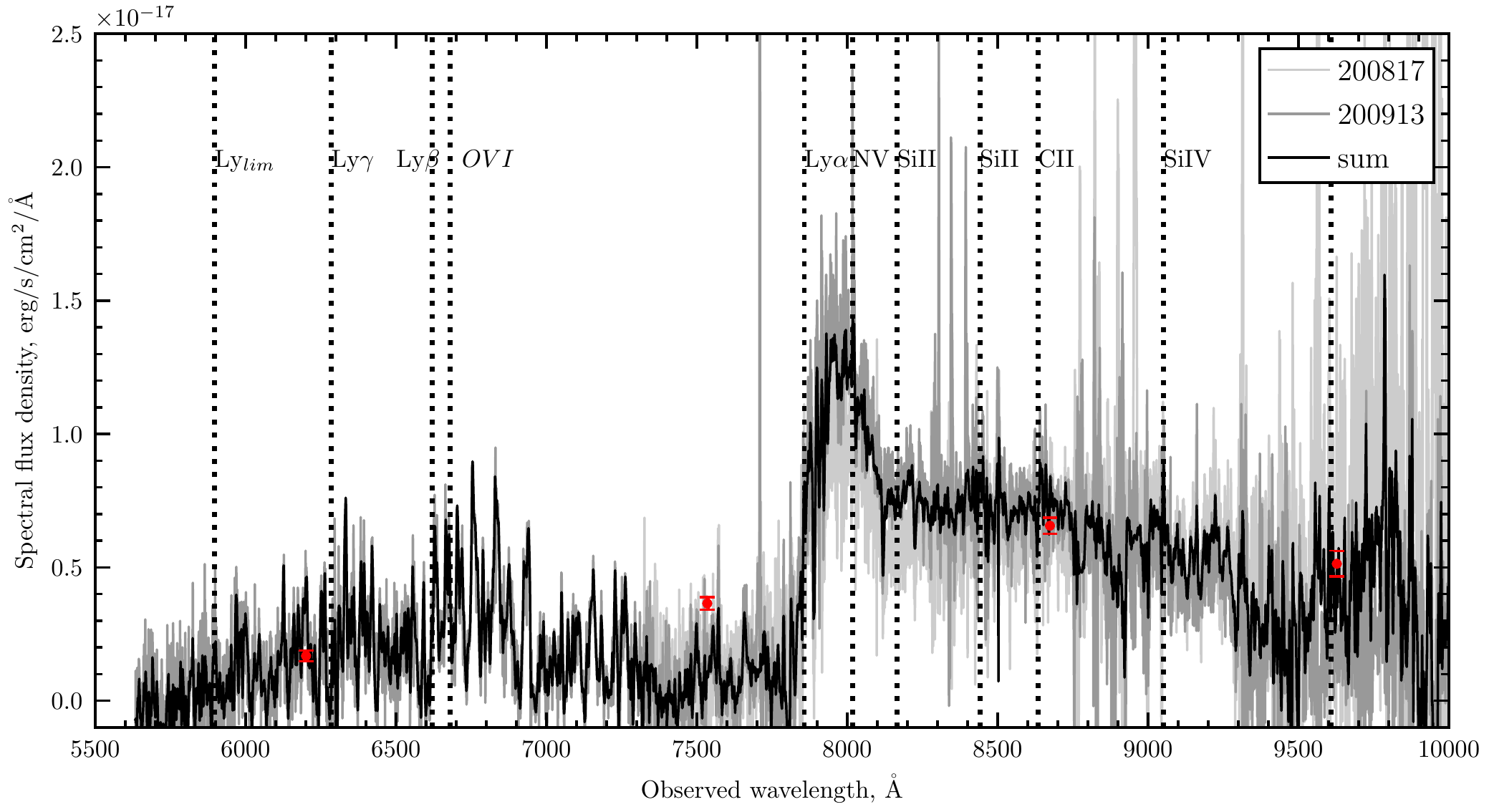}
\caption{Optical spectrum of quasar SRGE\,J170245.2+130107 (discovered by \srg/\erosita) obtained with the {BTA 6m telescope}. The light gray and dark gray lines show the spectra obtained on August 17, 2020, and September 13, 2020, respectively, while the combined spectrum is shown by the black line. The red dots show the source flux density in the Pan-STARRS $r, i, z, y$ filters. The vertical dashed lines show the expected positions of the peaks of the emission lines of the quasar at $z=5.466$. Adapted from \citet{khor21}.} 
\label{fig:qso55}
\end{figure}

The \erosita\ telescope has obtained the best map of the sky in the history of X-ray astronomy in the 0.3--2.3\,keV energy band (Fig.~\ref{fig:erosita_skylabels}) even after its first all-sky survey. The good angular resolution and high sensitivity of \erosita\ allowed it to detect over a million compact sources and map about 20 thousand extended sources. This huge number of sources cannot be displayed in a single image; only the brightest of them are visible on the map as dots. After scanning the sky for just six months, \erosita\ has constructed a map that is about four times more sensitive and contains almost eight times more sources than the previously best all-sky map that was obtained in 1990 by the {\it ROSAT} satellite. \erosita\ has already nearly doubled the total number of sources detected by all orbital observatories over the $\sim 60$ years of X-ray astronomy.

The \erosita\ map reveals spectacular giant bubbles of hot gas with temperatures up to 10 million K expelled from the plane of the Galaxy. These are the outcome of hundreds of thousands of supernova explosions and/or the intermittent activity of the supermassive black hole in the center of our Galaxy \citep{predehl20}. These X-ray bubbles are seen above and below the midplane of the image that encloses the well-known {\it Fermi} bubbles, which are associated with gamma-rays emitted through the interaction of cosmic rays with the ambient gas \citep{ackermann14}. 

About three-quarters of the objects in the \srg/\erosita\ map are AGN that are powered by the accretion of matter onto supermassive black holes residing in their centers. They are located far beyond the Milky Way. The quasar CFHQS~J142952+544717 at $z=6.2$ (corresponding to the age of the Universe of 900~million years) is particularly interesting. It was detected in X-rays for the first time by \erosita\ and proved to have the highest X-ray luminosity ($\sim 3\times 10^{46}$\,erg~s$^{-1}$, or $\sim 10^{13}$ bolometric luminosities of the Sun) of the quasars at $z>6$ \citep{Medvedev_2020,medvedev21}. During the calibration and performance verification (CalPV) phase and in the course of scanning the entire sky, \srg/\erosita\ has discovered a number of extremely luminous quasars at redshift $z > 5$ and a noticeable number at $z > 4$ \citep{khorunzhev20,dodin20,bikmaev20,wolf20}. In Fig.~\ref{fig:qso55} we show the spectrum of {such a quasar} discovered by \srg/\erosita\ at $z = 5.5$. It was obtained with the 6-meter BTA telescope in the North Caucasus \citep{khor21}. The majority of the $\sim 20000$ extended objects in the \erosita\ map are clusters of galaxies filled with dark matter and hot intergalactic gas, which shines in X-rays. Fewer than 50\% of these clusters were previously known from optical surveys or due to detection of the Sunyaev--Zeldovich effect \citep{sunyaev80} in their direction by the {\it Planck} spacecraft \citep{Planck_2016}, the Atacama Cosmology Telescope (ACT, \citealt{Hilton_2020}), or the South Pole Telescope (SPT, \citealt{Bleem_2015}) sky surveys. 

There will be much synergy and competition between different methods of observation of galaxy clusters over the next years through the continuation of sky surveys in different spectral bands, including the \srg\ survey in X-rays and ground-based surveys by ACT and SPT in microwaves. Not only will the results of these surveys be valuable for cosmological studies, but we should also expect the discovery of thousands of strong gravitational lenses due to the deep gravitational potential of clusters of galaxies.

Some two hundred thousand fairly close stars with coronae much more powerful than the corona of the Sun also contribute to the emission from the zones of relatively low temperature in the \erosita\ sky map (Fig.~\ref{fig:erosita_skylabels}). Interestingly, \erosita\ has detected X-ray emission from 150 stars with known exoplanets. This amounts to some 10\% of all nearby stars with known planetary systems (excluding the more distant stars with exoplanets in the field that were explored by the {\it Kepler} satellite). 

\begin{figure*}
\centering
\includegraphics[width=0.95\textwidth,clip]{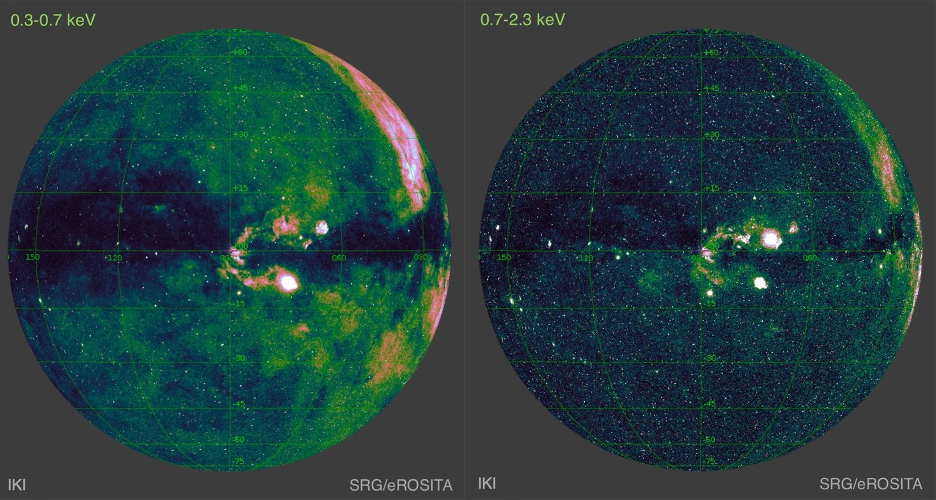}
\caption{\srg/\erosita\ maps of one half of the sky in the 0.3--0.7 and 0.7--2.3\,keV energy bands.} 
\label{fig:erosita_ru}
\end{figure*}

Figure~\ref{fig:erosita_ru} demonstrates the dramatic difference between images of the sky (one hemisphere) in the 0.3--0.7\,keV and 0.7--2.3\,keV energy bands. In the 0.3--0.7\,keV band, the emission is fairly homogeneous, bright, and diffuse. This might be the background radiation of a huge number of soft X-ray sources located at cosmological distances and/or emission from the hot ($10^5<T_{\rm e}<10^7$~K) gas in the halo of our Galaxy. An additional significant contribution to the sky brightness at these energies might be provided by relatively nearby sources such the hypothetical Local Bubble that was created by supernova explosions in the relative vicinity of the Solar System. The strong absorption of soft X-rays by the cold atomic and molecular gas and dust in the Galactic plane is also noteworthy. A very different picture is observed in the 0.7--2.3\,keV energy band. Here the image is dominated by many hundreds of thousands of extragalactic sources. The main contribution to the source counts and to the diffuse background is provided by AGN. The absorption by cold gas and dust in the Galactic plane is strongly reduced by the rapidly decreasing photoabsorption cross-section with increasing photon energy. Bright Galactic sources in the region of active star formation in the Cygnus constellation are clearly visible in both maps.

\begin{figure}
\centering
\includegraphics[width=\columnwidth,clip]{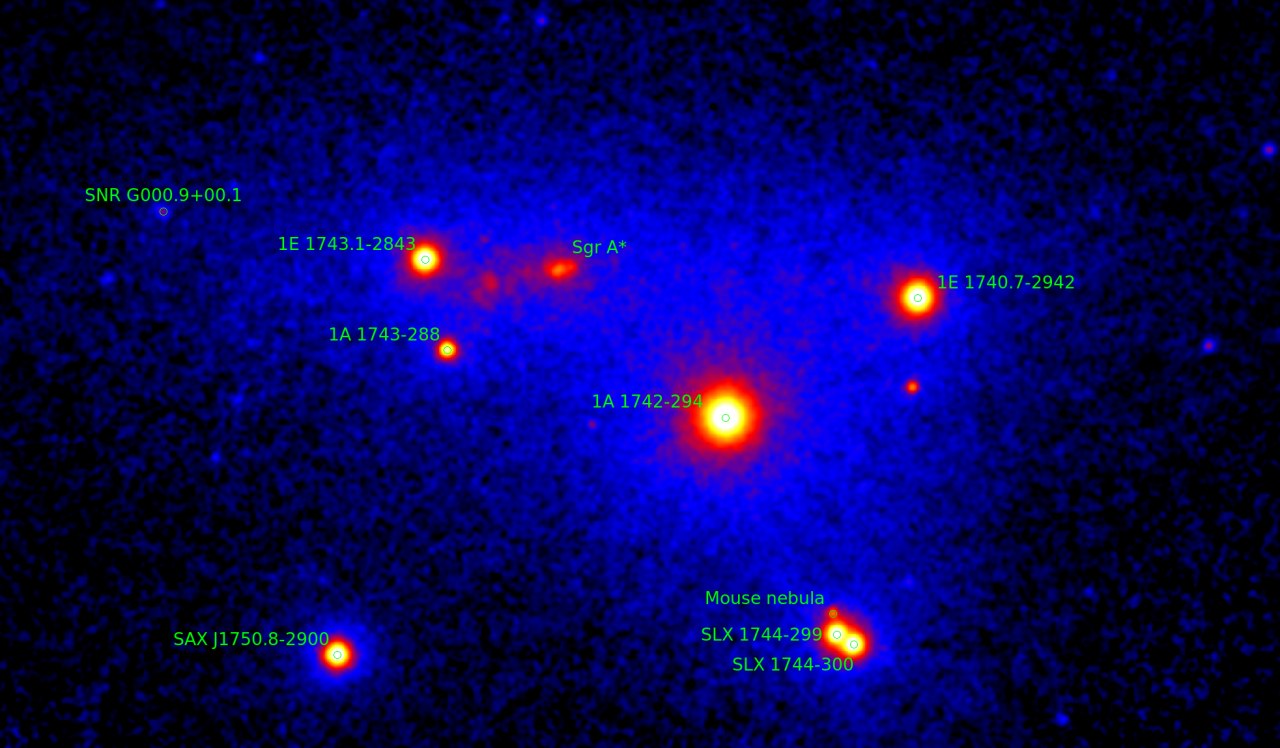}
\caption{Fragment ($\sim 3^\circ \times 2^\circ$) of an image of the central region of the Galaxy obtained by \art\ in the 4--12\,keV energy band during the CalPV phase.}
\label{fig:art_gc}
\end{figure}
\begin{figure}
\centering
\includegraphics[width=0.9\columnwidth,clip]{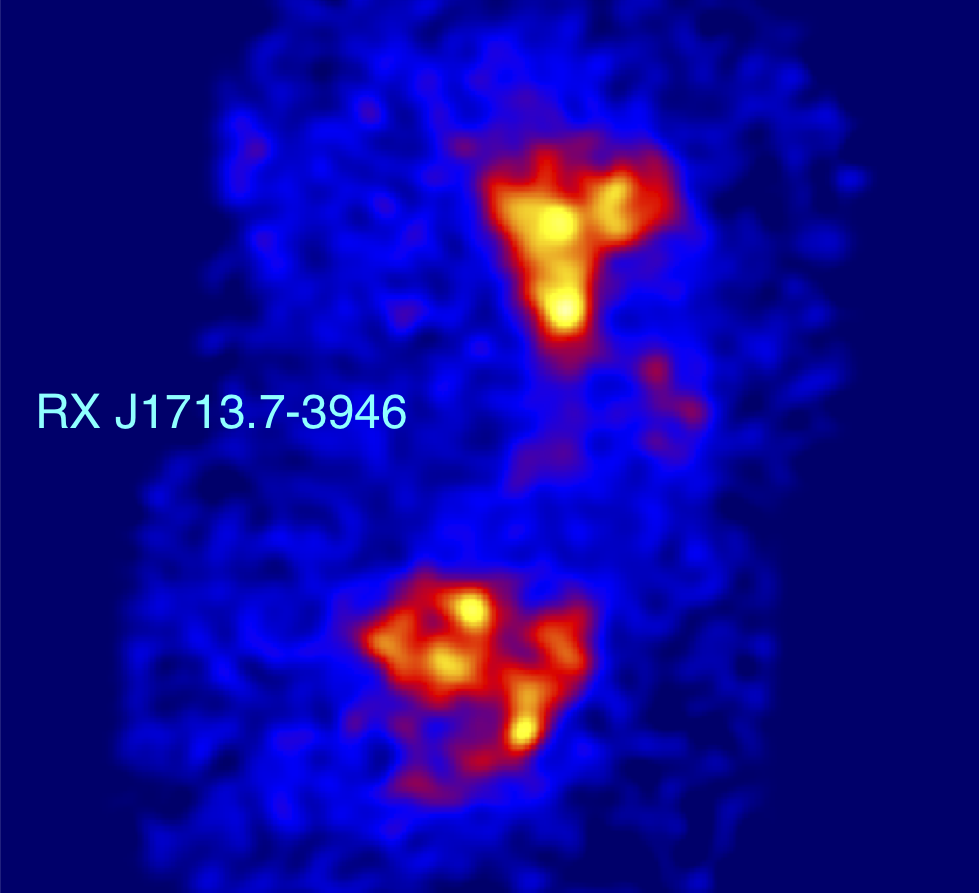}
\caption{Image ($\sim 40$\,arcmin on a side) of the SNR RX~J1713.7$-$3946 obtained by \art\ in the 4--12\,keV energy band during the CalPV phase.}
\label{fig:art_snr}
\end{figure}

\subsection{Examples of the results of deep surveys during the PV phase}
\label{s:deep}

\subsubsection{Galactic center}

During the CalPV phase, \art\ performed deep surveys of a number of extended fields in the sky, in particular, of a large ($\sim 40$\,square degrees) region in the center of our Galaxy. Figure~\ref{fig:art_gc} shows a fragment of the image obtained in the 4--12\,keV energy band, which reveals the high quality and richness of the data. 

All in-flight characteristics of \art\ have proved to be close to preflight expectations (see more details in \S\ref{s:art_flight}). In particular, the PSF averaged over the FoV is better than 1~arcmin (half-power diameter, HPD) in survey mode. The good angular resolution of the telescope is evident from the image (Fig.~\ref{fig:art_snr}) of the SNR RX~J1713.7$-$3946 that is clearly resolved by \art.

\subsubsection{Galactic Ridge}

\begin{figure*}
\centering
\includegraphics[width=0.9\textwidth,clip]{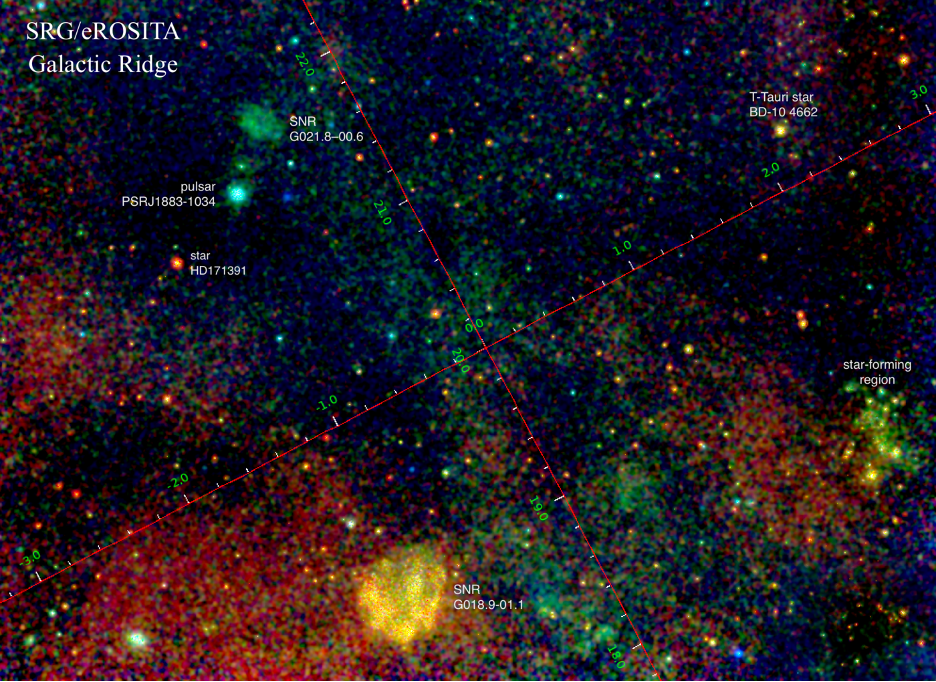}
\caption{Annotated X-ray RGB map of a 25\,square degrees region in the disk of the Milky Way (the so-called Galactic Ridge) obtained by \erosita\ in October 2019. In this field, thousands of Galactic X-ray sources are detected, as well as a number of quasars observed through the Galactic disk. In addition to many individual sources, the map also shows unresolved X-ray emission from hot gas and from a multitude of faint unresolved sources. Blue and green correspond to high photon energies emitted by a gas with a temperature of tens of millions of degrees, while red regions reveal colder gas of lower temperature. Adopted from Gilfanov, Medvedev, Sunyaev et al. 2022 (in preparation). Absorption in the cold interstellar gas near the Galactic plane attenuates soft X-ray radiation, but allows hard X-ray photons to leak through because their absorption cross-section is smaller. A significant contribution to the apparently diffuse emission from the Galactic disk and bulge is provided by the superposition of emission from numerous accreting white dwarfs, hot coronae of low-mass stars, and flares on them \citep{Revnivtsev2009}. \erosita\ cannot resolve individual contributions of these sources because of their great number and low luminosities.
} 
\label{fig:ridge}
\end{figure*}

Figure~\ref{fig:ridge} demonstrates a large variety of astrophysical objects of our Galaxy that are accessible for observation by the \srg\ observatory. During scans of the Galactic Ridge (in a field $\sim 20$~degrees away from the Galactic center), \erosita\ has detected stars with active X-ray emitting coronae, star-forming regions and clusters of young stars, X-ray pulsars (rapidly rotating magnetized neutron stars), and SNRs. In the latter, X-ray photons are emitted by gas that is compressed in shocks where the material of the exploded star collides with the surrounding interstellar matter. As is well known, hot gas occupies 80--90\% of the volume near the Galactic plane, and just 10--20\% of the volume is filled by dense clouds of cold molecular and atomic hydrogen. 

\subsubsection{Lockman Hole}

\begin{figure}
\centering
\includegraphics[width=0.95\columnwidth,clip]{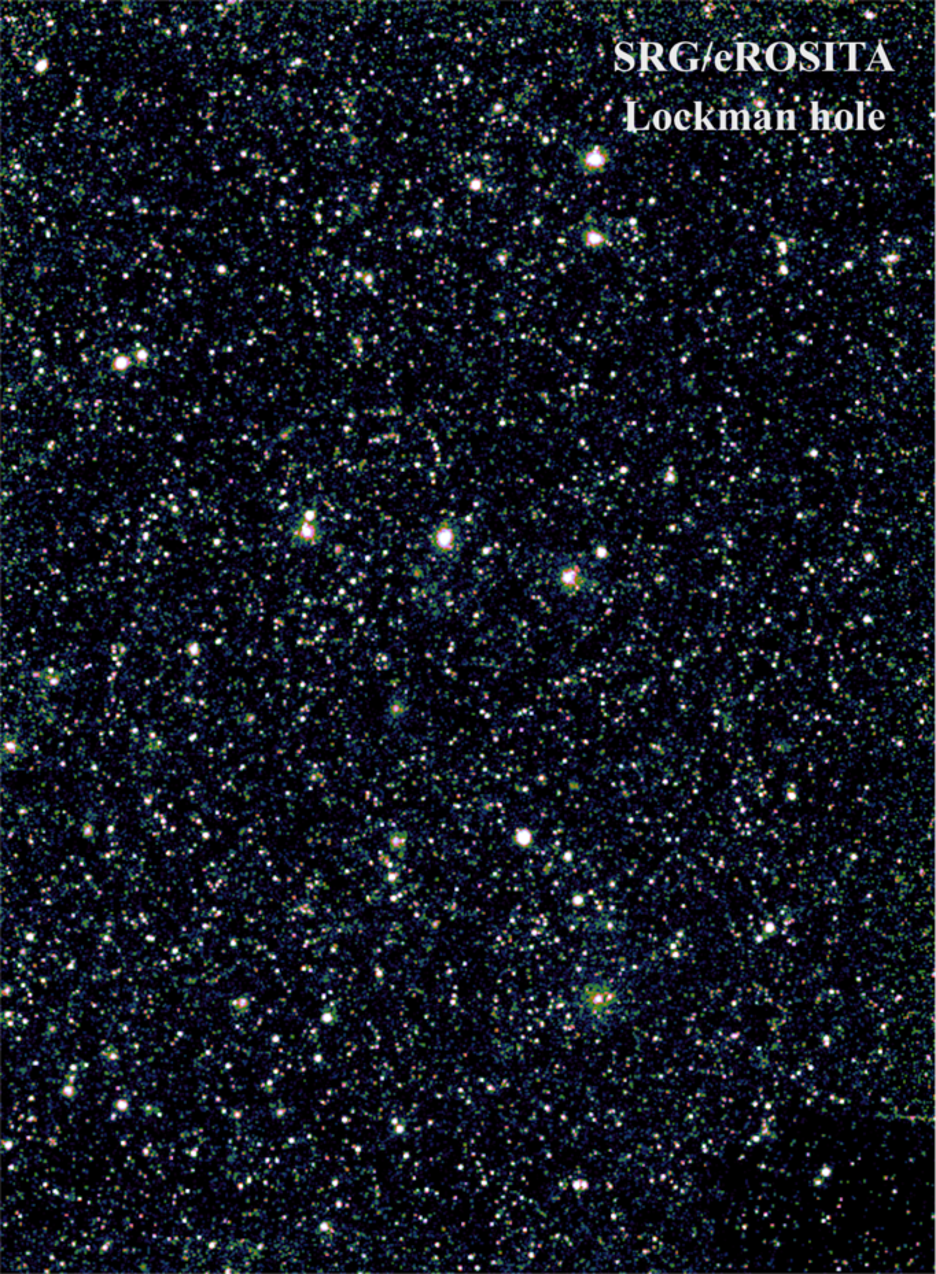}
\caption{X-ray image of the Lockman Hole obtained by \erosita: More than 8,500 X-ray sources in 18\,square degrees. Adopted from Gilfanov, Burenin, Sunyaev et al (in preparation).} 
\label{fig:lockman}
\end{figure}

Figure~\ref{fig:lockman} demonstrates the richness of the extragalactic X-ray sky revealed during a long scan (about 8 ks per pixel) of the Lockman Hole zone. In this unique region, the absorption of X-rays by the interstellar medium of the Galaxy is close to its minimum in the entire sky. This allows studies of distant quasars and clusters of galaxies in unprecedented detail.  In the $\approx 20$\,square degrees field, \erosita\ has detected over 8,500 point X-ray sources, that is, $\sim 400$ sources per square degree. This number corresponds to $\sim 16$ million objects when extrapolated to the whole sky. The vast majority of these sources are AGN. According to photometric redshift estimates, the most distant of the quasars detected by \erosita\ in the Lockman Hole are located at $z\sim 5$. Some  $\sim 200$  clusters and  rich groups of galaxies filled with hot gas were  detected. In addition, several hundred active Galactic stars are also seen in the Lockman Hole field. 

\subsubsection{Coma cluster}

\begin{figure*}
\centering
\includegraphics[width=0.99\columnwidth,clip]{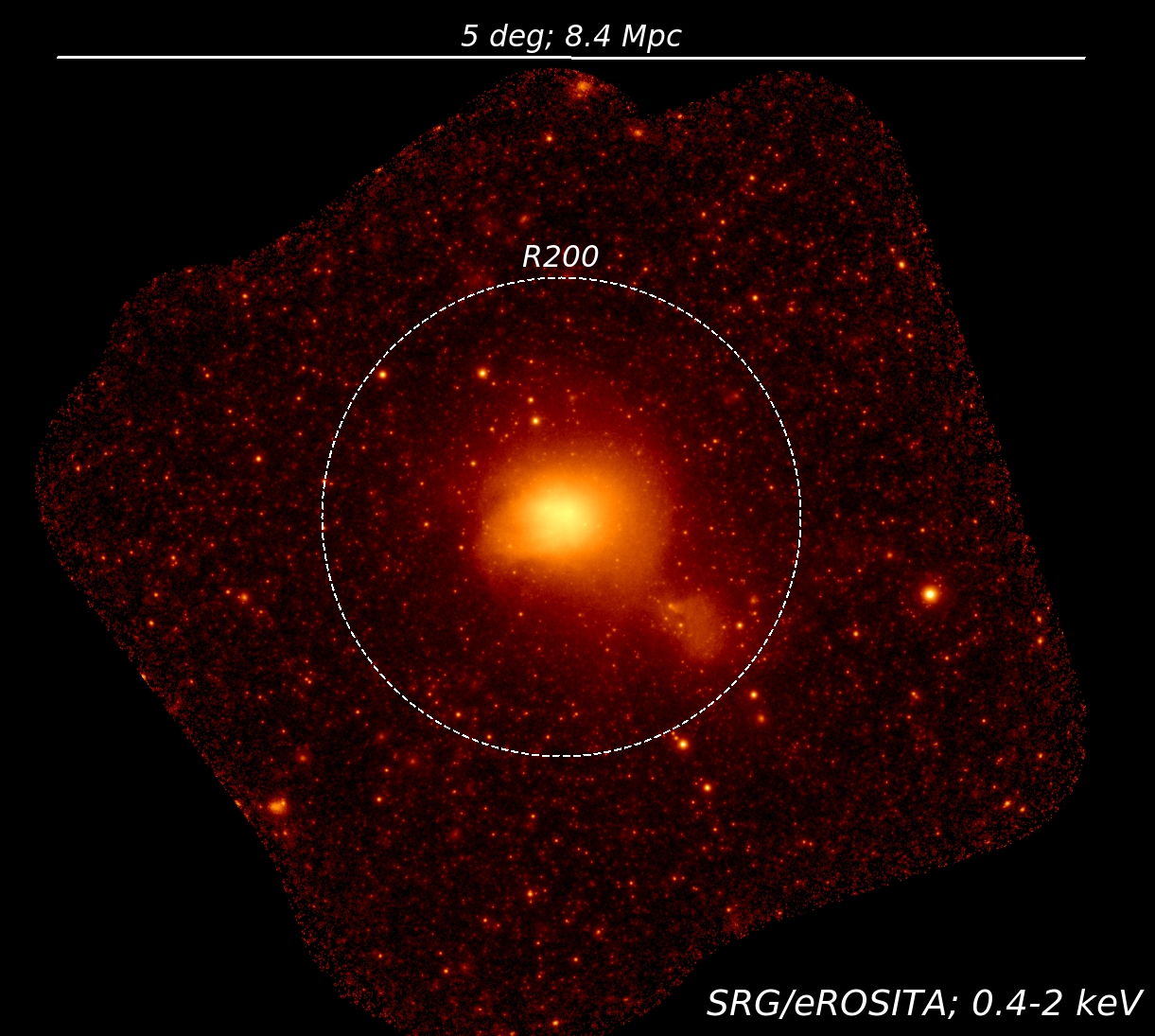}
\includegraphics[trim=0cm 0cm 0mm 0cm,width=0.935\columnwidth,clip]{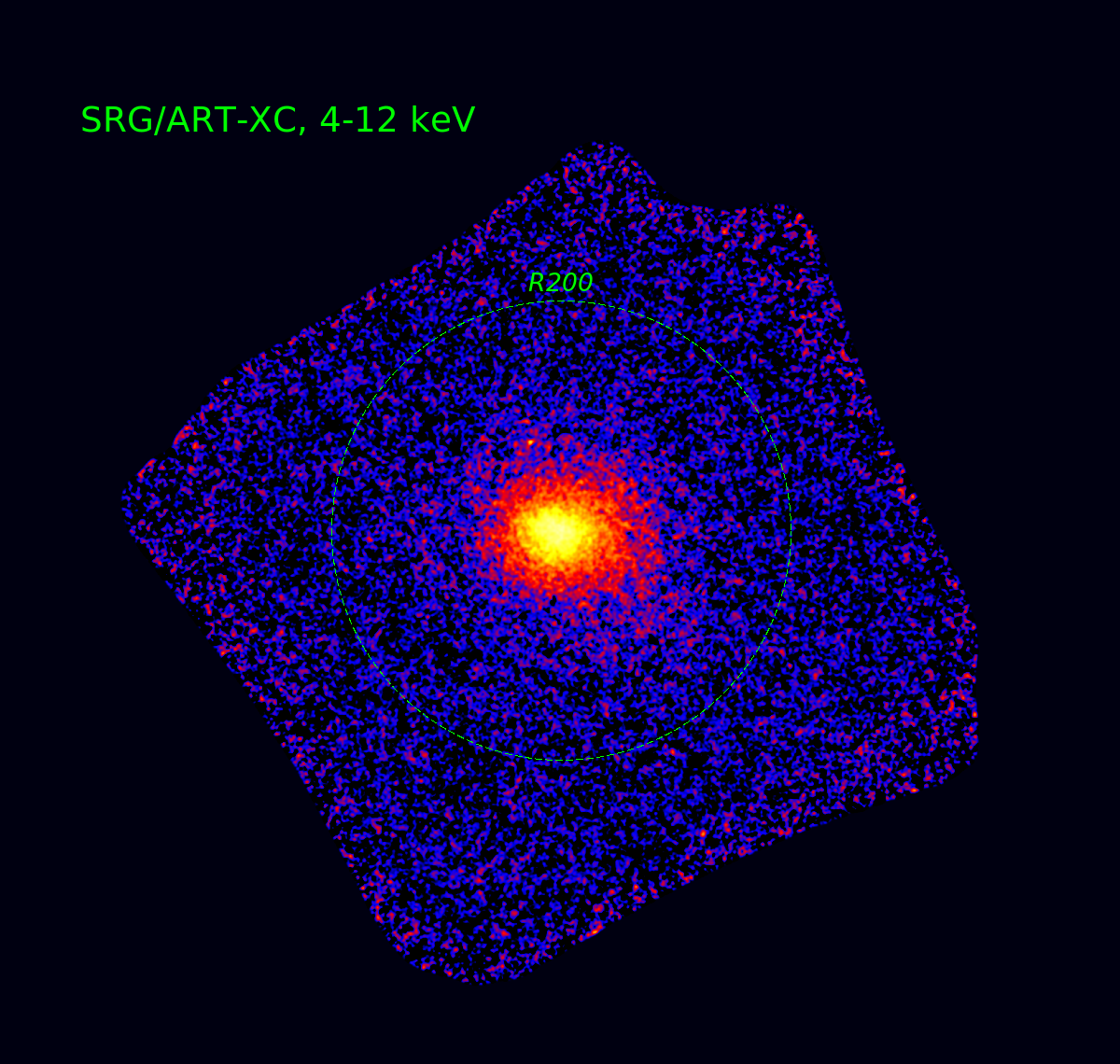}
\includegraphics[trim=0cm -2.3cm 0mm 0cm,width=0.99\columnwidth,clip]{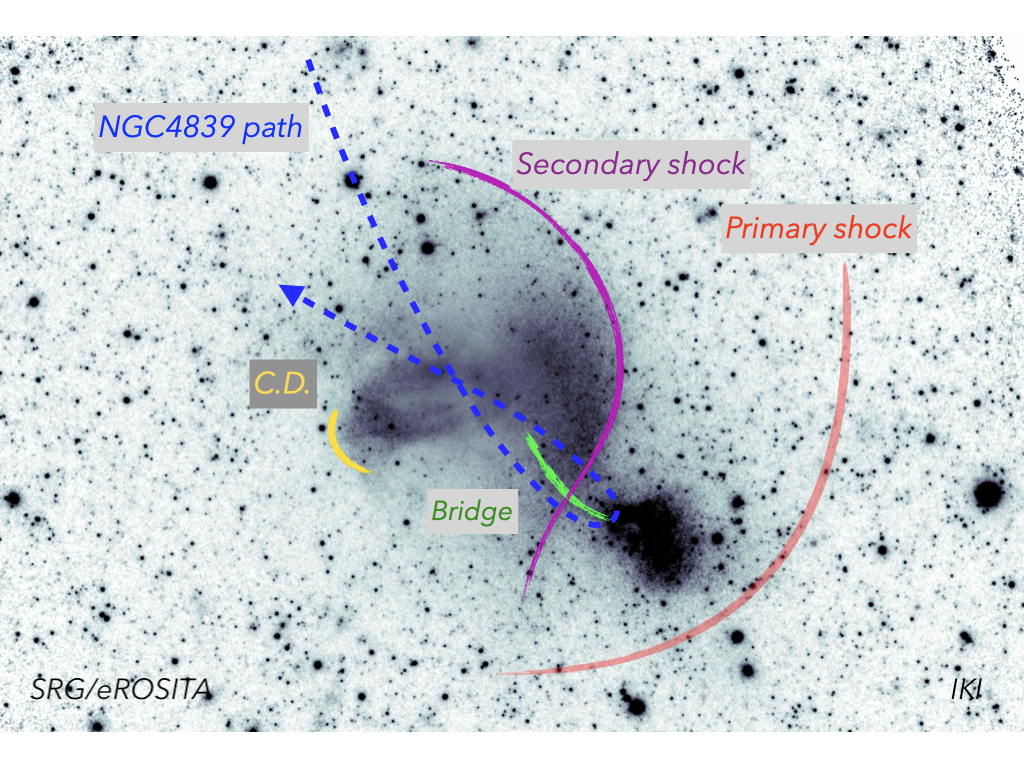}
\includegraphics[trim=0cm 0cm 0mm 0cm,width=0.935\columnwidth,clip]{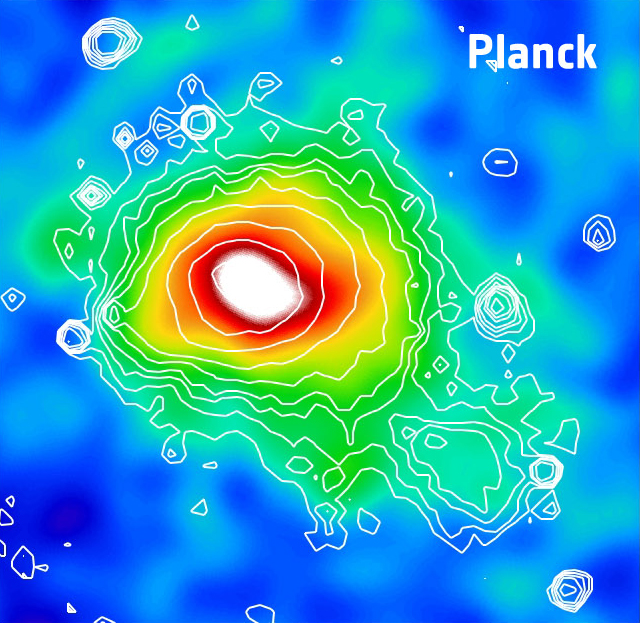}
\caption{{\it }  X-ray image of the Coma cluster \citep{churazov21} in the 0.4--2\,keV band obtained by \erosita\ in the course of the CalPV program (top left). The image is $\sim$6 degrees on a side, corresponding to 10\,Mpc at the distance of the cluster, with the logarithmic color-code spanning five orders of magnitude. The main cluster is in the process of merging with the NGC\,4839 group (a bright blob in the bottom right corner from the Coma cluster). Flattened X-ray image of the Coma cluster field (bottom left). The labels schematically mark some of the features that are presumably associated with the merger with the NGC\,4839 group. The dashed blue line is the suggested trajectory of the group, which enters the Coma cluster from the northeast direction, and is currently close to apocenter. The presumed positions of two shocks driven by the NGC\,4839 group are shown with the red and purple curves. The shock closer to the center is driven by the displaced gas that settles back to hydrostatic equilibrium. This is the most salient feature that can be directly seen in the image at the surface brightness edge. The green line shows the faint X-ray bridge connecting NGC\,4839 and the main cluster, which is a possible trace of the group passage through the Coma cluster. The yellow line shows the contact discontinuity, which is an interface between cold and hot gas patches with the same pressure. See \citet{churazov21} for details. ART-XC image of the Coma Cluster in the {4--12\,keV} energy band (top right).{\it } \textit{Planck} spacecraft $y$-parameter map of the Coma cluster, based on the data in the microwave spectral band (bottom right). The morphology of this map, which reflects the distribution of the hot gas electron pressure, is strikingly similar to the \erosita\ image in the X-ray band. 
} 
\label{fig:coma}
\end{figure*}

\begin{figure}
\centering
\includegraphics[width=0.9\columnwidth,clip]{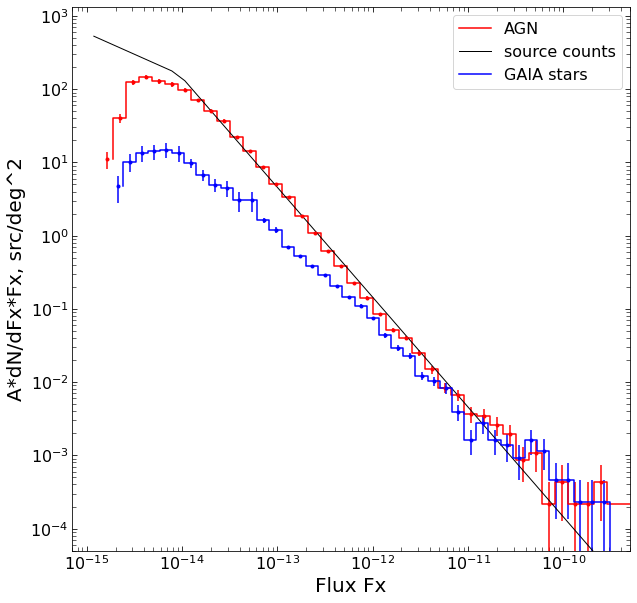}
\caption{Preliminary $\log(N)-\log(S)$  distributions of compact sources in the extragalactic part of the hemisphere, where the Russian consortium of scientists is responsible for the data analysis and interpretation.
The distribution of sources with and without Gaia stars among their optical counterparts are shown by blue and red histograms and are marked in the plot legend as GAIA stars and AGN, respectively.   The distributions are derived from partial data of the first two sky surveys and do not use the full potential of the first-year data. No incompleteness correction was applied at the faint flux end. The coefficient $A$ on the y-axis incorporates  approximations and inaccuracies of the preliminary analysis. It is reasonably close to unity and may weakly depend on flux,  which is not necessarily the same for GAIA stars and AGN. }
\label{fig:logn-logs}
\end{figure}

One of the main targets of the CalPV phase was the Coma cluster (Fig.~\ref{fig:coma}). \srg\ observations in the scanning mode are particularly well suited for a detailed mapping of this massive and nearby cluster well beyond its virial radius.
The obtained \art\ data enable measuring the temperature of the hot gas and estimating the contribution of cosmic rays to the observed X-ray emission, while \erosita\ images provide detailed information about the merger of the Coma cluster with its less massive companion, the galaxy group NGC4839. Fig.~\ref{fig:coma} clearly shows the rich substructure that is generated by the merger, including shock waves and contact discontinuities extending over a few megaparsec. The data from the {\it Planck} observatory (distribution of brightness in the microwave band due to the SZ effect) provide additional information about the distribution of pressure in the space between galaxies \citep{Planck_2013}. The comparison of the data from the three telescopes provides a detailed picture of the distribution of hot gas and dark matter in the cluster and its outskirts.

It is important to note that in the Coma cluster field, \erosita\ has unveiled many dozen massive clusters of galaxies and thousands of quasars located at cosmological distances far beyond the Coma cluster. This is a consequence of the deep scanning of a field with a size of $3^\circ\times 3^\circ$.

\subsection{Source catalog}

The source catalogs are maintained by the Russian (RU) and German (DE) consortia for the respective sky hemispheres. A preliminary version of the $\log(N)-\log(S)$ distribution of compact sources in the extragalactic part of the RU sky is shown in Fig.~\ref{fig:logn-logs}

\subsection{X-ray spectroscopy}

\begin{figure}
\centering
\includegraphics[width=0.7\columnwidth,clip]{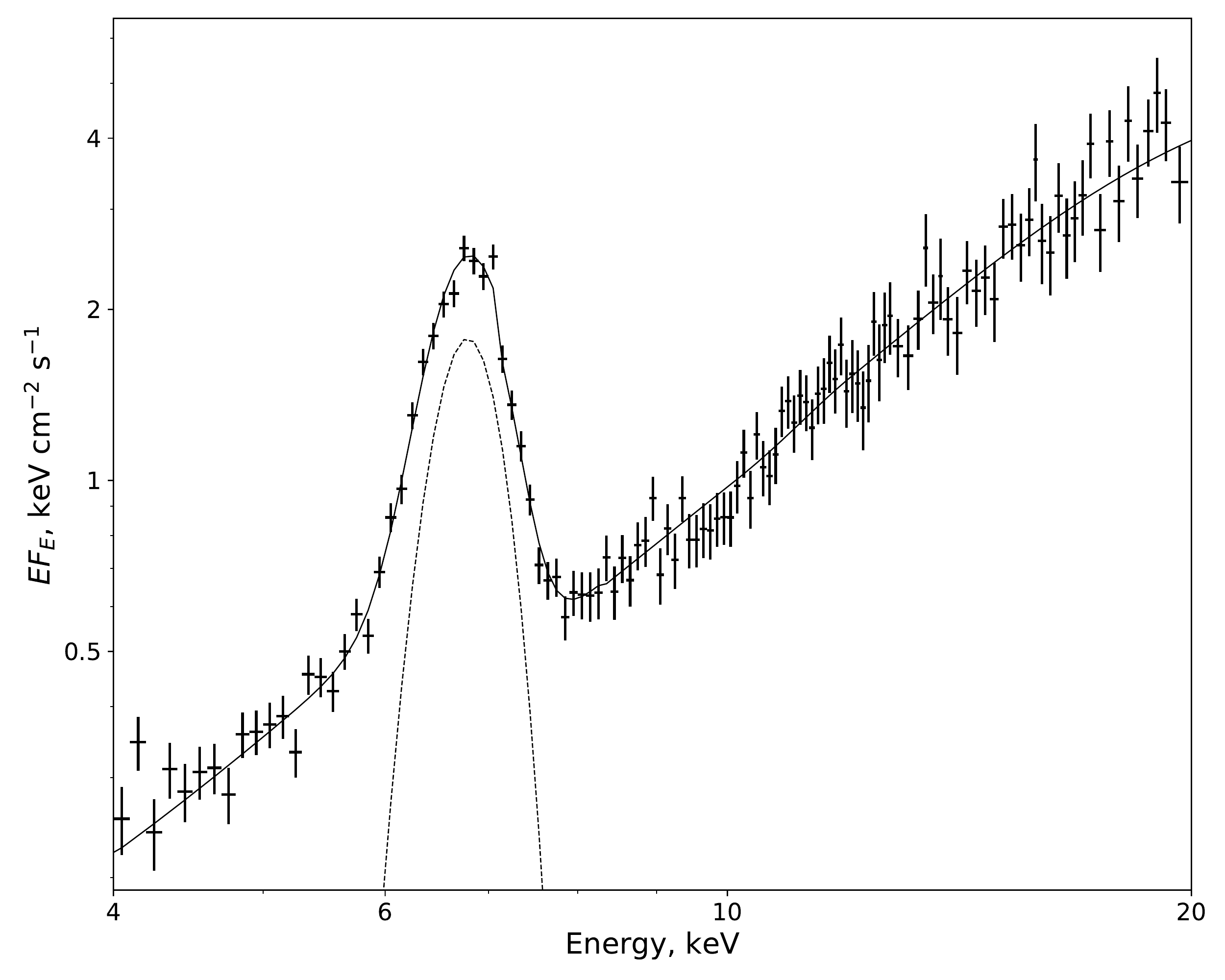}
\caption{X-ray spectrum of the Circinus galaxy measured by \art. The Fe-K$\alpha$ line is prominent on top of the complex continuum. The exposure is about 50~ks. Flux units are arbitrary.}
\label{fig:circinus}
\end{figure}

During the CalPV phase, \art\ performed long pointed observations of a number of bright X-ray sources, including the well-known nearby AGN Circinus galaxy. The obtained spectrum (Fig.~\ref{fig:circinus}) shows a complex continuum in the 4--20\,keV energy band along with a strong Fe-K$\alpha$ emission line. The estimated energy resolution of $\sim 1.3$\,keV at 6\,keV matches the preflight estimates.  

\begin{figure}
\centering
\includegraphics[width=0.95\columnwidth,clip]{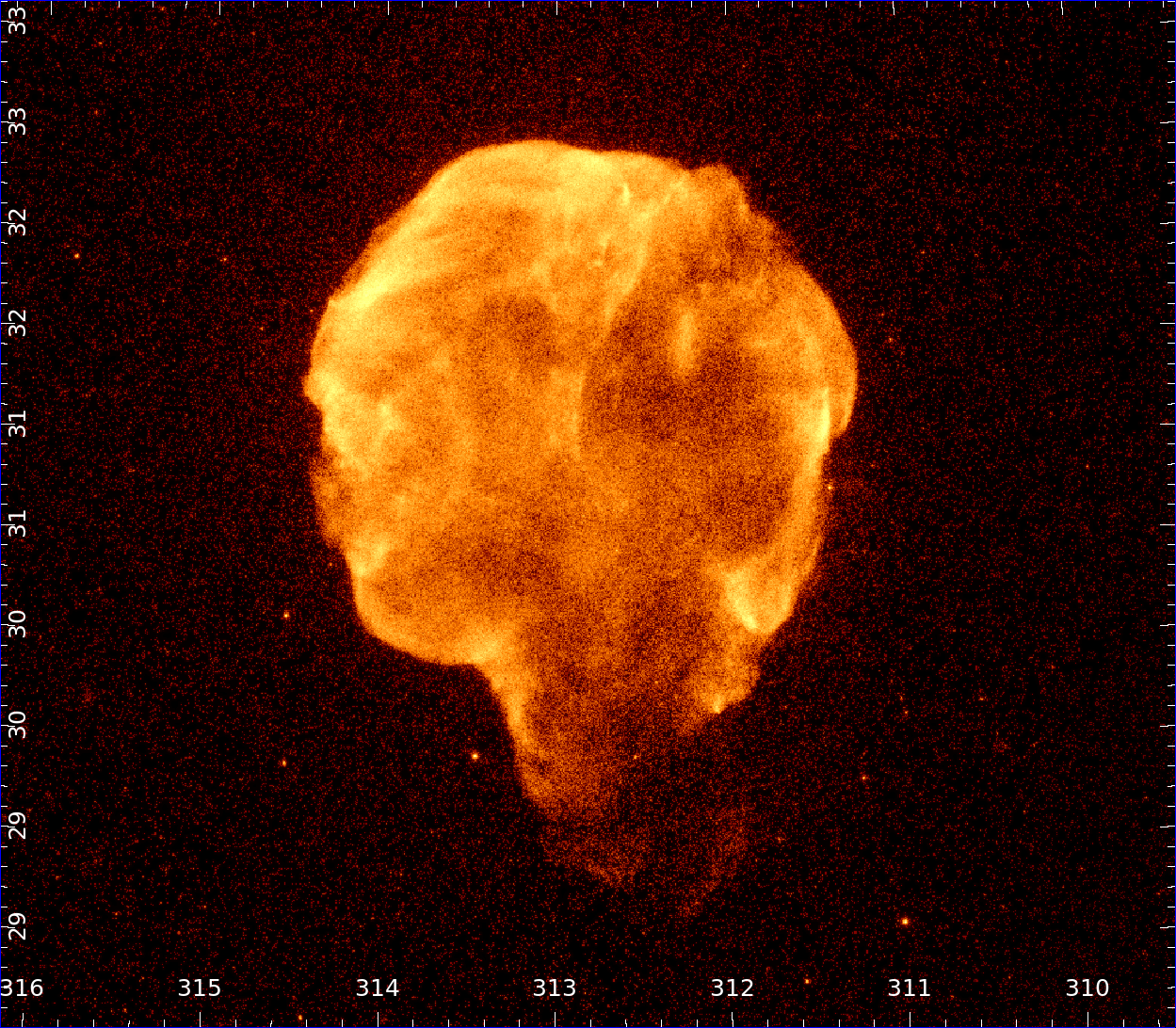}
\includegraphics[trim=0cm 4cm 0mm 6cm,width=1\columnwidth,clip]{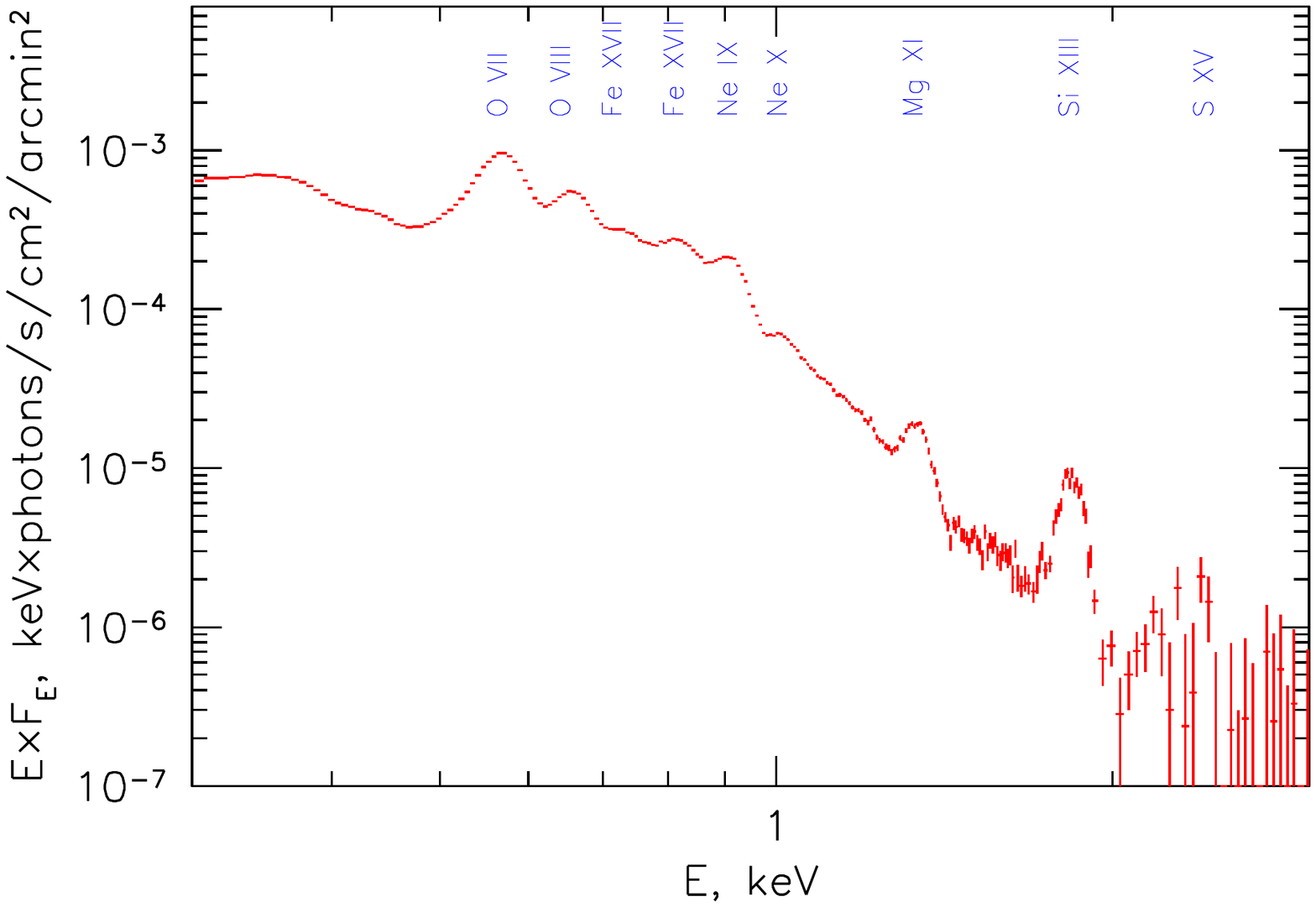}
\caption{0.4--2\,keV X-ray image of the SNR Cygnus Loop obtained by \erosita\ during the first two \srg\ all-sky surveys (upper panel). The longest exposure time (per point) across the image is $\sim$400\,s. The axes are labeled in degrees. X-ray spectrum of the entire Cygnus Loop (lower panel). The ions contributing to some of the brightest emission lines are indicated. The spectrum is normalized per square arcminute.} 
\label{fig:cygloop}
\end{figure}

\begin{figure}
\centering
\includegraphics[,bb=40 180 540 650,width=0.9\columnwidth]{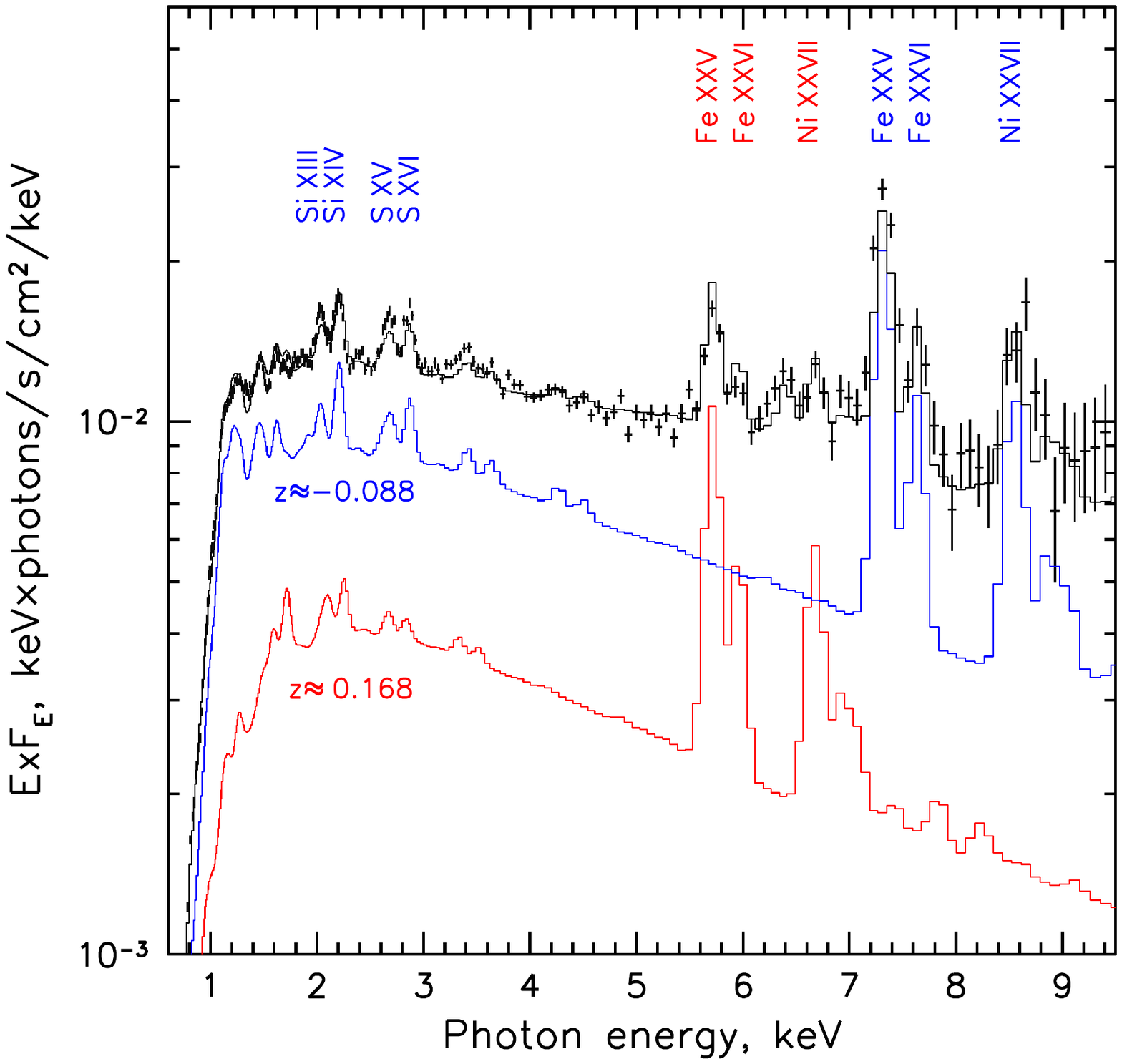}
\caption{{Spectrum of the Galactic microquasar SS 433 obtained during the \srg/\erosita\ PV phase. The data are well fit by the model of baryonic multitemperature jets \citep{2016MNRAS.455.1414K}, with contributions of the approaching (blueshifted) and receding (redshifted) jets shown in blue and red, respectively. The positions of the the lines of hydrogen- and helium-like silicon, sulphur, iron, and nickel are marked.}}
\label{fig:ss433pv}
\end{figure}
The excellent capabilities of the \erosita\ telescope in spectroscopy of hot astrophysical plasmas (temperatures of some million degrees) in bright X-ray sources during the all-sky survey are clearly demonstrated (Fig.~\ref{fig:cygloop}) by the spectrum of the central zone of the Cygnus Loop SNR obtained using an exposure of just 200\,s. X-ray emission lines of a number of ions of various chemical elements are clearly seen. The spectroscopic capabilities of \erosita\ in the 3--9\,keV band are demonstrated in Fig.~\ref{fig:ss433pv}, which shows the spectrum of the Galactic microquasar SS 433 obtained during the \srg/eROSITA PV phase. Emission lines of highly ionized atoms (silicon, sulphur, iron, and nickel) corresponding to the approaching (blueshifted) and receding (redshifted) baryonic multitemperature jets are clearly resolved. The line positions are fully consistent with the expectations based on the kinematic precession model with the bulk velocity of the jets equal to a quarter of the speed of light.  

\subsection{X-ray timing} 
\begin{figure}
\centering
\includegraphics[width=0.98\columnwidth,clip]{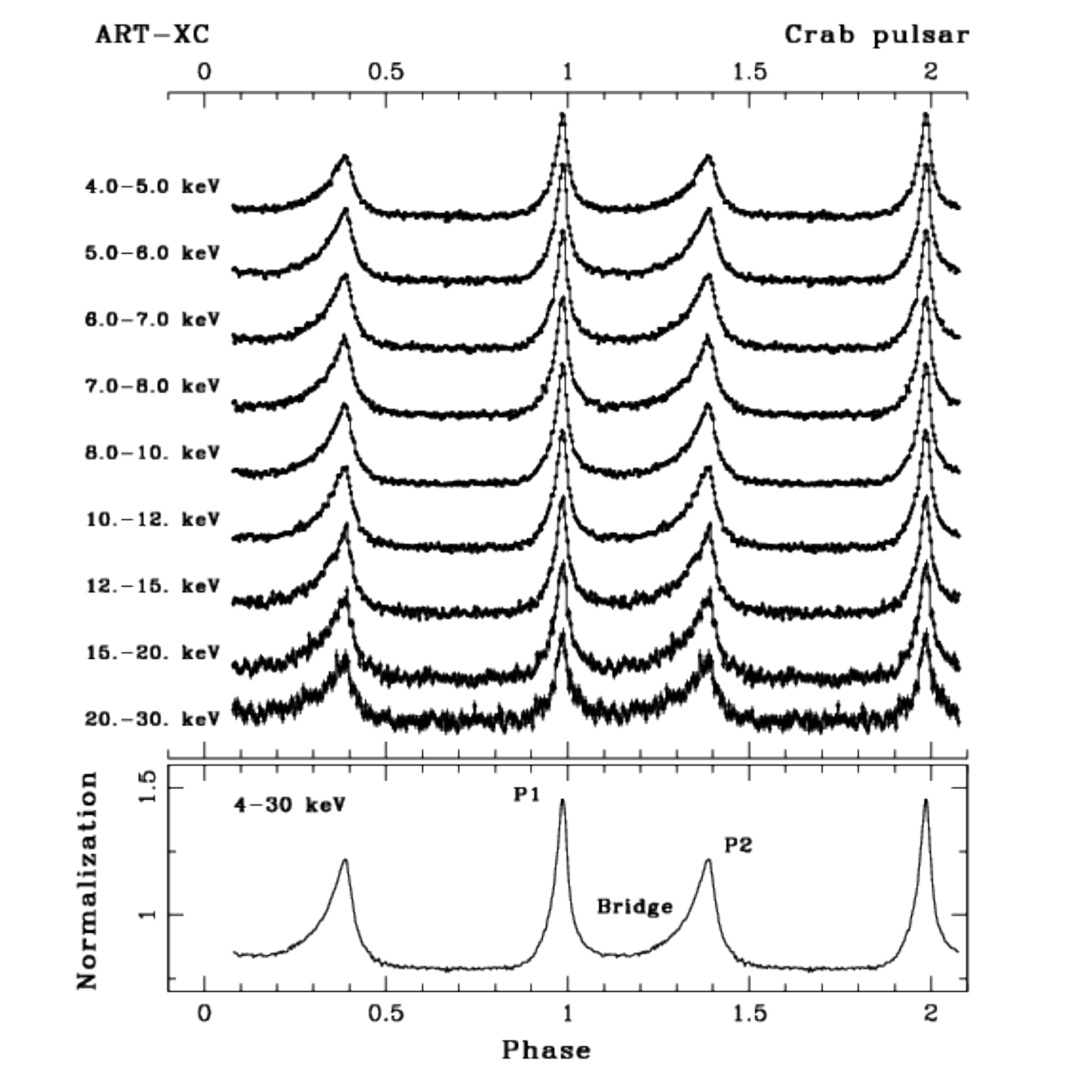}
\caption{Crab phase histograms reconstructed in several energy bands. The ART-XC exposure is about 45~ks.}
\label{fig:fold}
\end{figure}

\begin{figure}
\centering
\includegraphics[width=0.98\columnwidth,clip]{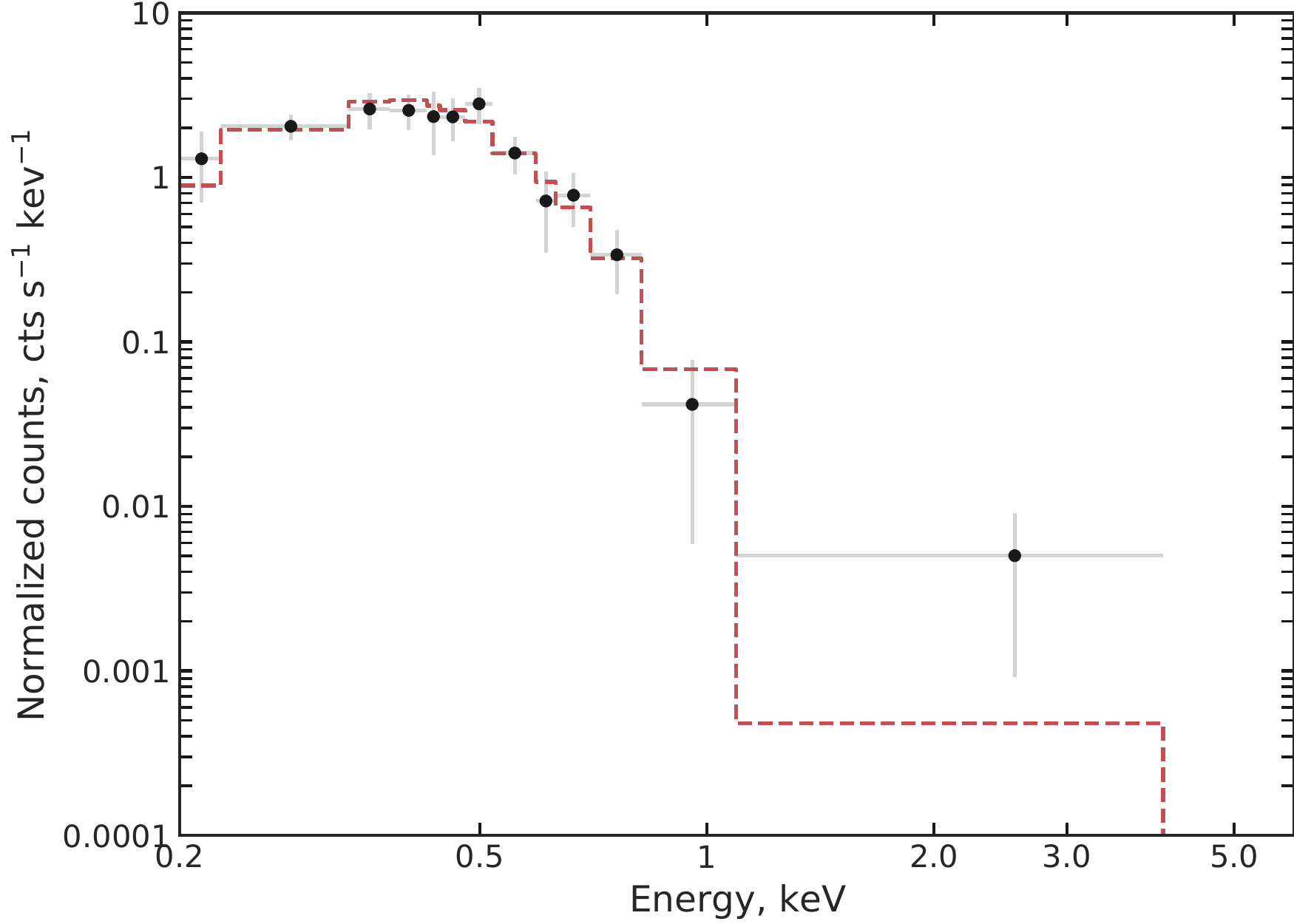}
\caption{\erosita\ spectrum of tidal disruption event SRGe J213527.3$-$181634/ZTF20abgbdpr. The best-fit multicolor accretion disk emission model with the inner disk temperature of $\sim 100$\,eV is shown by the dashed line  \citep{gilfanov20}.}
\label{fig:tde_spec}
\end{figure}

The \art\ telescope has an excellent time resolution of 23~$\mu s$, which is determined by the resolution of time stamps assigned to incoming events by the reading electronics. This permits a detailed study of the time behavior of millisecond pulsars in pointing mode. The timing capabilities of the telescope were tested during observations of the Crab pulsar. Figure~\ref{fig:fold} shows the light curve of this pulsar in several energy bands as measured by \art\ and folded with the Jodrell Bank radio ephemeris. The $\sim33$~ms pulsations are clearly detected up to 30\,keV, and the energy dependence of the pulse morphology is as expected.

The nominal integration time of the \erosita\ CCDs is 50~ms. This integration time is set by the requirement that smearing of X-ray images in the all-sky survey is to be avoided. With the nominal rotation rate (360 degrees in four hours), the sources in the telescope FoV shift by $3.5"$ in 50~ms, which is $\text{about three}$ times smaller than the CCD pixel size. While this integration time precludes a systematic study of the variability on millisecond scales, the stability of the background and instrument characteristics provides an excellent opportunity for monitoring objects on longer timescales in three important intervals: 0.05--40 s (for sources crossing the FoV during a single scan), four hours to one day (during six intra-day scans of the same object), and on scales from six months to four years (over the entire \srg\ all-sky survey). In addition, the variability of sources located close to the intersections of individual scans near the ecliptic poles can be tracked nearly continuously on scales longer than four hours.

\subsection{ART-XC Galactic transients}

In its daily scans (exposure of $\sim 60$\,s per source), the \art\ telescope reaches a sensitivity ($\sim 5\sigma$) of $\sim 8\times10^{-12}$\,erg~cm$^{-2}$~s$^{-1}$ in the 4--12\,keV energy band (about 0.6\,mCrab). By surveying about $1\%$ of the sky every day, \art\  provides rapid alerts about new X-ray transients (e.g., \citealt{atel13571,atel14206,atel14219}) or new outbursts from known or poorly studied historical sources (e.g., \citealt{atel13606,atel14051}). These events are typically also detected by \erosita, which provides a more accurate localization for newly discovered sources and detailed spectral information below $\sim 2$--8\,keV for a broadband spectral analysis. 

The ART-XC telescope provides the unique possibility of studying the population of faint transients that would be otherwise missed because they are too weak for wide FoV telescopes and all-sky monitors [such as the {\it INTErnational Gamma-Ray Astrophysics Laboratory} ({\it INTEGRAL})/IBIS, {\it Swift}/BAT, and {\it MAXI}]. The relatively hard X-ray band of the \art\ telescope also makes it less dependent on the source intrinsic or Galactic absorption and allows the detection of highly absorbed sources, which could be missed by soft X-ray instruments. For example, \art\ transients  often appear quite unremarkable in \erosita\ data. Follow-up campaigns have already enabled establishing the nature of several \art\ sources, such as the new microquasar SRGA\,J043520.9+552226/AT2019wey \citep{yao21a,yao21b,Mereminskiy2021}, the new nova-like CV SRGt\,J062340.2$-$265715 \citep{Schwope2021}, several new Be systems \citep{doroshenko21,Lutovinov2021}, and a few others.

\subsection{Extragalactic transients and stellar flares} 

The approach of repeated sky surveys when every (typical) location on the sky is visited every six months has proven to be an efficient tool for studying the long-term variability of sources and for discovering various types of Galactic and extragalactic transients. Every 24 hours, \erosita\ detects between half a dozen and a dozen sources that have changed their luminosity by more than an order of magnitude compared to the previous visit six months earlier (e.g., \citealt{gilfanov20,sazonov20}; Medvedev et al. in preparation). About half of these sources are associated with {\it Gaia} stars, and the remaining half are presumably of extragalactic origin. The requirement of at least a tenfold flux increase corresponds to an effective transient detection threshold of $\sim 2\times 10^{-13}$\,erg/s/cm$^2$. 

As every scan path crosses the ecliptic poles, sources in these regions are repeatedly scanned every four hours. For these sources, we have already accumulated light curves covering over 16 months, which opens unique prospects of detailed variability studies of stars and AGN on timescales from several hours to a few years (Medvedev et al. in preparation). The sky scans are planned to proceed in an unchanged manner at least until the end of the third sky survey. 

Tidal disruption events (TDEs) are being actively searched for among extragalactic \erosita\ transients based on the optical and infrared properties of their hosts and on the shape of their X-ray spectra (Fig.~\ref{fig:tde_spec}). The Russian consortium identifies about one relatively bright TDE candidate every week \citep{khabibullin20,khabibullin20b,gilfanov20,Gilfanov2021}. They are followed up with optical spectroscopy on various telescopes in Russia (the 6 m BTA telescope in the Caucasus; the 1.5m Russian-Turkish telescope, RTT-150, in Turkey; the 1.6m AZT-33IK telescope of the Sayan Observatory; and the 2.5m telescope of the Caucasus Mountain Observatory of the Sternberg Astronomical Institute of the Moscow State University). An active collaboration with the Zwicky Transient Facility (ZTF) team facilitates timely classification of detected transients and their optical follow-up on Palomar and Keck telescopes. During the period from June 10 to December 14, 2020, spanned by the second \srg/\erosita\ all-sky survey, 16 TDEs have already been detected and optically confirmed in the $0<l<180^\circ$ hemisphere (\citealt{Sazonov2021}, Gilfanov et al., in preparation). The first \erosita\ TDEs have also been reported in the other half of the sky \citep{Liu2021}.

The \textit{SRG} scanning strategy means that more rapid transients may be found that vary on a timescale of $\text{about one}$ day, in particular, X-ray afterglows of gamma-ray bursts even without an observable gamma-ray trigger  \citep{Khabibullin_2012}. A systematic search for these events is continuously ongoing \citep[e.g.,][]{Wilms_2020}.

\section{SRG spacecraft}
\label{s:spacecraft}

Design of the structure and configuration of the {\it Spektr-RG} spacecraft continued until 2008 (see \S\ref{s:history} below about the history of the \srg\ project). At that time, the concept design was defined and accepted.

\subsection{{\it Navigator-SRG} platform}
\label{s:platform}

\begin{figure}
\centering
\includegraphics[width=0.98\columnwidth,clip]{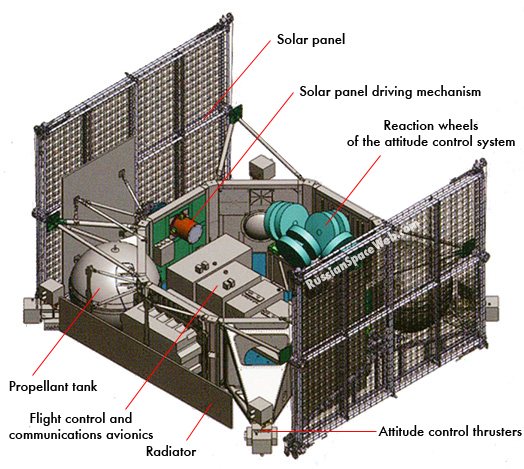}
\caption{Navigator platform. Credit: NPO Lavochkin, see also http://www.russianspaceweb.com/navigator.html (credit: A.Zak).}
\label{fig:platform}
\end{figure}

The spacecraft (Figs.~\ref{fig:srg_npol}, \ref{fig:flight}, and \ref{fig:srg_spacecraft}) is based on the {\it Navigator-SRG} space platform (Fig.~\ref{fig:platform}), which has been developed by NPOL for spacecraft of hydro-meteorological and scientific designation. By the time the \srg\ design was accepted, the flight models of the Navigator platform for the {\it Spektr-R} ({\it Radioastron}) and {\it Electro-L} No. 1 spacecraft had been produced and tested. In 2011 the platform started its flight qualification. This meant an opportunity to gather invaluable experience of the platform operation, which was considered during the creation of {\it Spektr-RG}. Modifications aimed at enhancing the mission reliability and achieving the required technical parameters of the spacecraft as a whole were made. A principal change made to the Navigator platform was the installation of a new X-band radiocomplex, induced by the necessity of maintaining ground contact with the spacecraft at large distances.

The Navigator-SRG platform includes the following units: an onboard control system, an onboard radiocomplex, solar panels with the orientation mechanism, a propulsion unit, a thermal control system, a telemetry system, an antenna-feeder system, a cable network, construction, a power supply system, and miscellaneous supplementary systems and units.

The onboard control system performs the following basic tasks: it controls operation of the scientific payload and service systems of the spacecraft, it controls the motion of the spacecraft around the center of mass and the motion of the center of mass, it controls the spacecraft orientation by choosing the correct omnidirectional antenna for linkage with ground stations, it controls the technical state and diagnostics of onboard systems with subsequent transfer of information to Earth via the onboard radiocomplex, it manages contingency cases, and performs automatic transition of the spacecraft into stand-by mode in contingency cases that cannot be managed by the onboard systems.

The onboard control system includes the following elements: an onboard computer, power automation units, a gyroscopic angular velocity sensor, two SDP-1 solar position sensors, two POS 347K solar orientation devices, two SED26 star trackers, and four Agat-15M reaction wheel sets.

The main orientation control actuators of the onboard control system are reaction wheels (RWs). Three RWs are in operation, and one RW is in cold reserve. A set of three RWs ensures an angular velocity up to 0.07\,deg/s. For RWs desaturation (momentum dump), 16 stabilization thrusters are used (8 operating and 8 in cold reserve, 0.5~N each). These thrusters are also used during the initial flight phase to establish the solar orientation and in contingency cases.

In addition to the stabilization thrusters, the propulsion unit also includes larger correction thrusters (5~N each) that are used for trajectory correction. All the thrusters are powered by hydrazine contained in two tanks, each of which is filled with 180\,kg of fuel.

Electric power supply for the onboard service systems and scientific instruments is provided by solar panels. The total power consumption of all onboard systems does not exceed 1700~W, which is less than the expected capacity of the solar panels at the end of their service lifetime (1870~W). The power supply system also includes a 55~Ah battery that provides power for onboard systems during the initial phase after separation of the spacecraft from the launch vehicle or in contingency cases.

Thermal control of the spacecraft is established by axial and contour heat pipes, as well as by electrical heaters that are controlled by the onboard control system. In addition, both telescopes have their own thermal control systems.

The \erosita\ and \art\ telescopes are placed on a geometrically stable frame mounted on the top joint of the spacecraft base module. A number of electronic units of \art\ and the gyroscopic angular velocity sensor device of the onboard control system are installed on a thermo-stabilized platform. 

\subsection{Orientation and operation regimes of the observatory}

Except during the initial flight phase, the onboard control system operates in inertial orientation mode, which ensures triple-axis stabilization and spacecraft rotation into a programmed orientation with respect to the inertial frame. Orientation data from the star trackers are continuously coupled with the gyroscopic sensor data, resulting in an estimated orientation quaternion.

In the inertial orientation mode, the following operations can be performed: spacecraft rotation for pointing telescopes (\srg\ +X~axis) to a predefined target on the celestial sphere and triple-axis stabilization with respect to the inertial frame; spacecraft rotation at a constant angular velocity (typically 0.025\,deg/s) around the \srg\ +Z axis, in which the rotation axis is constantly rotated at a low angular velocity ($\sim 1$\,deg/day) to ensure a smooth all-sky survey; and spacecraft triple-axis stabilization during trajectory-correction impulses.

\begin{figure}
\centering
\includegraphics[width=0.98\columnwidth,clip]{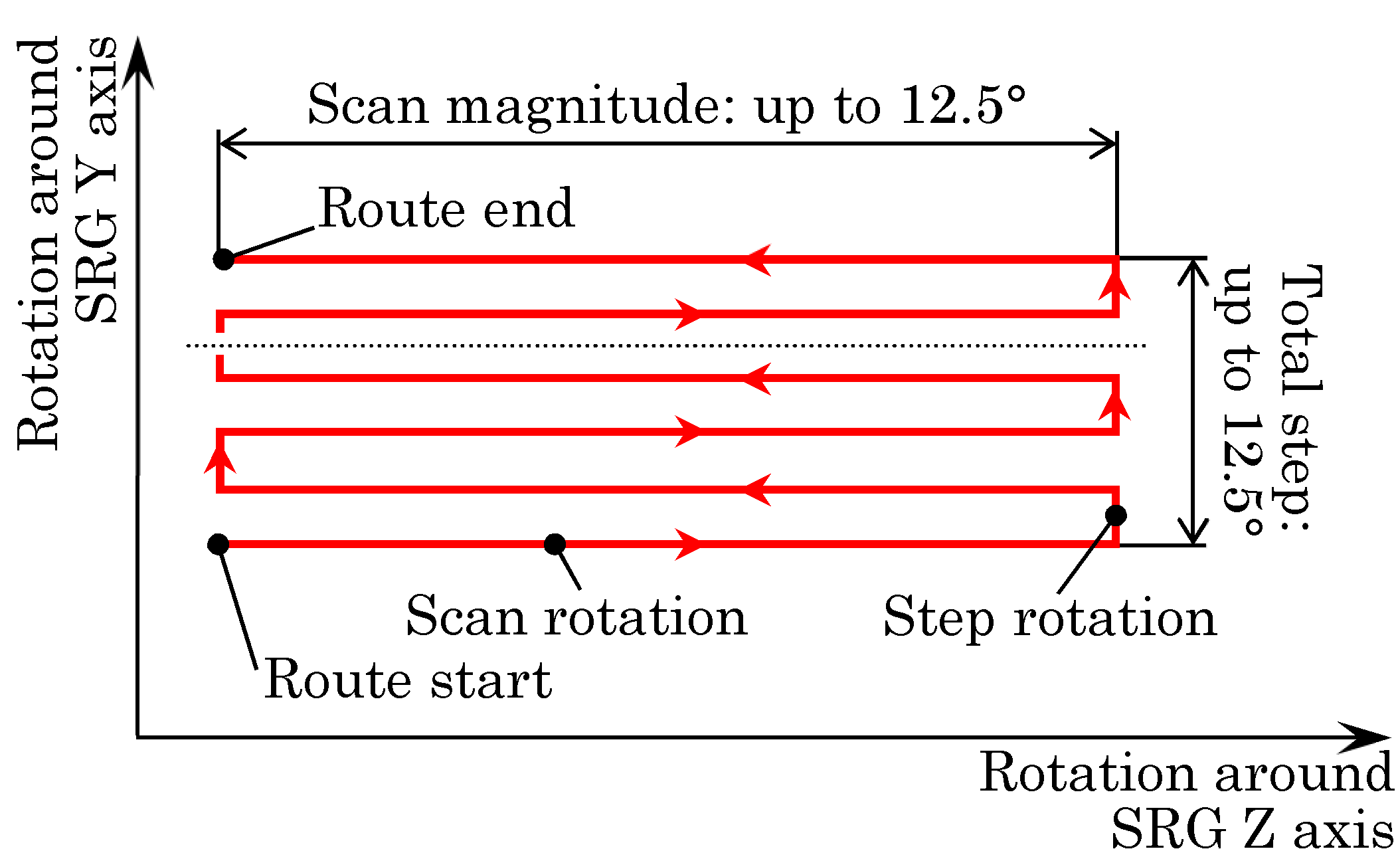}
\caption{Typical route of the X-axis in scan mode.}
\label{fig:srg_scan}
\end{figure}

Several consequent spacecraft rotations can be combined into series to scan fields on a celestial sphere by predefined routes. In this mode, the onboard control system points the \srg\ X-axis to the required starting point of a scan field (with respect to the inertial frame) and performs a program of relative rotations around the Y- and Z-axes of the spacecraft. During the relative rotations, the coupling of the star trackers and gyro data does not interrupt the precise route tracking with respect to the inertial frame. The scan field size can be up to $12.5^\circ\times 12.5^\circ$. A typical scan sequence consists of several S-turns (Fig.~\ref{fig:srg_scan}). The main scanning turns are performed around the Z-axis and smaller auxiliary turns (steps) are performed around the Y-axis. Step rotation can be arbitrarily low and is usually chosen from the 4--12~arcmin range. The highest angular velocity of the scanning turns may be as high as 0.04\,deg/s.

\begin{figure}
\centering
\includegraphics[width=0.98\columnwidth,clip]{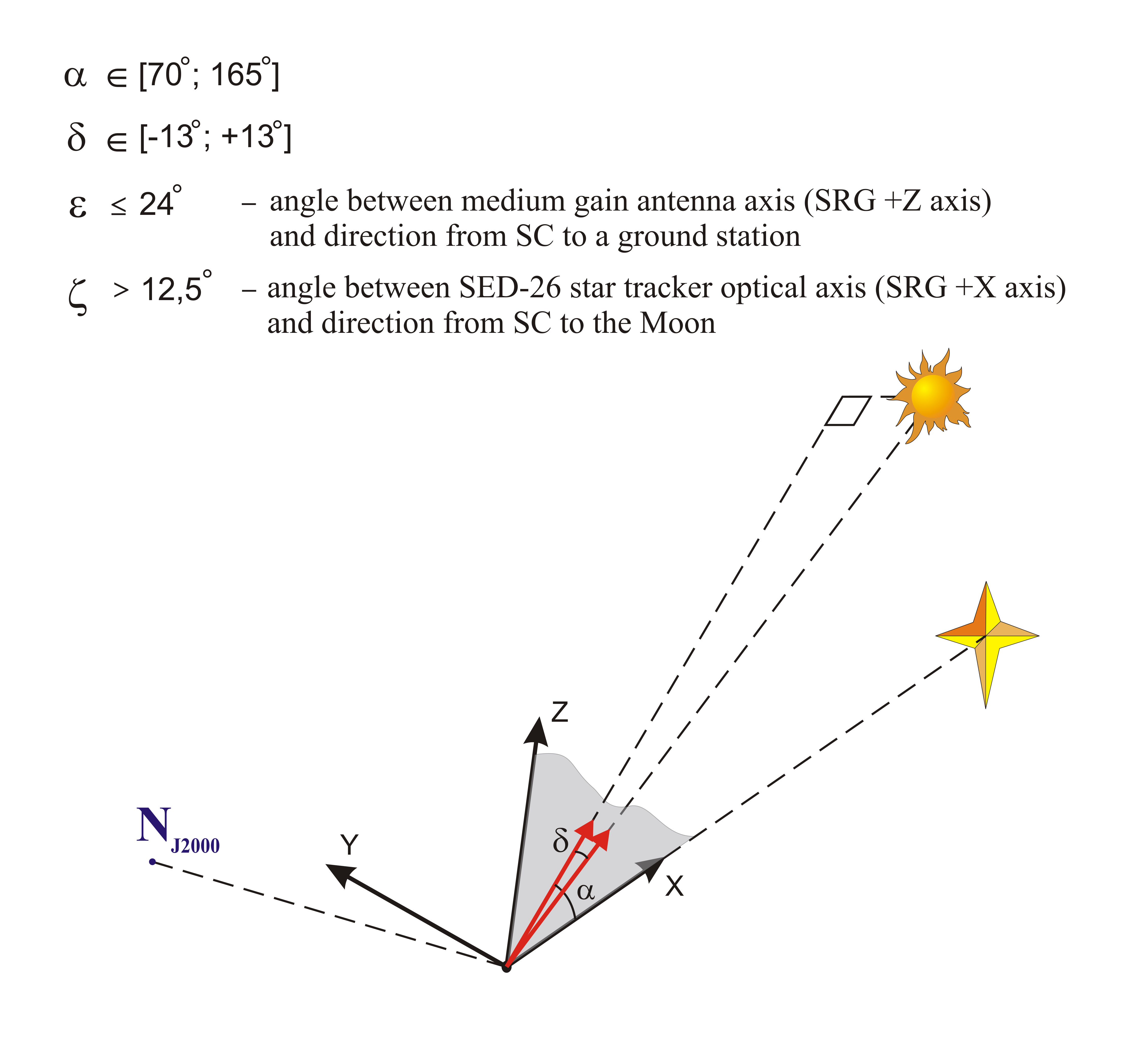}
\caption{Main orientation restrictions for observations during scans and pointings.}
\label{fig:main_restrictions}
\end{figure}

The spacecraft attitude-control error has been confirmed to be less than 10~arcsec. The angular velocity stabilization error does not exceed 0.72~arcsec/s. Figure~\ref{fig:main_restrictions} illustrates the main orientation restrictions on pointing and scanning observations of the \srg\ observatory.

\subsection{Radiocomplex and data transmission}

In addition to receiving commands and performing trajectory measurements, the onboard radiocomplex is also capable of combining scientific data and service telemetry into one data stream. The nominal rate of transmission of service telemetry data is 16~Kbit/s (real-time transmission of TM-data) and 32~Kbit/s (stored TM-data dump). 

The nominal rate of scientific data transmission is 512~Kbit/s, but it can also be performed at 64, 128, and 256~Kbit/s. Telecommands from ground stations are uploaded at a rate of 500~bits/s (typical) or 125~bits/s (contingency cases).

The radiocomplex receives telecommands through an omnidirectional antenna. It transfers data through a medium-gain antenna with a beam width of $\pm 24^\circ$.

The onboard telemetry system collects service telemetry sensor data and digital telemetry data from the onboard control system and writes it into the internal memory device. During ground contact, it is possible to dump the stored telemetry data or run live telemetry transfer mode.

\subsection{Launch and operational orbit insertion}
\label{s:launch}

\begin{table*}
\caption{Trajectory-correction schedule for the \srg\ flight to the vicinity of the L2 point.}
\label{tab:trajectory_flight}
\begin{tabular}{c|c|c|c|c|c}
\hline
No. & Date/time & Days in flight & Number of PU ignitions & Total characteristic speed & Consumed propellant \\
 & (Moscow) & & & m/s & kg \\
\hline
1 & Jul. 22, 2019   & 10 & 2 & 13.59 & 15.98\\
  & 17:30:00.000 & & & & \\
2 & Aug. 06, 2019  & 25 & 2 & 3.49 & 4.09\\
  & 17:30:00.000 & & & & \\
\hline 
\end{tabular}
\end{table*}

The \srg\ observatory operates in a quasi-periodic orbit around the L2 libration point of the Sun--Earth system. This orbit has a number of clear advantages for the mission related to the remoteness (1.5 million km, corresponding to a light travel time of 5\,s) of the L2 point from Earth in the direction away from the Sun, in particular, long visibility periods from ground stations for control and receipt of scientific data, and the prevention of periodical entries of the spacecraft into Earth’s radiation belts.

The location of the Russian ground-control stations in the northern hemisphere required choosing a launch date that would ensure maximum daily radio visibility of the spacecraft. In addition, the choice of an optimal operational orbit had to minimize the characteristic velocity required to keep the spacecraft near the L2 point. The formal solution based on these two criteria implied the necessity of preparing separate flight plans for the upper stage for each possible launch date. 

Taking into account the anticipated time frame of the readiness of the spacecraft segments, two possible launch windows were eventually chosen: June 21-22, 2019, and July 12-13, 2019. The launch of the \srg\ spacecraft from the Baikonur Cosmodrome successfully occurred on July 13, 2019, at 15:30:57 Moscow time.

The {\it Proton-M} launch vehicle placed the orbital unit (OU), including the upper stage (US), the adapter system, and the spacecraft, in a falling trajectory with an apogee altitude of $\sim 200$~km. The first ignition of the propulsion unit (PU) of the US placed it in a transfer orbit $\sim 168\times 2013$~km. The second ignition placed the OU in the nominal flight orbit toward the L2 point $\sim 500\times 1450000$~km. The spacecraft later separated from the US, followed by the deorbiting of the US. The placement into orbit took two hours. During this process, the US spent 34~minutes in the Earth’s shadow.
The flight toward L2 lasted approximately 100 days without any shadow intervals.

To ensure placement of the spacecraft in the nominal orbit, three flight trajectory corrections on the way to the L2 point had been planned: on days 10, 20, and 40 of the mission. The corresponding reserve dates were also defined: on days 15, 25, and 45. A characteristic velocity budget of 100~m/s was allocated for these corrections. The first correction was executed in accordance with the schedule. The second correction was moved to the reserve date to acquire more orbit measurements after the first correction. By the time the third correction was due, the calculations indicated that it was not needed, so that there was no correction on day 40 (see Table~\ref{tab:trajectory_flight}).

The operational orbit of the \srg\ observatory was defined by the following constraints: the maximum distance of the spacecraft from the L2 point in the ecliptic plane $\sim 920000$~km, the exit from the ecliptic plane $\sim 700000$~km and $\sim 550000$~km toward the north and south ecliptic poles, respectively.
After placement of the spacecraft into the nominal orbit near the L2 point (on day 100), orbit corrections are executed approximately every 50 days during the mission (see \S\ref{s:orbcorr} below). 

\begin{figure}
\centering
\includegraphics[width=0.95\columnwidth,clip]{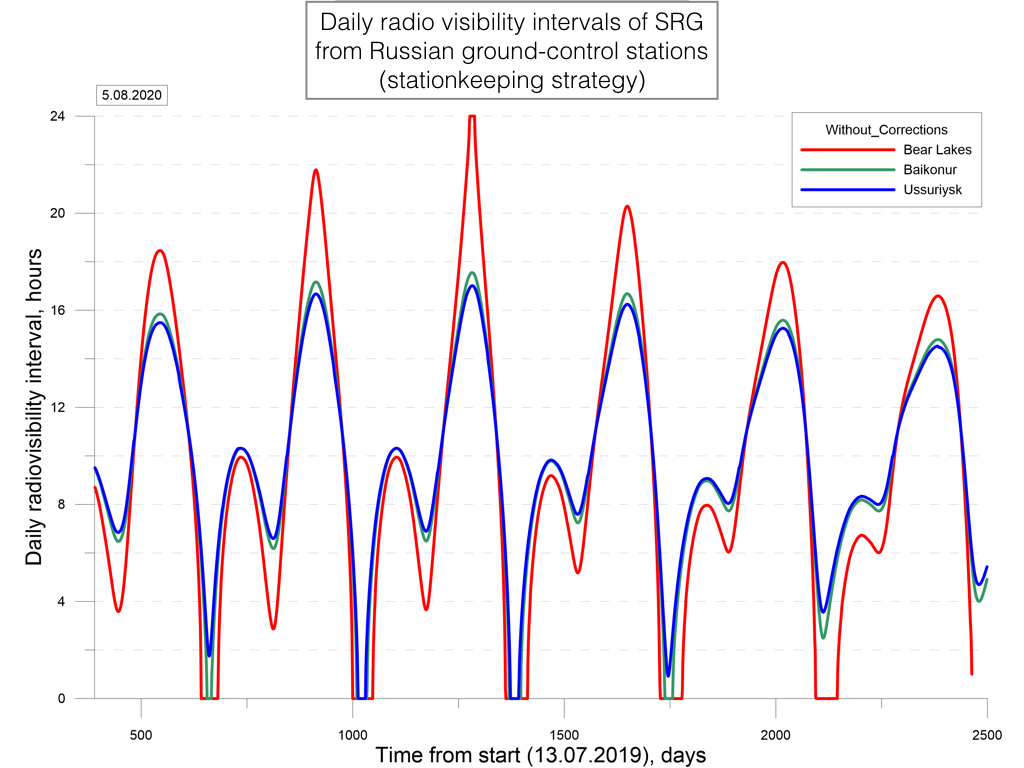}
\caption{Minimum visibility without corrections.}
\label{fig:vis_nocorr}
\end{figure}

\begin{figure}
\centering
\includegraphics[width=0.95\columnwidth,clip]{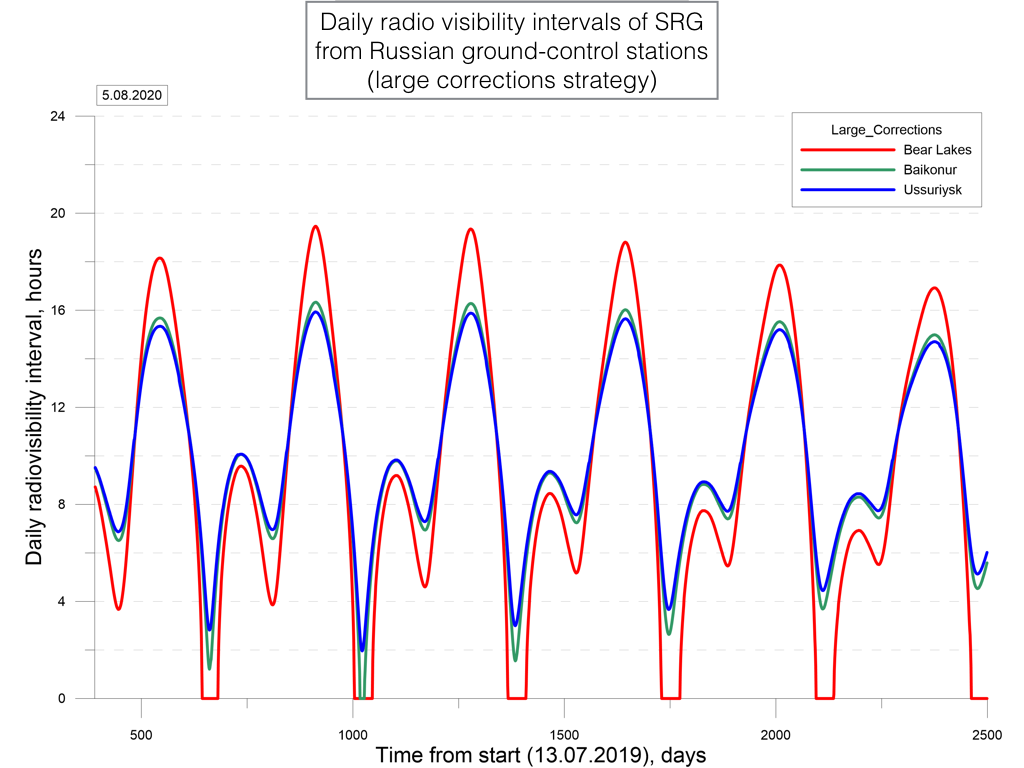}
\caption{Minimum visibility with large corrections.}
\label{fig:vis_corr}
\end{figure}

\subsection{Spacecraft control}

Spacecraft operation planning is performed on three timescales (long term, medium term, and short term). The degree of detail is to be increased when proceeding to the next stage.

Long-term mission planning is determined by science tasks formulated by the scientific community.
At the medium-term level, operations of the spacecraft and ground stations are planned for the following calendar month. Approximately two weeks before its initiation, NPOL performs a priori calculations of the spacecraft trajectory for this month, schedules preliminary intervals for spacecraft control sessions, and, if necessary, schedules technological operations for the spacecraft (which usually include only trajectory correction). NPOL delivers these data to the science operation and data center of IKI, where the month plan is completed with scientific operations in accordance with the long-term plan. The plan produced at this stage determines the sequence and parameters of the spacecraft orientation modes. The plan undergoes preliminary verification of consistency with the spacecraft and ground-stations operation limitations both on the side of IKI and NPOL. The spacecraft orientation limitations are the main verification criteria. In case of changes in the ground stations or in the onboard systems' condition, a monthly plan can be adjusted even during its implementation. In these cases, secondary verification and decision making usually takes no longer than three days.

At the short-term level, exact operations of the ground station and spacecraft during an upcoming ground-contact session are planned. The main task of short-term planning is to ensure necessary spacecraft operations starting from the beginning of a session until at least the next session. Operations scheduled in a month plan are detailed in the form of a ground-contact session program, which is a sequence of commands for ground stations, spacecraft service systems, and the telescopes. To compose a ground-contact session program, NPOL collects requests for commands from supervisors of the corresponding systems. In particular, a sequence of commands to ensure the telescopes' operation (with specific command parameters and issue times or intervals) is generated at IKI based on requests of the \art\ and \erosita\ teams. 

Commands for the spacecraft and scientific instruments may be executed immediately during a ground-contact session or stored in the onboard control system memory for execution at a preset time (time-tagged commands). NPOL performs a simulation of a prepared ground-contact session on the onboard control system's logical testing bench. It provides means of precise verification of a program consistency with spacecraft operation limitations, and detects logical planning errors and incorrect commands. Simulation is performed for an interval from the start of a planned control session until the start of the next session. This technology ensures end-to-end simulation during the whole mission, which increases control reliability.

Provided the simulation results are positive and all the parties endorse a control session plan, it is executed. The  execution of the commands is controlled using telemetry data. If necessary, the program may be adjusted during a session, including issuance of commands initially not planned by the \art\ and \erosita\ teams.

Ground-contact sessions take place daily. Each session lasts approximately 4--5 hours, which makes it possible to dump all scientific and service telemetry data gathered throughout the day, conduct trajectory measurements by at least two ground stations, check the status of onboard systems and scientific instruments, and upload a time-tagged command sequence for a subsequent period of up to several days. During a control session, the full real-time data stream from the spacecraft is received by NPOL through ground stations and transmitted to IKI.

\subsection{Mission ground-control complex}

Spacecraft control is provided by the Russian ground-control complex of the \srg\ spacecraft, including the NPOL mission control center, ground stations, ballistic centers, and means of communication. The central authority responsible for the flight tests and spacecraft control is NPOL, which provides a platform for the cooperation of specialists participating in flight tests and the implementation of the scientific program. 

Radio communication with the spacecraft is provided by radio-technical ground complexes located at Baikonur (TNA-57 antenna of 12~m diameter), Bear Lakes (TNA-1500 antenna of 64~m diameter), and Ussuriysk (P-2500 antenna of 70~m diameter). Currently, only the Bear Lake and Ussuriysk ground complexes are used for receiving scientific data. The radio-technical ground complexes had been modernized for the implementation of the \srg\ project and currently fully ensure the mission requirements. During periods of short daily radiovisibility intervals from the Russian Deep Space Network antennas (which usually happen in April and May), the Malargue, Cebreros, and New Norcia stations of the ESTRACK network are involved when necessary in receiving scientific data, according to the agreement between Roskosmos and ESA.

Ballistic maintenance of the mission is provided by two ballistic centers located at the Keldysh Institute of Applied Mathematics of the Russian Academy of Sciences and the Central Research Institute of Machine Building (TsNIIMash). All elements of the ground-control complex are united into one data system by means of a high-bandwidth data network, which ensures data exchange between the elements of the ground-control complex and the scientific ground complex. 

\subsection{Large corrections of the SRG orbit}
\label{s:orbcorr}

\begin{table*}
\caption{Trajectory-correction schedule for the orbiting of \srg\ near the L2 point.}
\label{tab:trajectory_l2}
\begin{tabular}{c|c|c|c|c|c}
\hline
No. & Date/time & Days in flight & Number of PU ignitions & Total characteristic speed & Consumed propellant \\
 & (Moscow) & & & m/s & kg \\
\hline
1 & Oct. 21, 2019 & 100 & 1 & 0.21 & 0.25\\
  & 19:00:00.000 & & & & \\
2 & Dec. 10, 2019 & 150 & 1 & 0.18 & 0.21\\
  & 19:00:00.000 & & & & \\
3 & Jan. 30, 2020 & 201 & 1 & 0.27 & 0.32\\
  & 19:00:00.000 & & & & \\
4 & Apr. 01,2020 & 263 & 1 & 1.00 & 1.13\\
  & 19:00:00.000 & & & & \\
5 & Jun. 16, 2020 & 339 & 1 & 0.95 & 1.07\\
  & 21:00:00.000 & & & & \\
6 & Aug. 05, 2020 & 389 & 1 & 0.94 & 1.08\\
  & 20:00:00.000 & & & & \\
7 & Oct. 05, 2020 & 450 & 1 & 3.01 & 3.38\\
  & 19:00:00.000 & & & & \\
8 & Nov. 23, 2020 & 499 & 1 & 6.23 & 6.99\\
  & 19:00:00.000 & & & & \\
9 & Feb. 28, 2020 & 588 & 1 & 6.24 & 6.94\\
  & 17:00:00.000 & & & & \\
  \hline
\multicolumn{6}{c}{Remaining propellant: 319.63\,kg} \\
\hline 
\end{tabular}
\end{table*}

After placement of the spacecraft into the nominal orbit near the L2 point (on day 100), station keeping corrections were executed approximately every 50 days until September 2020 (see Table~\ref{tab:trajectory_flight}). Because the launch date was chosen using many criteria, there is a gap of about a month every Spring in the radiovisibility of \srg\ in its nominal operational orbit from the Russian ground-control stations (see Fig.~\ref{fig:vis_nocorr}). In the Fall of 2020, a strategy of large manoeuvres began to be implemented (see Table~\ref{tab:trajectory_l2}) to alleviate this problem. This strategy was developed in accordance with \citet{canalias04} and consists of one test manoeuvre (3~m/s) and ten manoeuvres (6~m/s each). As demonstrated by calculations (see Fig.~\ref{fig:vis_corr}), it should enhance the visibility of \srg\ from the Russian Deep Space Network antennas during Spring in the next years. This will significantly improve the implementation of control operations with the Navigator platform and telescopes as well as the dumping of science data.

\section{Mikhail Pavlinsky ART-XC}
\label{s:art}

\art\ on board the \srg\ spacecraft is an X-ray grazing-incidence mirror telescope array. It was developed by the Space Research Institute (IKI) and the All-Russian Scientific Research Institute for Experimental Physics (VNIIEF). The NASA Marshall Space Flight Center (MSFC) produced the flight modules of the X-ray mirror systems. \art\ is designed for conducting an all-sky survey in the 4--12\,keV energy band and pointed observations of selected astrophysical objects in the 4--30\,keV energy band. 

\subsection{Design} 

\begin{figure}
\centering
\includegraphics[width=.98\columnwidth,clip]{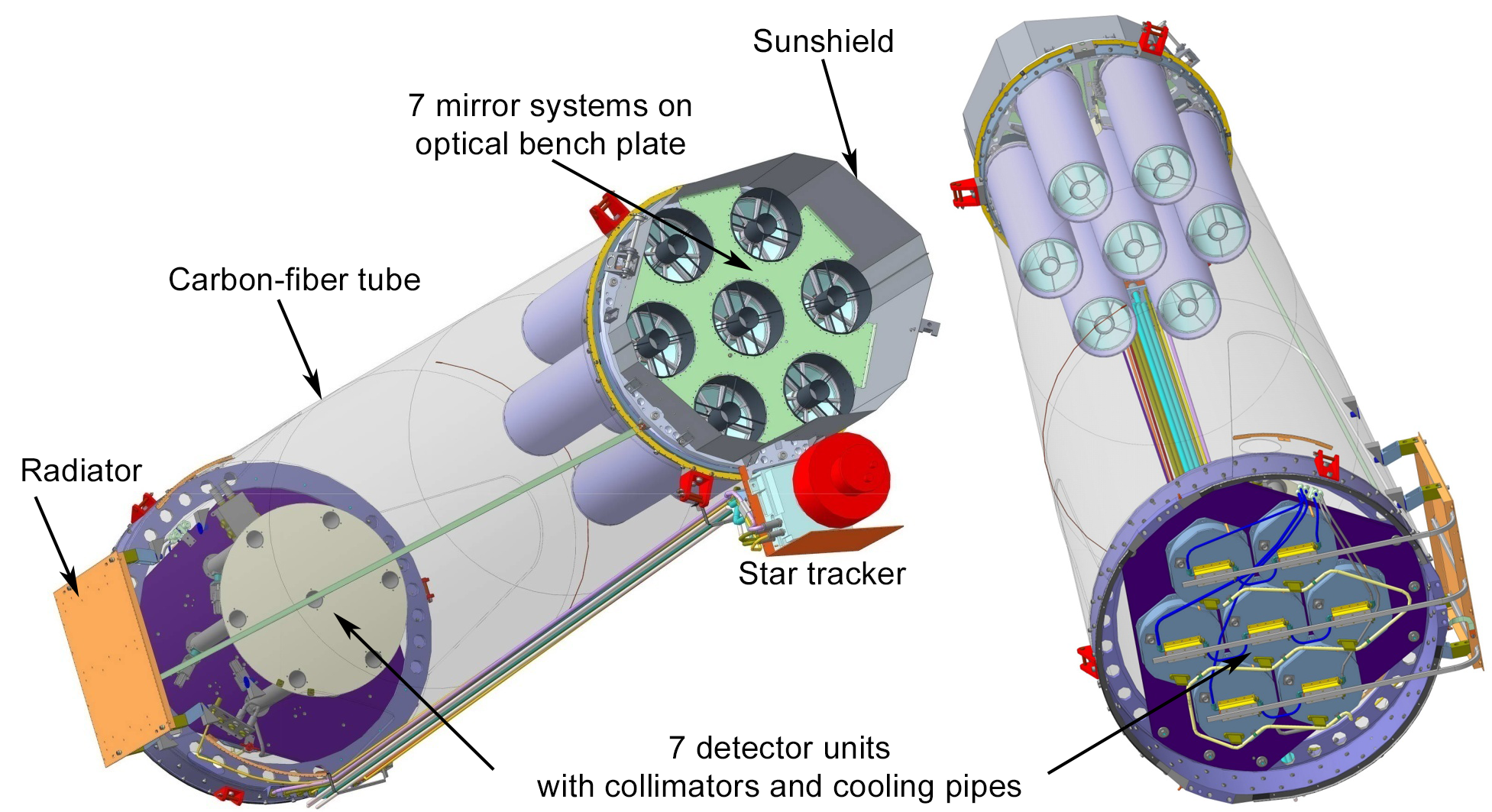}
\caption{Internal arrangement of the \art\ telescope: Cone-shaped carbon fiber tube with seven identical X-ray mirror systems at the top. The mirror systems focus X-rays onto seven detector units.}
\label{fig:artxc}
\end{figure}

\art\ consists of seven identical mirror systems (MS) paired with the unit of the Rontgen detector (URD) (Fig.~\ref{fig:artxc}). Each MS plus URD pair forms a telescope. All seven of them are pointing in the same direction. 

The basic structure of \art\ is a cone-shaped carbon fiber tube with a height of three meters. The MSs are mounted on top of this tube and focus X-ray photons onto their respective URDs. A sunshield protects the MSs from direct sunlight. The upper part of this tube is covered with copper shielding to reduce the stray-light background in the detectors. Each URD is equipped with a collimator to reduce stray-light background. The collimator includes a block of calibration X-ray sources ($^{55}$Fe + $^{241}$Am) for in-flight calibration. Heat pipes and a radiator are used to maintain the operating temperature of the detector at about {$-20\deg$C}. The star tracker is mounted near the MSs.

\begin{figure}
\centering
\includegraphics[width=0.8\columnwidth,clip]{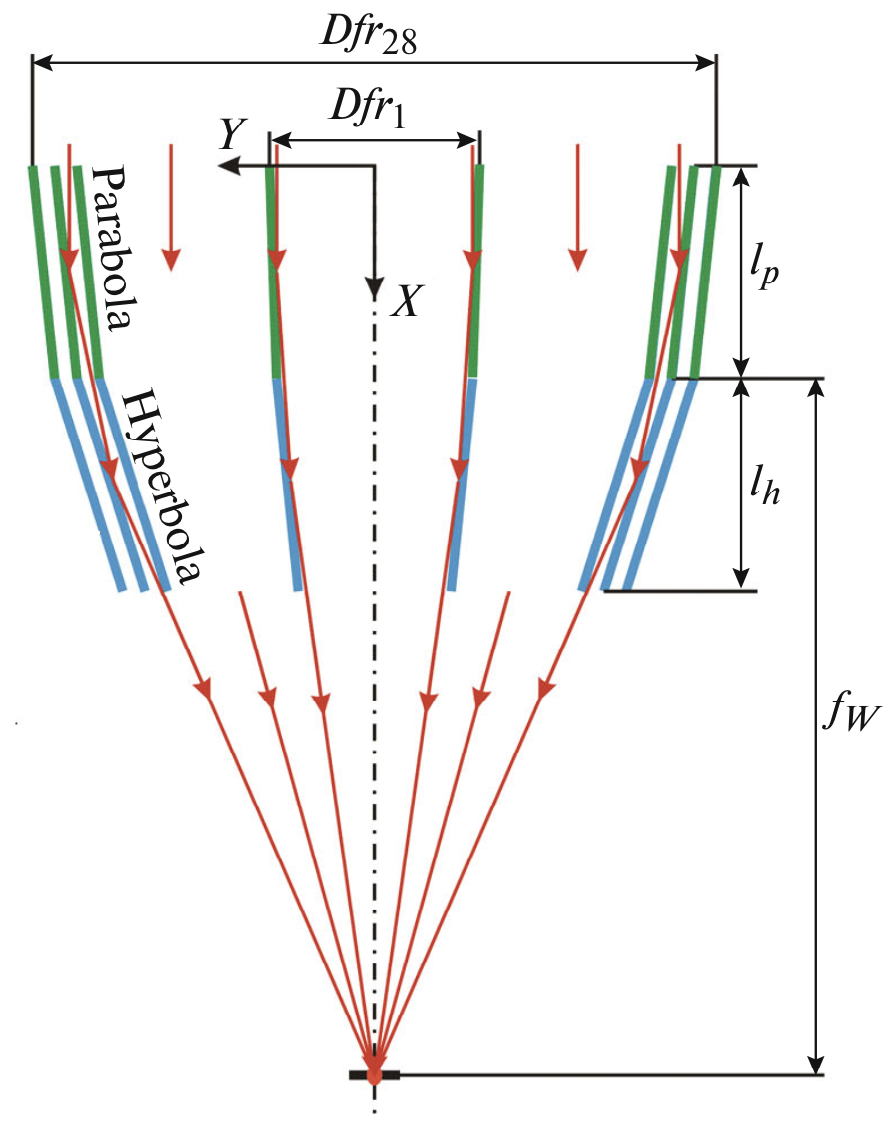}
\caption{{Wolter-I optical scheme of \art. $Dfr_{28}$ and $Dfr_1$ are the entrance apertures of the outermost and innermost shells, respectively, $l_h$ and $l_p$ are the heights of the paraboloids and hyperboloids, respectively, and $f_W$ is the focal length of the mirror system. Adapted from \citet{Pavlinsky2020}.}
}
\label{fig:wolter}
\end{figure}

The \art\ MSs were produced and calibrated by MSFC \citep{Gubarev_2012,Gubarev_2014,Krivonos_2017}. Each MS contains 28 Wolter-I (Fig.~\ref{fig:wolter}) nested-mirror shells. The shells were fabricated using an electroformed-nickel-replication technique and coated with a $\sim10$~nm layer of 90\% bulk density iridium. The shell thickness varies slightly with radius: the thickness of the outer shells is larger than the nominal thickness to make them stiffer and thus to improve the angular resolution of the module. The upper ends of the shells are glued to the supporting spider. The nominal focal length of the MSs is 2700~mm. They were defocused by 7~mm during installation on the telescope to provide a more uniform angular resolution across the FoV.

The detector system of \art\ consists of seven URDs, two electronic modules, and a serial interface connection module. The position-sensitive X-ray detector for \art\ was developed by IKI \citep{Levin_2014,Levin_2016}. The sensitive element is a double-sided strip detector (DSSD) based on a CdTe die with dimensions of $29.953\times29.953\times1$~mm$^3$. The high-quality CdTe dies were manufactured specifically for IKI by Acrorad Co. Ltd. (Japan). Coordinate resolution is provided by two mutually perpendicular sets of 48 strips on the two sides of the crystal. For signal acquisition, two VA64TA1 ASICs per detector are used, one on each side. The ASICs were manufactured by Ideas (Norway). More than 30 URDs were produced and tested at IKI. Seven of them were installed in the flight model of the \art\ telescope.

\subsection{Characteristics}
\label{s:art_char}

\begin{figure}
\centering
\includegraphics[width=\columnwidth,clip]{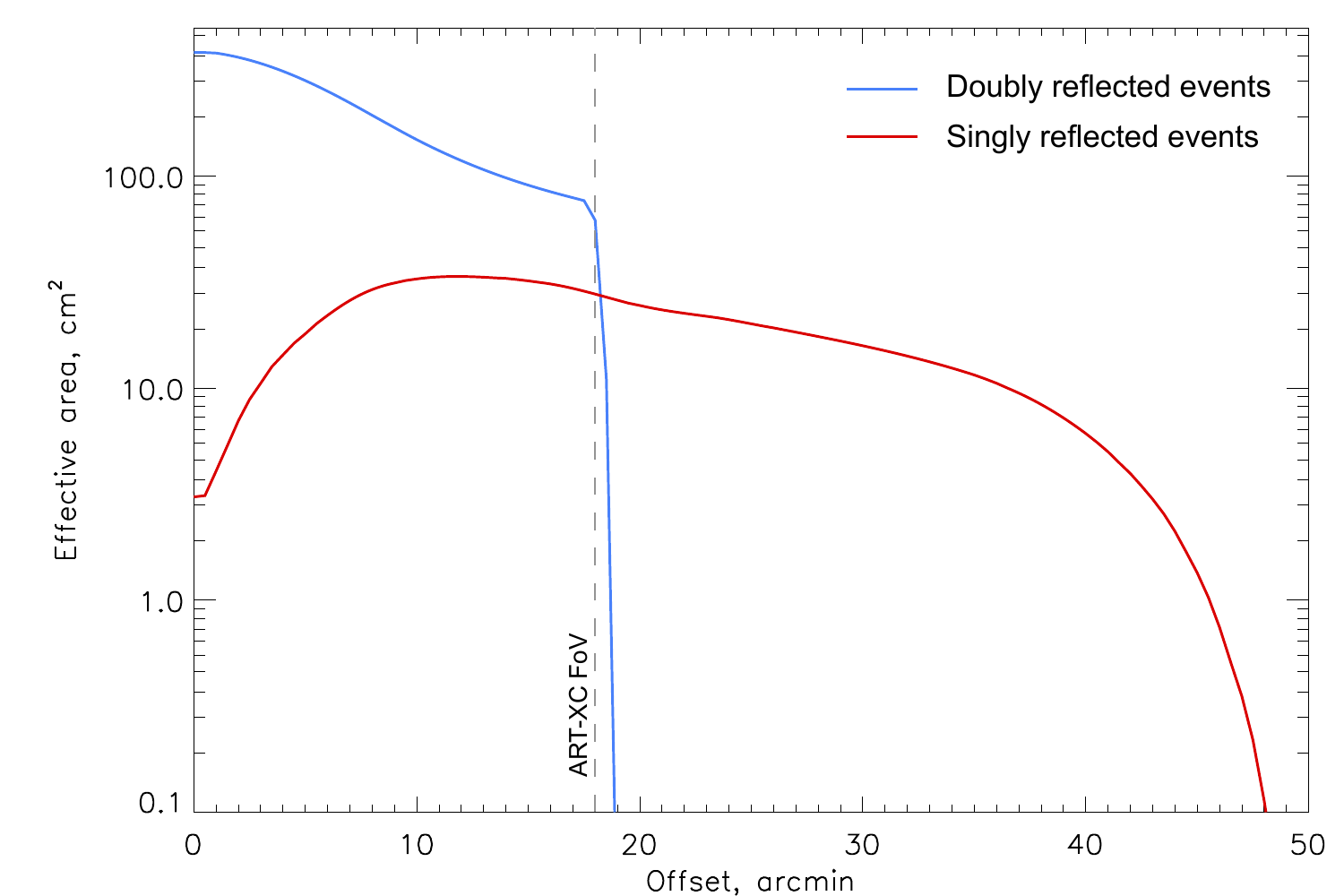}
\includegraphics[width=\columnwidth,clip]{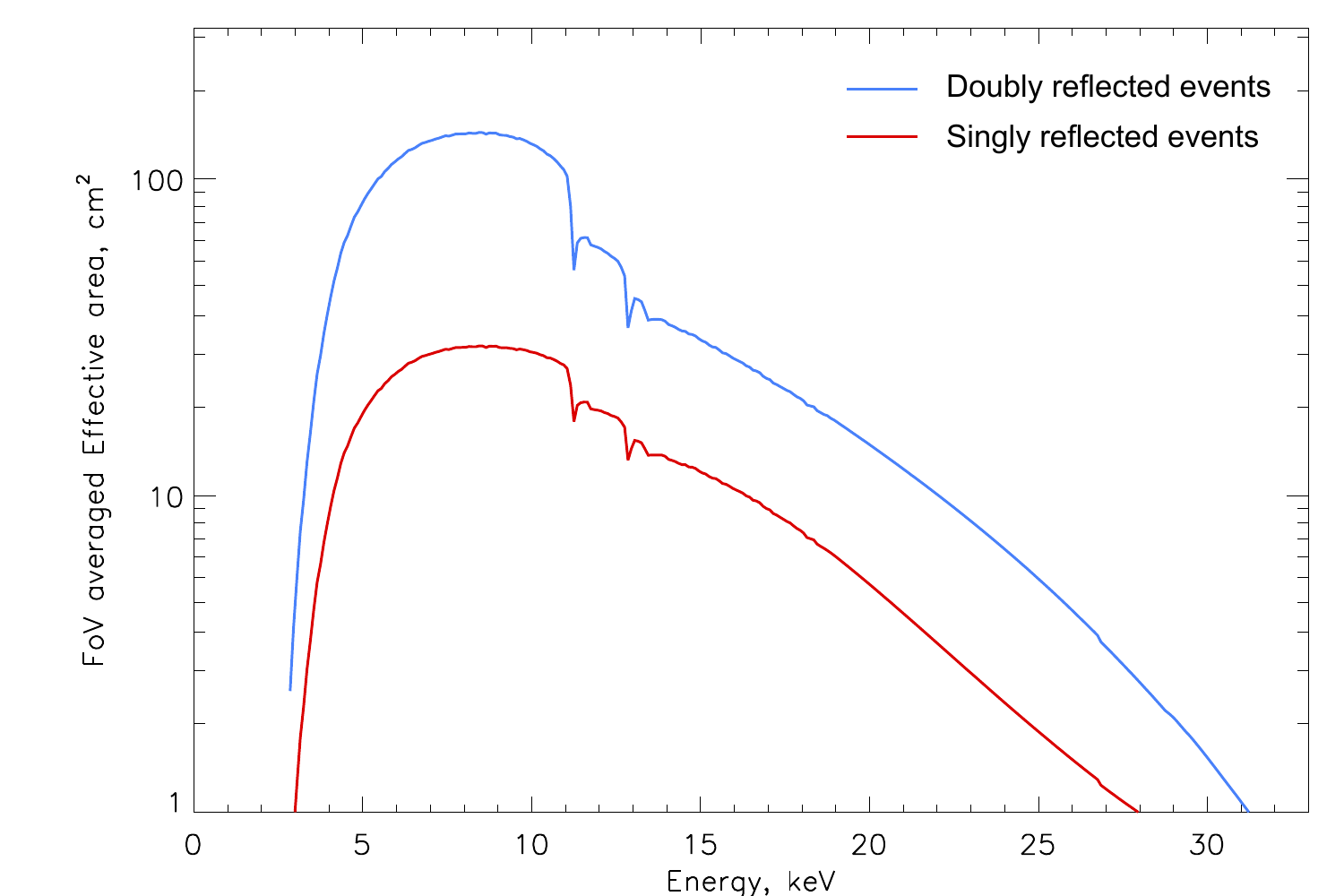}
\caption{\art\  effective area for doubly reflected events (blue) and singly reflected events (red) at 8.1\,keV as a function of offset angle (left). \art\  FoV-averaged effective area for doubly (blue) and singly reflected events (red) in the 4--35\,keV energy band as a function of energy (right). The graphs for the effective area are based on the effective area of the simulated mirror system and on the efficiency of the spare detector.}
\label{fig:art_effar}
\end{figure}

The \art\  FoV is $\sim36$ {arcmin} ($\sim0.3$\,sq.~deg), within which the MSs provide an angular resolution better than 1\arcmin. The effective area (on-axis) is substantial up to energies $\sim 30$\,keV. In addition, there are photons that are only once reflected from the MSs, and as a result, they can fall onto the detector with offset angles up to $\sim 50'$. Figure~\ref{fig:art_effar} illustrates the effect of these singly reflected photons on the \art\  FoV and their contribution to the FoV-averaged effective area. For this type of events, the \art\ \ FoV is $\sim2$\,square~degrees. Although true imaging cannot be done in this extended FoV and singly reflected photons generally cause an increase in the background, they can be used to measure the X-ray fluxes of bright sources. Therefore, the \art\ telescope can also be exploited as a concentrator. In survey mode, \art\ can therefore monitor bright transient sources for at least 28--32~hours \citep{Pavlinsky_2019b}.

{The URD can register photons with energies up to $\sim 100$\, keV, with an energy resolution of 9\% at 13.9 keV.}
The detector strip size corresponds to an angular size of 45\arcsec. During a calibration campaign at the IKI X-ray test facility \citep{Pavlinsky_2018,Pavlinsky_2019a,Pavlinsky_2019b}, it was determined that the efficiency of the \art\ spare URD reaches 50\% in the 4.47--4.76\,keV energy range and 90\% in the 9.43--10.04\,keV energy range.

Based on the model for the MS effective area and on the telescope ground calibrations, the on-axis effective area, vignetting, and grasp of \art\ have been estimated.
The \art\ on-axis effective area at 8.1\,keV is 385\,cm$^2$. The grasp at 8.1\,keV is 43.8\,cm$^2$~deg$^2$.
Table~\ref{tab:art_keypar} summarizes the performance parameters of the \art. 

\begin{table}[hb]
\caption{ART-XC characteristics.}
\label{tab:art_keypar}
\resizebox{0.95\columnwidth}{!}{
\centerline{\begin{tabular}{l|c}
\hline
Telescope mass & 350\,kg \\
Dimensions & 3.5~m $\times \diameter 0.9$~m \\
Power & 150~watts \\
Number of modules & 7 \\
Nominal focal length & 2700~mm \\
Operational energy range & 4--30\,keV \\
FoV & 0.3\,sq. deg \\
Eff. area for pointed observations & 385\,cm$^2$ @ 8.1\,keV \\
Grasp & 43.8\,cm$^2$~deg$^2$ @ 8.1\,keV \\
Ang. resolution (FWHM) in survey & $53\arcsec$ \\
Energy resolution & 9\% @ 13.9\,keV \\
Time resolution & 23\,$\mu$s \\
\hline
\end{tabular}}
}
\end{table}

\subsection{Scientific goals}
\label{s:art_goals}

The main goal of \art\ is to survey the whole sky in the broad X-ray energy range of 4--30\,keV with a sensitivity of $\sim 10^{-12}$\,erg~s$^{-1}$~cm$^{-2}$ ($\sim 10^{-13}$\,erg~s$^{-1}$~cm$^{-2}$ near the ecliptic poles) in the 4--12\,keV band and an angular resolution better than an arcminute. The \art\ survey will thus be the most sensitive all-sky survey ever conducted at these energies. The \art\ survey complements the all-sky survey performed by \erosita\ in the overlapping energy band of 0.3--8\,keV (with the highest sensitivity below 2\,keV). The harder energy band of the \art\ survey is particularly valuable for studying populations of heavily obscured astrophysical objects.

Preliminary estimates show that during the four-year all-sky survey, \art\ will detect $\sim 5000$ X-ray sources, mostly AGN, including heavily obscured ones (with hydrogen column densities $N_{\rm H}\gtrsim 10^{23}$~cm$^{-2}$). This will provide a rich database for explorations of the AGN population at $z\lesssim 0.3$. 

\art\ will also provide valuable information about the temperature of the intracluster gas in rich low-redshift clusters of galaxies. This will help tighten constraints on the cosmological parameters inferred from the \erosita\ all-sky survey. 

The \art\ all-sky survey can make breakthroughs in studying various classes of Galactic X-ray sources, such as X-ray binaries and cataclysmic variables (CVs). As many as $\sim 10^3$ CVs can be found, compared to $\lesssim 100$ known from previous X-ray surveys. 

With its unique combination of broad energy coverage, good angular resolution, and wide FoV, \art\ should make major advances in the study of SNRs and the Galactic Ridge X-ray emission. 

Finally, \art\ is well suited for discovering and monitoring transient and variable X-ray sources, such as X-ray and gamma-ray bursts, Galactic X-ray transients, and AGN. Although the probability of catching GRBs in the \art\  FoV is not high, bright GRBs can penetrate the shielding material of the telescope and induce a signal on the detectors. For these events, \art\ can provide precise timing information that can be used to localize them through triangulation with other space observatories.

\subsection{In-flight performance}
\label{s:art_flight}

\art\ has been operating in orbit for almost two years at the time of writing. The results obtained during the CalPV phase and the all-sky survey so far fully confirm the expected unique capabilities of the instrument. 

In particular, in order to calibrate the \art\ effective area, a series of observations of the Crab nebula was performed during the CalPV phase. The effective area proved to agree well with the results of simulations and ground calibrations  \citep{Pavlinsky_2018,Pavlinsky_2019a,Pavlinsky_2019b}. 


\section{eROSITA}
\label{s:erosita}

\subsection{History}
\label{s:erosita_history}

The concept of the \erosita\ telescope is based on a long series of previous scientific and technological developments dating back to the very successful German, US, and UK \textit{ROSAT} mission \citep[1990-1999;][]{Truemper1982}, which was developed and managed under the leadership of the Max Planck Institute for Extraterrestrial Physics (MPE). \textit{ROSAT} carried out the first complete survey of the sky with an imaging X-ray telescope in the energy range between 0.1 and 2.4\,keV, and it performed tens of thousands of pointed observations.  

The following generation of large X-ray telescopes, the NASA-led \textit{Chandra} Observatory and the ESA mission \textit{XMM-Newton}, required the development of mirror systems with longer focal length (7.5-10 meters) in order to bring harder X-ray radiation into focus. These observatories, both launched in 1999, were only able to perform pointed observations, {however, their} limited FoV meant that it was not possible to perform large-area surveys. This led to the proposal of a hard X-ray imaging telescope with all-sky surveying capabilities. The Astrophysikalisches Institut Potsdam (AIP), the MPE, and the University of T\"ubingen (IAAT) therefore proposed the mission called  A BRoad-band Imaging X-ray All-sky Survey (ABRIXAS). 

From the very beginning, the \textit{ABRIXAS} mission concept was developed through the coherent adaptation of mirror and detector technologies developed for \textit{XMM-Newton} for a national, small-scale satellite mission. Consequently, the development timescale for the project was relatively short ($\text{about three}$ years) and the overall costs moderate. The seven mirror modules shared in the focal plane an identical copy of the pn-CCD camera developed for \textit{XMM-Newton}, and with their small focal length of 1.6 meters, they were ideal for a small satellite. Because of a design failure in the power supply, however, the satellite lost its main battery soon after launch, in April 1999, and could never be used for scientific purposes. The hard- and software developments enabled by the \textit{ABRIXAS} mission have been extremely useful in the framework of a number of subsequent projects, however.

\begin{figure*}
\centering
\includegraphics [width = 0.45\textwidth,clip]{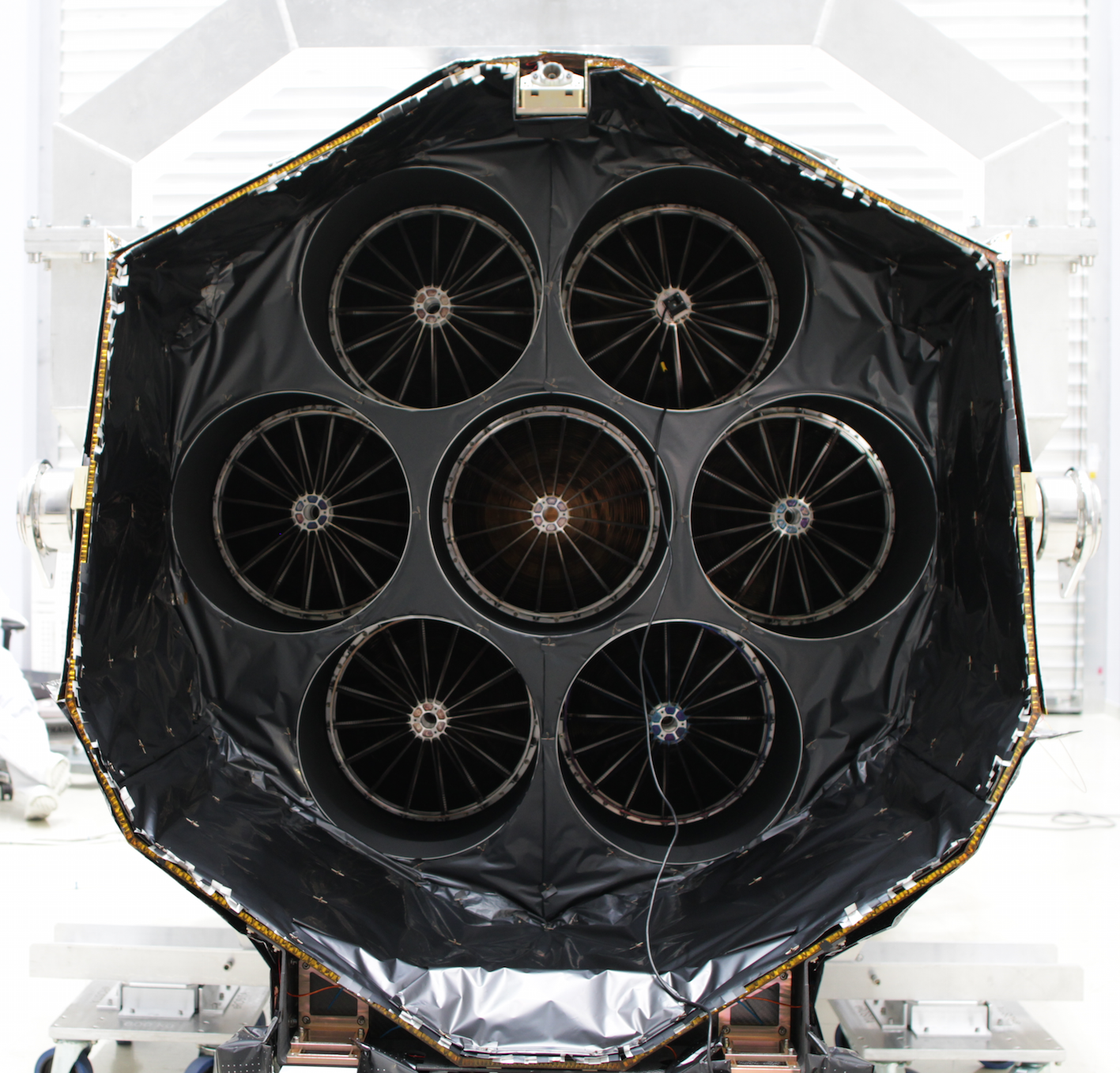}
\includegraphics [width = 0.45\textwidth,clip]{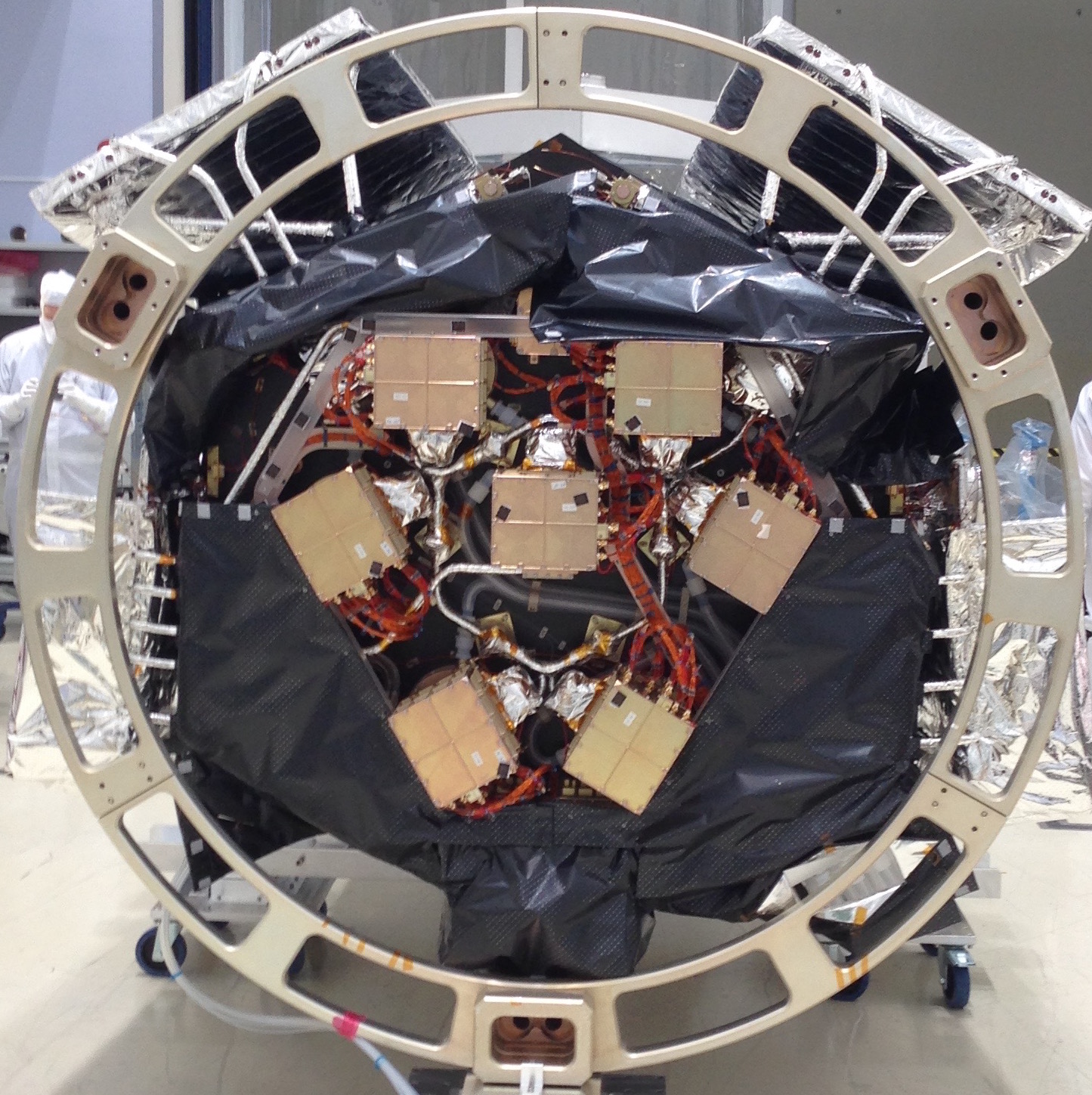}
\caption{Pictures of the fully integrated \erosita\ telescope taken during the final space qualification campaign in December 2016. Front view of the telescope (with open cover) with all seven mirror assemblies installed (left). Rear view of the telescope with all seven camera assemblies  installed (right).}
\label{fig:erosita}
\end{figure*}

Despite the unfortunate outcome of the \textit{ABRIXAS} mission, the appeal of the original scientific goal of an imaging hard X-ray all-sky survey remained high, with no other planned astrophysical mission with similar objectives on the horizon. With the help of the Semiconductor Laboratory (HLL) for the production of high-sensitivity detectors of the MPE, a new project to further develop the highly successful \textit{XMM-Newton} pn-CCD technology was launched. In March 2002, all participating institutes proposed to ESA to accommodate the ROentgen Survey with an Imaging Telescope Array (ROSITA) telescope on an external platform of the International Space Station (ISS). The seven mirror modules would have been built identically to those of \textit{ABRIXAS,}  but each of them would have been equipped with its own newly developed, frame-store pn-CCD in the focal plane. In September 2002, this proposal was supported by ESA with the highest scientific priority, and was recommended for phase A study. Launch would only have been possible from 2011 onward, however, because the external platforms of the ISS was occupied before then. Subsequently, it emerged that the ISS would not have been a viable option for ROSITA, first of all because NASA decided to terminate its shuttle flights to the ISS in 2010, and furthermore, because a contamination experiment on the ISS revealed that its environment was not safe from contamination for the sensitive X-ray mirrors and detectors of ROSITA.

At the turn of the millennium, the observations of supernovae type Ia by two independent groups (awarded with the Nobel Prize for Physics in 2011 for their discoveries) revealed that the expansion of the Universe is  accelerating, a fact that may suggest the existence of a cosmological constant. The subsequent measures, taken by Boomerang and \textit{WMAP}, of the tiny temperature fluctuations of the microwave background radiation out of which galaxies, clusters of galaxies, and the overall large-scale structure of the Universe originated, pointed toward a flat Universe, whereby the total matter plus energy content of the cosmos attains almost exactly the critical value. In fact, studies of the baryon fraction in X-ray selected (mainly from \textit{ROSAT}) clusters of galaxies throughout the 1990s had already shown compelling evidence that the matter density was lower than unity \citep{2003Schuecker}.

The fact that a very large sample of clusters of galaxies can be particularly useful for precision cosmology stimulated many different groups to conceive dedicated large-area cluster surveys. In April 2003, members of the ROSITA team participated in the proposal of DUO, a NASA SMEX-mission, based on a modification of the ROSITA telescope design. Together with another four missions, DUO was selected among 36 competing proposals for a phase-A study, carried out in 2004. DUO was to survey an area of the sky of about {6,000} deg$^2$, overlapping that explored by the optical Sloan Digital Sky Survey (SDSS). In this way, about {10,000} clusters of galaxies could have been detected, providing constraints on the dark energy density to an accuracy of within 10\%. NASA did not select the DUO-project for further development, however, and the only mission of the five that was finally executed within the SMEX program was the hard X-ray focusing telescope \textit{NuSTAR}, which was launched in 2012.

\begin{table*}
\caption{Basic eROSITA instrument parameters in launch configuration}             
\label{tab:erosita}
\centering          
\begin{tabular}{|c c|c c|c c|}     
\hline                    
\multicolumn{2}{|c|}{Instrument} & \multicolumn{2}{|c|}{7 Mirror Assemblies} & \multicolumn{2}{|c|}{7 Camera Assemblies}\\
\hline                    
Size & 1.9 $\diameter$ $\times$ 3.5\,m & Diam. of outer shell   & 358\,mm    & CCD image  & $2.88 \times 2.88$ cm$^2$ [$1\fdg{}03 \times 1\fdg{}03$] \\  
     Mass & 808\,kg              & Number of shells       & 54       & pixelsize & $75\mu\mathrm{m}\times 75\mu\mathrm{m}$ [$9\farcs{}6 \times 9\farcs{}6$]\\
     Power & 522\,W max.           & focal length           & 1600\,mm   & Time Resol. & 50 msec \\
         {Datarate} & 600\,MB/day max.   & HEW on axis (1.5\,keV)  & $18''$ & Energy Resol & 70\,eV [1\,keV] \\
                  &                  & HEW FoV averaged       & $26''$ & Qu-Efficiency & 95\% \\ 
\hline                  
\end{tabular}
\end{table*}

In February 2005, the Astronomy and Astrophysics Advisory Committee (AAAC), founded by the National Science Foundation (NSF), NASA, and the Department of Energy (DOE), and the High Energy Physics Advisory Panel (HEPAP), founded by NSF and DOE, established a Dark Energy Task Force (DETF) with the task of advising NSF, NASA, and DOE on the optimal strategies for the future of dark energy research. In particular, the DETF evaluated and compared various ground- and space-based instrumentation and observational methods. A dedicated white paper \citep{Haiman_2005} reached the conclusion that within available technologies at the time, it would have been possible to gather a sample of about 100,000 X-ray selected clusters of galaxies that would have provided very stringent constraints on the fundamental parameters of the cosmological model of the Universe. The construction of a sample of galaxy clusters of this magnitude is the primary objective of the \erosita\ telescope. To achieve this goal, the \textit{ABRIXAS} mirror design had to be modified to include 27 additional shells, doubling the diameter of each telescope module and increasing the effective area at low energies by a factor of five. With this fundamental rescope, \erosita\ will likely be the first stage IV dark energy probe to be realized and will outperform the DUO-like stage IV probe originally considered by the DETF.

In June 2006, the funding proposal for \erosita\ was submitted to the German Space Agency (DLR). Five German institutes (MPE, IAAT, AIP, the Hamburg Observatory, and the Dr. Remeis-Sternwarte in Bamberg, the Astronomical Institute of the Erlangen-N\"urnberg University) agreed to work together to develop, build, and organize the scientific exploitation of the instrument and formed the German \erosita\ Consortium. They were later joined by three more institutes: the Argelander-Institut f\"ur Astronomie at the University of Bonn, the Max-Planck-Institut f\"ur Astrophysik (MPA), and the Universit\"ats-Sternwarte M\"unchen (LMU). In March 2007, DLR approved the project and funded the \erosita\ telescope, while a Memorandum of Understanding was signed between DLR and the Russian Space Agency (Roskosmos) to ensure that \erosita\ would be launched in the framework of the {\it Spektrum-Roentgen-Gamma} mission. Soon afterward, the construction of the telescope began. In September 2008, Roskosmos came to a final decision on the payload orbit and launcher. \erosita\ would be launched by a Zenit-Fregat launcher (later in 2016 to be replaced by a Proton-M and Block-DM combination) together with the Russian hard-X-ray telescope \art\ on an L2 orbit to ensure maximum efficiency for a survey instrument. Eventually, additional funds from the Max Planck Society and DLR were secured for \erosita\ in July 2009, to finance the cost-driving L2 orbit. A detailed agreement between Roskosmos and DLR was signed the following month.

In the following years, the teams in Germany and Russia were intensively kept busy in the development and construction of the instrument. The mirror modules, smaller versions of the \textit{XMM-Newton} mirrors, were supposed to be built using the same 
technology and by the same industrial consortium. However, there were significant problems with the shorter focal length and the smaller mirror shells, as figuring errors have a significantly greater effect on the image quality. The mirror engineering model had only a 42 arcsec HEW, almost three times lower than required. Fixing these issues took more development than expected, but was ultimately successful. Another major challenge was the development of the electronics: after the decision for the L2 orbit, the basically completed design had to be changed to use radiation hard components. All qualification tests were carried out in the MPE test facilities or in the facilities of companies in the Munich area.

\erosita\ was finally delivered to  Russia in January 2017. A series of tests was then conducted at NPOL before integration of the instruments into \textit{SRG} could begin. This was then completed in April 2019 with the transport to the Baikonur Cosmodrome.

\begin{figure}
\centering
\includegraphics [width =0.95\columnwidth,clip]{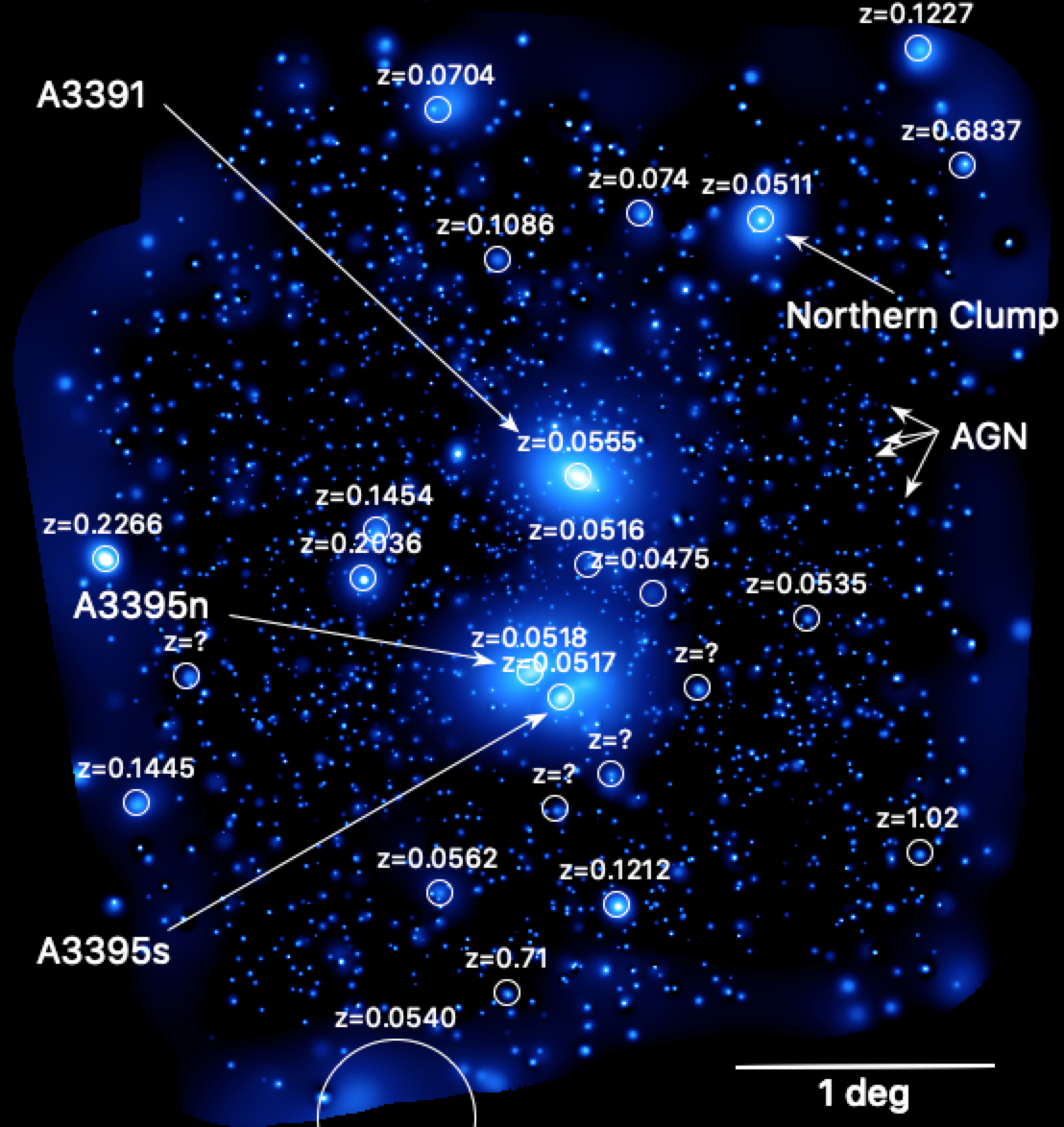}
\caption{\erosita\ 0.3--2.3\,keV wavelet-filtered X-ray image of the Abell 3391/95 system, also showing several clumps of diffuse gas at the same distance as well as several background galaxy clusters. One degree corresponds to 3.9\,Mpc at the redshift of A3391. From  \citet{reiprich21}.}
\label{fig:a3391}
\end{figure}

\subsection{Instrument}
\label{s:erosita_instrument}

\erosita\ (Fig.~\ref{fig:erosita}) consists of seven identical and coaligned X-ray telescopes  housed in a common optical bench in a hexagonal shape. A system of carbon fibre honeycomb panels connects the seven mirror assemblies on the front side with the associated seven camera assemblies on the focal plane side. The optical bench is connected to the S/C bus through a hexapod structure. 

Each of the mirrors comprises 54 paraboloid or hyperboloid mirror shells in a Wolter-I geometry, with an outer diameter of 360 mm and a common focal length of 1\,600 mm \citep{Friedrich2008,Arcangeli2017}. The average on-axis resolution of the seven mirror modules as measured during the on-ground calibration is $16.1''$ half-energy width (HEW) at 1.5\,keV. The unavoidable off-axis blurring typical of Wolter-I optics is compensated for by a 0.4~mm shift of the cameras toward the mirrors. This places each telescope slightly out of focus and leads to a slight degradation of the on-axis performance ($18''$), but improves the angular resolution averaged over the FoV. Each mirror assembly has a CCD camera in its focus \citep{Meidinger2014}. The \erosita\ CCDs are advanced versions of the EPICpn CCDs on \textit{XMM-Newton}. They have $384\times384$ pixels in an image area of $28.8\,\mathrm{mm} \times 28.8\,\mathrm{mm}$, yielding a square FoV of $1\fdg{}03 \times 1\fdg{}03$. Each pixel corresponds to a sky area of $9\farcs{}6 \times 9\farcs{}6$. The nominal integration time for all \erosita\ CCDs is 50\,msec. The additional presence of a frame-store area in the CCD  substantially reduces the amount of so-called out-of-time events, which are recorded during the CCD read-out, a significant improvement with respect to the PN camera on \textit{XMM-Newton} \citep{Strueder2001}. For optimal performance during operations, the CCDs are cooled down to about $-85^\circ$ by means of passive elements \citep{Fuermetz2008}. 

For calibration purposes, each camera has its own filter wheel with a radioactive $^{55}$Fe source and an {aluminum--titanium} target. This provides three spectral lines at 5.9\,keV (Mn-K$\alpha$), 4.5\,keV (Ti-K$\alpha$), and 1.5\,keV (Al K$\alpha$). 

The electronics for onboard-processing of the camera data is provided by seven sets of camera electronics (CE), each one mounted and interfaced to the cameras. Each of the CEs provides the proper voltage control and readout timing of the associated camera and performs the onboard data processing within the time constraints of the camera integration time.

\begin{figure*}
\centering
\includegraphics [width = 0.45\textwidth,clip]{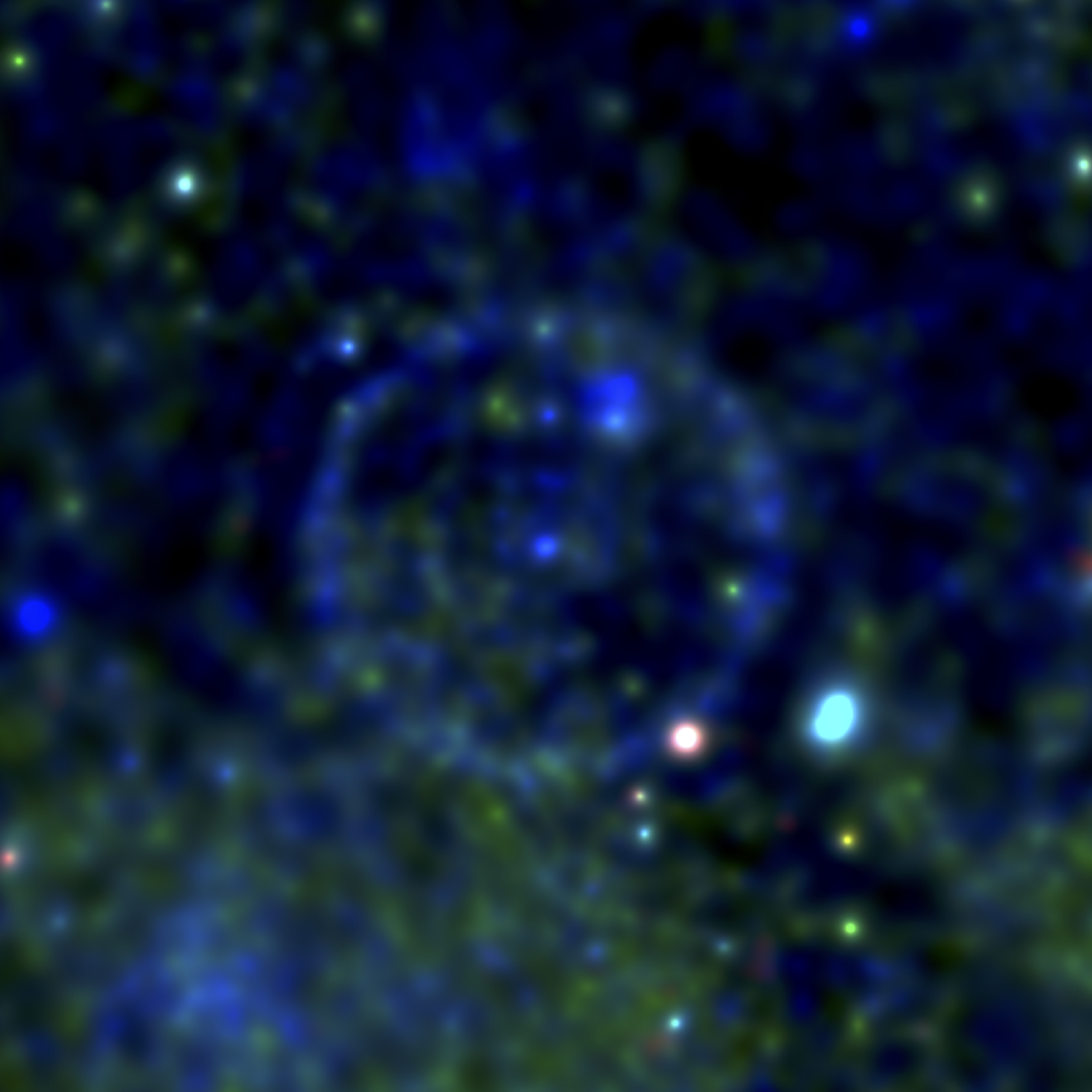}
\includegraphics [width = 0.45\textwidth,clip]{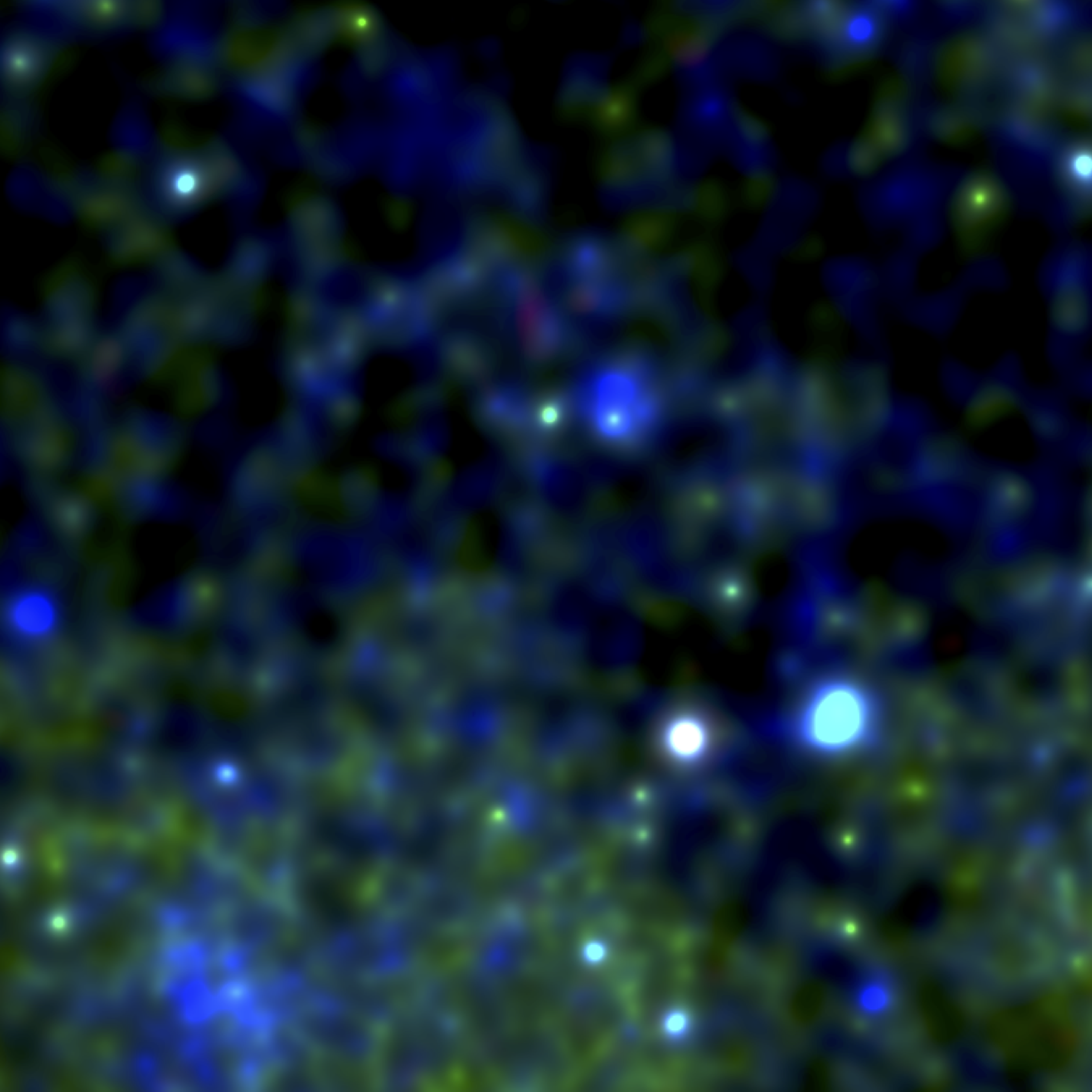}
\caption{False-colour images of the {the first \erosita\ all-sky survey (eRASS1, left) and the second \erosita\ all-sky survey (eRASS2, right)} observations in the energy bands 0.2--0.6\,keV (red), 0.6--1.0\,keV (green), 1.0--2.3\,keV (blue), adaptively smoothed. The size of the images is 3 degrees $\times$ 3 degrees. The circular structure is a dust scattering ring, echoing the outburst of the X-ray binary MAXI J1348-630. From \citep{lamer21}.}
\label{fig:ring}
\end{figure*}

The Interface and Thermal Controller (ITC) receives the telemetry generated by each CE and stores it in the mass memory, commands each of the CEs, and controls the power distribution and the temperatures of the mirrors and cameras.
Given its criticality, the ITC is a cold redundant unit \citep{Coutinho2018}.

Finally, two (redundant) star trackers are mounted on \erosita\ for the determination of an accurate boresight. The dimensions of the telescope structure are approximately 1.9\,m (diameter) $\times$ 3.2\,m (height in launch configuration, with closed front cover). The total weight of \erosita\ is 808\,kg. Table~\ref{tab:erosita} summarizes the basic \erosita\ instrument parameters.

\begin{table*}
\caption{Summary of performance characteristics of the \erosita\ telescope and its survey sensitivity. The background counts are based on the first all-sky survey data. For eRASS:1 and the PV eFEDS 140 deg$^2$ survey { \citep{2021arXiv210614517B}}, the flux sensitivity in each band has been computed by taking all sources detected above a likelihood of 8 (soft band) or 10 (hard band), and measuring the flux below which the logarithmic number counts start declining. For eRASS:8 the predictions are based on detailed simulations that include all instrumental effects and particle background intensity consistent with that measured at L2. For each field or region, we quote the total (unvignetted) exposure in seconds. The corresponding effective (vignetted) exposures can be computed by dividing the total exposure by 1.8 and {3.31} for the soft and hard band, respectively.
}             
\label{tab:erosita_sens}
\renewcommand{\arraystretch}{1.3}        
\begin{tabular}{|c|c|c|c|}     
\hline
    \multicolumn{2}{|c|}{}    & \multicolumn{2}{c|}{Energy Range} \\
    \cline{3-4}
    \multicolumn{2}{|c|}{} & Soft Band & Hard Band \\
    \hline
    \multicolumn{2}{|c|}{} & 0.2--2.3\,keV & 2.3--8\,keV \\
    \hline
    \multicolumn{2}{|c|}{FoV averaged effective area [cm$^2$]} & 1,365 @ 1keV & 139 @ 5\,keV \\
    \hline
    \multicolumn{2}{|c|}{Total Background [10$^{-3}$ cts/s/arcmin$^2$]} & $\approx$ 3.7 & $\approx$ 2.1 \\
    \hline
  \multicolumn{4}{l}{Point source sensitivity eRASS:1} \\
  \hline
  Ecliptic Equatorial region & Total exposure [s] = 200 & $5 \times 10^{-14}$\,erg/s/cm$^2$ & $7 \times 10^{-13}$\,erg/s/cm$^2$ \\
  \hline
  Ecliptic Polar region & Total exposure [s] = 4000 & $7 \times 10^{-15}$\,erg/s/cm$^2$ & $9 \times 10^{-14}$\,erg/s/cm$^2$ \\
    \hline
    \multicolumn{4}{l}{Point source sensitivity eFEDS} \\
  \hline
  eFEDS field & Total exposure [s] = 2500 & $9 \times 10^{-15}$\,erg/s/cm$^2$ & $1.3 \times 10^{-13}$\,erg/s/cm$^2$ \\
    \hline
  \multicolumn{4}{l}{Point source sensitivity eRASS:8 (predicted)} \\
  \hline
    Ecliptic Equatorial region & Total/Effective exposure [s]= 1600 & $1.1 \times 10^{-14}$\,erg/s/cm$^2$ & $2.5 \times 10^{-13}$\,erg/s/cm$^2$ \\
    \hline
  Ecliptic Polar region &  Total exposure [s] = 30000 & $2.5 \times 10^{-15}$\,erg/s/cm$^2$ & $4 \times 10^{-14}$\,erg/s/cm$^2$ \\
\hline
\end{tabular}
\end{table*}

\subsection{Early operations, first results, and performance}
The commissioning phase of \erosita\ had the objective of switching on all subsystems to verify they were functional following the launch and that they performed as expected to fulfill the scientific objectives. This phase served not only to test and commission the complete \erosita\ telescope, but also for the  teams in Khimky and Moscow (NPOL, IKI) and Garching (MPE) to learn and update the procedures on how to safely operate the spacecraft and the telescopes in space. The first mission-critical event was the switch-on of the ITC, which had to happen less than four hours after launch, to enable the thermal control of mirrors and electronics. The second event was the opening of the telescope cover, which occurred on July 22. Camera switch-on, instead, had to wait several days before being activated to avoid excess contamination from the first two spacecraft burns, which occurred on days 10 and 25 of the mission. In addition, camera cooling could not be performed without a minimum of 21 days of out-gassing following the cover opening. The commissioning of the cameras lasted about two months, including time necessary to perform a series of tests of the functionality of the camera electronics and thermal balance system.

All seven \erosita\ X-ray telescope modules had been observing the sky simultaneously since October 13. Over the following eight weeks, \erosita\ collected its first-light images and performed a series of observations designed to accurately calibrate the instruments and verify that the performance of the telescope met pre-launch expectations. The Russian and German science teams jointly defined this CalPV program, which included a combination of pointed observations, field scans, and full-circle scan tests.

Figures~\ref{fig:a3391} and \ref{fig:ring} show two examples of images captured by \erosita\ during the CalPV and early survey phase. They highlight the main properties of this unique X-ray telescope, namely, its ability to take deep images, which are highly sensitive to both point-like and diffuse emission, over very large areas of the sky. As a preview of the \erosita\  capabilities, a mini-survey called \erosita\ Final Equatorial Depth Survey (eFEDS) was devised as part of the PV plan, imaging a 140 deg$^2$ patch of the sky to a depth comparable to that expected at the end of the all-sky survey (see Table~\ref{tab:erosita_sens}). These data confirm the sensitivity of the X-ray telescope to its main target classes with high accuracy. This mini-survey revealed more than 20,000 point-like X-ray sources, about 80\% of which are distant AGN harboring growing supermassive black holes. Most of the remainder are X-ray active stars.

Finally, on December 13, 2019, the all-sky survey began. It was completed on June 12, 2020, after 182 days of almost continuous scanning of the sky. Eight all-sky surveys are planned in total, each delivering an average exposure with \erosita\ of about $200 \mathrm{s}/cos(lat)$, where $lat$ is the ecliptic latitude, while the area of 1\,square degree around the ecliptic poles is revisited every four hours. This accumulates an exposure of about 30~ks per survey. 

During the course of the all-sky survey, the spacecraft has been rotating continuously with a scan rate of 90 deg hr$^{-1}$, giving a four-hour period per revolution. The rotation axis is oriented to the neighborhood of the sun, with an average progression in ecliptic longitude of about 1 deg day$^{-1}$, thus completing one all-sky survey in half a year. The scan speed guarantees that the angular resolution is not degraded by smearing of photons during the 50~ms CCD read-out cycle, and it provides sufficient overlap between individual scans to enable source variability analysis and homogeneous survey exposure.

A preliminary analysis indicates that more than one million X-ray point sources and about 20,000 extended sources are detected in the survey. This is comparable to and may indeed exceed the total number of X-ray sources known before \erosita. About 80\% of the point sources are distant AGN (comprising ~80\% of all known blazars, among others), and 20\% are coronally active stars in the Milky Way. 

In summary, during its first year of operations in space, most technical, operational, and scientific design characteristics of the \erosita\ instrument onboard \srg\ have been validated. Table~\ref{tab:erosita_sens} describes the main performance characteristics of \erosita\ based on the data collected in this period during the PV phase and the all-sky survey. Compared to the pre-launch estimates of \citet[table 4.4.1 there]{Merloni2012}, the performance closely matches the expectations in the soft energy band, while it is slightly poorer in the hard band, mainly because the level of particle background is higher.

\section{History of the SRG project in Russia} 
\label{s:history}

\subsection{International X-ray astronomy projects in Russia}

The history of international X-ray astronomy space projects in the USSR and Russia began with the {\it Roentgen} observatory on the {\it Kvant} module of the {\it Mir} space station. This project had been proposed by the Space Research Institute (IKI) of the USSR Academy of Sciences for implementation in the framework of the Soviet Intercosmos program. A number of European institutions were invited to this project: (i) the University of Birmingham (UK) and the SRON Netherlands Institute for Space Research, which built the TTM coded-mask-aperture X-ray telescope, sensitive in the 2--25\,keV energy band, (ii) the MPE (Germany), which provided the HEXE hard X-ray spectrometer, sensitive in the 20--120\,keV energy band, (iii) the  European Space Research and Technology Center (ESTEC) of ESA, which built the GSPS high-pressure-gas spectrometer. IKI constructed the Pulsar X-1 hard X-ray detector for this project.

The {\it Kvant} module was put into orbit by a Proton rocket and successfully docked with the space station in April 1987. The main scientific result of the {\it Roentgen} observatory was the discovery in August 1987 of hard X-ray emission with an unusual spectrum from the very bright supernova SN 1987A, which had exploded five months earlier in the Large Magellanic Cloud \citep{Sunyaev_1987a,Sunyaev_1987b}. The detected photons were emitted as gamma-ray lines associated with the decay of radioactive cobalt 56. During the passage through the optically thick shell of the supernova, these photons experienced multiple Compton scatterings off relatively cold electrons and lost their energy due to recoil. As a result, an extremely hard power-law X-ray spectrum formed, and this spectrum was detected for several months by HEXE and Pulsar X-1. At energies below 20\,keV, photoabsorption of photons by ions of heavy elements, first of all iron and cobalt, came into play. This explained the absence of a detectable signal by TTM. 

In 1987--1995, TTM discovered many transient and persistent sources, which now wear the names ``KS'' ({\it Kvant} source, \citealt{sunyaev91a}). The availability of instruments sensitive in different bands of the X-ray spectrum allowed broadband spectra in the 2--200\,keV energy range to be constructed for very many bright transient and persistent sources in binary stellar systems, including black holes and neutron stars (X-ray pulsars and neutron stars with weak magnetic fields) \citep{sunyaev91b,sunyaev94}. It was demonstrated for the first time how strongly the spectra of X-ray binaries depend on the nature of the accreting object. TTM provided excellent timing observations of X-ray pulsars \citep{gilfanov89} and high-quality X-ray images of the Galactic center region.

The second X-ray orbital observatory of IKI within the Intercosmos program was the {\it GRANAT} satellite, built by NPO Lavochkin and put into an elongated elliptical orbit around Earth with a four-day period by a Proton rocket. It operated in orbit from December 1989 to May 1999. The payload of {\it GRANAT} included the SIGMA hard X-ray and gamma-ray telescope with a germanium spectrometer, the PHEBUS gamma-ray burst detectors developed in France, the WATCH all-sky monitor (Denmark), and the ART-P instrument constructed by IKI and its subsidiary in Frunze (now Bishkek, Kyrgyzstan), sensitive in the 2--25\,keV energy band. Both SIGMA and the ART-P telescopes had a coded mask and position-sensitive detectors, permitting construction of the images.   

The main results of the {\it GRANAT} mission include the maps of the central region of the galaxy in hard X-rays by the SIGMA (40--100\,keV) and ART-P (3--30\,keV) telescopes \citep[e.g.,][]{Churazov_1994,1993ApJ...418..844G,Pavlinsky_1994}. WATCH discovered the extremely interesting black hole X-ray binary GRS~1915$-$105 (``GRS'' means that it is a {\it GRANAT} source, \citealt{Castro_1992}). 
The spatial correlation of the 8--22\,keV X-ray emission observed by ART-P with the distribution of molecular gas in the Galactic center region became the first indication of the X-ray echo of the past Sgr~A* X-ray activity \citep{1993ApJ...407..606S} that is scattered by H$_2$ molecules in the dense gas surrounding it, according to the prediction by \cite{1980SvAL....6..353V}. PHEBUS observations helped to clearly demonstrate \citep{Tkachenko_2002} the existence of two types of gamma-ray bursts: short and normal, which, as we know now, is closely related to the parameters of the stellar objects that are responsible for their appearance (merger of neutron stars in the case of short bursts and collapse of massive stars in the case of normal bursts). 

In 2002, according to the agreement between ESA and Roscosmos, a Proton rocket placed the {\it International Gamma-ray Laboratory} ({\it INTEGRAL}) into a high-apogee orbit. {\it INTEGRAL} continues to provide scientific data to this day. Russian scientists are granted 25\% of the observing time of this mission. The prominent  results of {\it INTEGRAL} include the detailed spectroscopy of the electron-positron annihilation radiation \citep{churazov05} and the first detection of gamma-ray lines from the type Ia supernova SN2014J \citep{churazov14}.

\subsection{Beginning of the \textit{SRG} project in the USSR}

\begin{figure}
\centering
\includegraphics[width=0.98\columnwidth,clip]{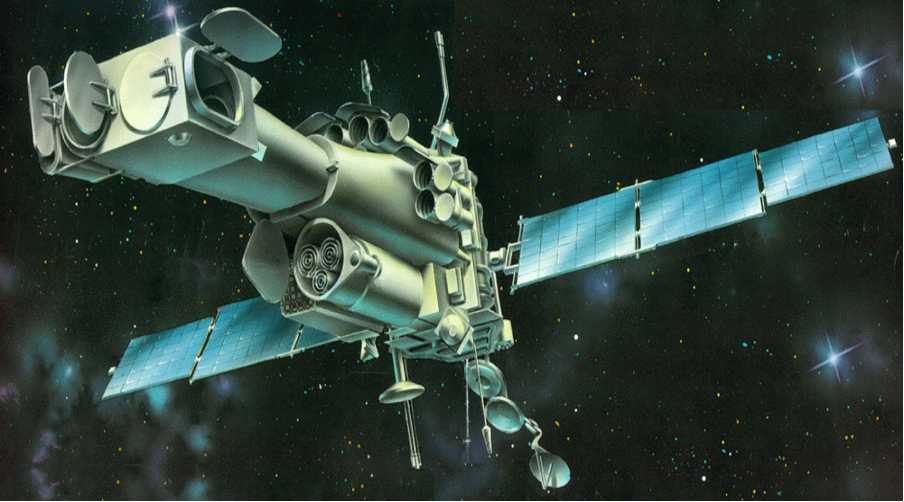}
\caption{Former \srg\ project.}
\label{fig:old_srg}
\end{figure}

In 1987, the thirtieth anniversary of the launch of the first artificial satellite of Earth ({\it Sputnik}) was celebrated. On this occasion, a large international meeting was held at IKI, where Intercosmos announced a competition for the payload of two orbital observatories. This resulted in a decision to support projects {\it Radioastron} and {\it Spectrum-Roentgen-Gamma} (\srg). The \srg\ project was supported by a number of renowned Soviet physicists, including Ya.B. Zeldovich, A.D. Sakharov, and the IKI director R.Z. Sagdeev. In addition to the High Energy Astrophysics Department of IKI, which proposed the \srg\ project, it was joined by Denmark, the UK, Italy, MPE (Germany), NASA (USA), Finland, Switzerland, Israel, and Turkey. The work on the project was in full swing (see Fig.~\ref{fig:old_srg}), but huge changes in the USSR led to a slowdown in the works on the spacecraft and instruments in 2001--2002, after which the project was terminated.

In 2003, the Space Council of the Russian Academy of Sciences considered and supported a more modest project with a launch by a Soyuz rocket (instead of a Proton), a smaller number of instruments, and with a shift in the scientific goals toward cosmology, fine X-ray spectroscopy, and monitoring the whole sky in X-rays. The idea of attracting international cooperation to the project was also supported. The project was then considered by Roscosmos, and a key decision was made to elaborate it further. It was also decided to retain (for the sake of continuity) the former project's name \srg, although the new project did not foresee the presence of instruments sensitive in the gamma-ray range.

Among the first invited to discuss a possible participation in the project were scientists from the MPE in Germany, with whom the High Energy Astrophysics Department of IKI has closely worked during the {\it Roentgen} mission on {\it Mir}/{\it Kvant} and on the first version of {\it SRG}. Moreover, MPE had developed and built the X-ray telescope for the extremely successful {\it ROSAT} satellite, which obtained wonderful X-ray maps of the sky in 1990. 

MPE proposed using a modification of the telescope that was developed by MPE for the unfortunately failed {\it ABRIXAS} mission for observations of clusters of galaxies (the most massive objects in the Universe, which are of interest to the Russian side as well). This telescope and its detector were designed for exploring the sky in a significantly harder energy band (from 0.5 to 10\,keV) compared to {\it ROSAT} (see a more detailed discussion in \S\ref{s:erosita_history}). By this time, the number of massive clusters of galaxies required for the detection of baryonic acoustic oscillations in their spatial distribution had been calculated at the Max Planck Institute for Astrophysics (MPA) and was proved to be close to 100,000 \citep{Hutsi_2006}. This would require finding virtually all massive clusters of galaxies in the observable Universe. This question was actively discussed at MPA in connection with searches for the baryonic acoustic oscillations predicted by theorists \citep{Peebles_1970,Sunyaev_1970} and with emerging hopes to find a large number of galaxy clusters through the Sunyaev--Zeldovich effect using ground-based telescopes and spacecraft operating in the microwave band. Therefore the question arose about a telescope with a higher sensitivity and working in a softer X-ray energy band compared to the existing modifications of the instrument for the {\it ABRIXAS} satellite, which could not provide the necessary sensitivity to fulfill this task, nor the majority of other tasks that were of interest for cosmologists at the time. 

After long discussions, the Russian side agreed on drastic increases in the size, mass, (800\,kg) and power consumption of the German instrument to fulfil the requirements (according to elaborations by G. Hasinger and P. Predehl) posed by the planned scientific results of the sky survey. The parameters of the Russian ART-XC instrument were also growing in parallel.

On March 23, 2007, Roscosmos and DLR signed a memorandum about the inclusion of the \erosita\ instrument into the payload of the \srg\ orbital observatory. This opened the way to intensive work on the project and the telescope. 

\subsection{Data sharing between the participants of the \srg\ project}

All data obtained by the \art\ telescope belong to the scientists of IKI, who have developed the telescope. The data within an area of 200\,square~degrees around the north ecliptic pole (where the sensitivity of the all-sky survey is highest) are processed together by scientists at IKI and the MSFC, in exchange for the delivery by the latter of a subset of the \art\ X-ray mirrors.

\begin{figure}
\centering
\includegraphics[width=0.98\columnwidth,clip]{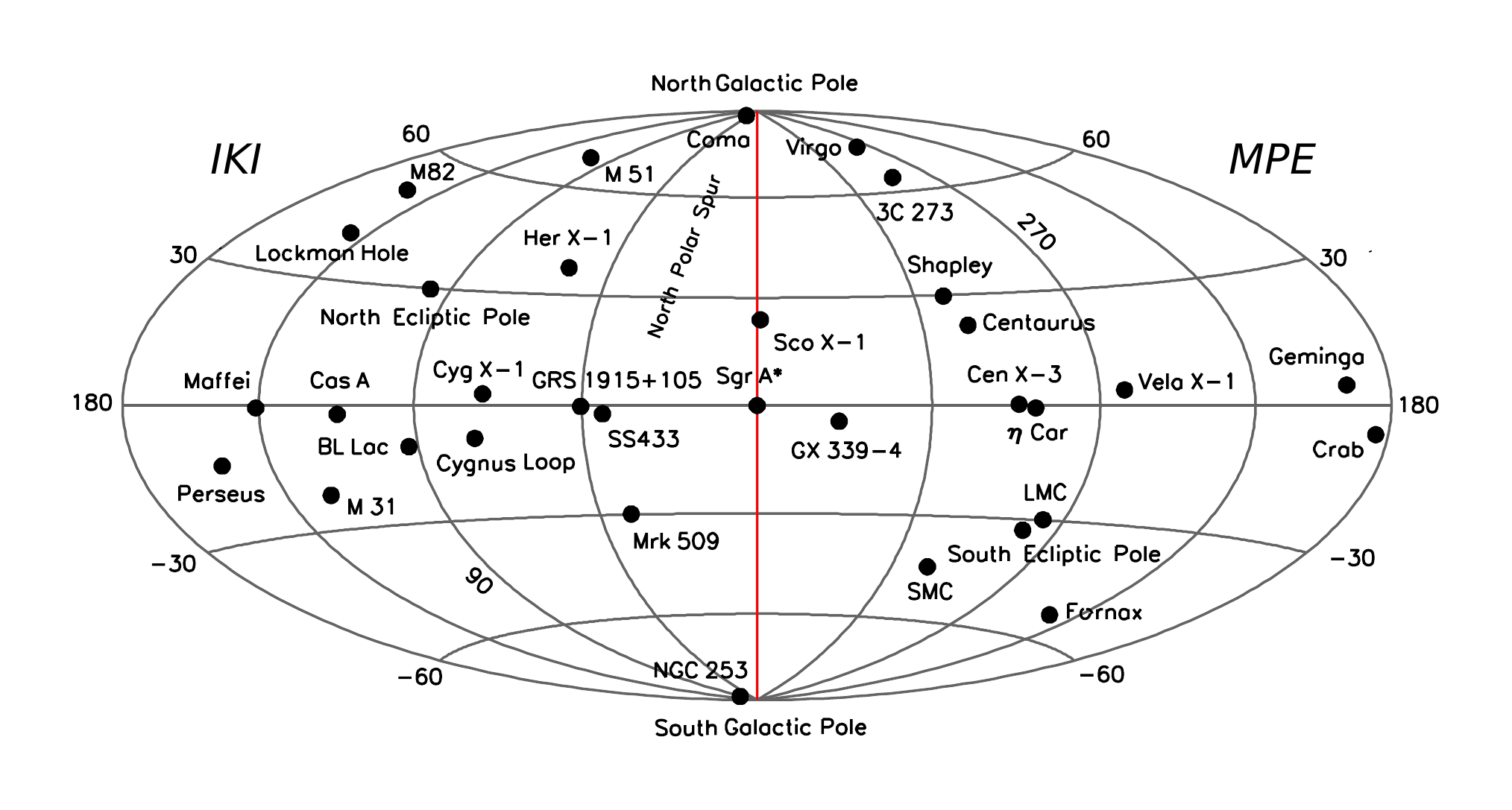}
\caption{Division of the \srg/\erosita\ data between the German and Russian consortia (denoted MPE and IKI, respectively). The border vertically follows in Galactic coordinates through the central supermassive black hole Sgr~A*.}
\label{fig:division}
\end{figure}

According to the memorandum of 2007 between Roscosmos and DLR, the \srg/\erosita\ data belong in equal parts to the German and Russian scientists. Because the primary goal of \srg/\erosita\ consists of constructing all-sky X-ray maps and creating X-ray source catalogs, a decision has been made that the German scientists are responsible for the processing and publication of data in one hemisphere of the sky, and the Russian scientists provide this in the other hemisphere (see Fig.~\ref{fig:division}).

\section{Conclusion}

The \srg\ observatory has been operating in orbit for almost two years at the time of writing. The \art\ and \erosita\ telescopes have already scanned 40\% of the entire sky for the fourth time. The experience of 25 months of operation of \srg\ allows us to say that the initial plans of the observatory are being successfully implemented and the number of extragalactic objects being discovered corresponds to the expectations before the launch. Using the new data obtained in the fourth scan of the sky, it is possible to search for sources whose X-ray brightness has strongly changed over the period of 6--18 months that passed since their previous appearances in the telescopes' FoV during the first three all-sky surveys. Every day, the observatory discovers about$\text{ } \text{five to ten}$ objects within a $360^\circ \times 1^\circ$ strip (i.e., less than 1\%) of the sky whose brightness has increased more than ten times over 6 months. The depth of the X-ray maps and the number of detected sources continues to increase with increasing exposure. The observatory is continuing its round-the-clock mission. 

\section{Acknowledgements}

This work is based on observations with the \erosita\ and \art\ telescopes aboard the \srg\ observatory. The \srg\ observatory was built by Roskosmos in the interests of the Russian Academy of Sciences represented by its Space Research Institute (IKI) in the framework of the Russian Federal Space Program, with the participation of the Deutsches Zentrum f\"ur Luft- und Raumfahrt (DLR). The \srg/\erosita\ X-ray telescope was built by a consortium of German Institutes led by MPE, and supported by DLR. The \srg\ spacecraft was designed, built, launched, and is operated by the Lavochkin Association and its subcontractors. The science data are downlinked via the Deep Space Network antennas in Bear Lakes, and Ussurijsk, funded by Roskosmos. The \erosita\ data used in this work were processed using the eSASS software system developed by the German \erosita\ consortium and proprietary data reduction and analysis software developed by the Russian \erosita\ Consortium.

\bibliographystyle{aa} 
\bibliography{srg_review.bib}

\begin{thebibliography}{100}
\expandafter\ifx\csname natexlab\endcsname\relax\def\natexlab#1{#1}\fi

\bibitem[{{Ackermann} {et~al.}(2014){Ackermann}, {Albert}, {Atwood}, {Baldini},
  {Ballet}, {Barbiellini}, {Bastieri}, {Bellazzini}, {Bissaldi}, {Blandford},
  {Bloom}, {Bottacini}, {Brandt}, {Bregeon}, {Bruel}, {Buehler}, {Buson},
  {Caliandro}, {Cameron}, {Caragiulo}, {Caraveo}, {Cavazzuti}, {Cecchi},
  {Charles}, {Chekhtman}, {Chiang}, {Chiaro}, {Ciprini}, {Claus},
  {Cohen-Tanugi}, {Conrad}, {Cutini}, {D'Ammando}, {de Angelis}, {de Palma},
  {Dermer}, {Digel}, {Di Venere}, {Silva}, {Drell}, {Favuzzi}, {Ferrara},
  {Focke}, {Franckowiak}, {Fukazawa}, {Funk}, {Fusco}, {Gargano}, {Gasparrini},
  {Germani}, {Giglietto}, {Giordano}, {Giroletti}, {Godfrey}, {Gomez-Vargas},
  {Grenier}, {Guiriec}, {Hadasch}, {Harding}, {Hays}, {Hewitt}, {Hou},
  {Jogler}, {J{\'o}hannesson}, {Johnson}, {Johnson}, {Kamae}, {Kataoka},
  {Kn{\"o}dlseder}, {Kocevski}, {Kuss}, {Larsson}, {Latronico}, {Longo},
  {Loparco}, {Lovellette}, {Lubrano}, {Malyshev}, {Manfreda}, {Massaro},
  {Mayer}, {Mazziotta}, {McEnery}, {Michelson}, {Mitthumsiri}, {Mizuno},
  {Monzani}, {Morselli}, {Moskalenko}, {Murgia}, {Nemmen}, {Nuss}, {Ohsugi},
  {Omodei}, {Orienti}, {Orlando}, {Ormes}, {Paneque}, {Panetta}, {Perkins},
  {Pesce-Rollins}, {Petrosian}, {Piron}, {Pivato}, {Rain{\`o}}, {Rando},
  {Razzano}, {Razzaque}, {Reimer}, {Reimer}, {S{\'a}nchez-Conde}, {Schaal},
  {Schulz}, {Sgr{\`o}}, {Siskind}, {Spandre}, {Spinelli}, {Stawarz}, {Strong},
  {Suson}, {Tahara}, {Takahashi}, {Thayer}, {Tibaldo}, {Tinivella}, {Torres},
  {Tosti}, {Troja}, {Uchiyama}, {Vianello}, {Werner}, {Winer}, {Wood}, {Wood},
  \& {Zaharijas}}]{ackermann14}
{Ackermann}, M., {Albert}, A., {Atwood}, W.~B., {et~al.} 2014, \apj, 793, 64

\bibitem[{{Arcangeli} {et~al.}(2017){Arcangeli}, {Borghi}, {Br{\"a}uninger},
  {Citterio}, {Ferrario}, {Friedrich}, {Grisoni}, {Marioni}, {Predehl},
  {Rossi}, {Ritucci}, {Valsecchi}, \& {Vernani}}]{Arcangeli2017}
{Arcangeli}, L., {Borghi}, G., {Br{\"a}uninger}, H., {et~al.} 2017, in Society
  of Photo-Optical Instrumentation Engineers (SPIE) Conference Series, Vol.
  10565, \procspie, 1056558

\bibitem[{{Ascenzi} {et~al.}(2020){Ascenzi}, {Oganesyan}, {Salafia},
  {Branchesi}, {Ghirlanda}, \& {Dall'Osso}}]{Ascenzi2020}
{Ascenzi}, S., {Oganesyan}, G., {Salafia}, O.~S., {et~al.} 2020, \aap, 641, A61

\bibitem[{{Bikmaev} {et~al.}(2020){Bikmaev}, {Irtuganov}, {Nikolaeva},
  {Sakhibullin}, {Gumerov}, {Sklyanov}, {Glushkov}, {Borisov}, {Burenin},
  {Zaznobin}, {Krivonos}, {Lyapin}, {Medvedev}, {Meshcheryakov}, {Sazonov},
  {Sunyaev}, {Khorunzhev}, \& {Gilfanov}}]{bikmaev20}
{Bikmaev}, I.~F., {Irtuganov}, E.~N., {Nikolaeva}, E.~A., {et~al.} 2020,
  Astronomy Letters, 46, 645

\bibitem[{{Bleem} {et~al.}(2015){Bleem}, {Stalder}, {de Haan}, {Aird}, {Allen},
  {Applegate}, {Ashby}, {Bautz}, {Bayliss}, {Benson}, {Bocquet}, {Brodwin},
  {Carlstrom}, {Chang}, {Chiu}, {Cho}, {Clocchiatti}, {Crawford}, {Crites},
  {Desai}, {Dietrich}, {Dobbs}, {Foley}, {Forman}, {George}, {Gladders},
  {Gonzalez}, {Halverson}, {Hennig}, {Hoekstra}, {Holder}, {Holzapfel},
  {Hrubes}, {Jones}, {Keisler}, {Knox}, {Lee}, {Leitch}, {Liu}, {Lueker},
  {Luong-Van}, {Mantz}, {Marrone}, {McDonald}, {McMahon}, {Meyer}, {Mocanu},
  {Mohr}, {Murray}, {Padin}, {Pryke}, {Reichardt}, {Rest}, {Ruel}, {Ruhl},
  {Saliwanchik}, {Saro}, {Sayre}, {Schaffer}, {Schrabback}, {Shirokoff},
  {Song}, {Spieler}, {Stanford}, {Staniszewski}, {Stark}, {Story}, {Stubbs},
  {Vand erlinde}, {Vieira}, {Vikhlinin}, {Williamson}, {Zahn}, \&
  {Zenteno}}]{Bleem_2015}
{Bleem}, L.~E., {Stalder}, B., {de Haan}, T., {et~al.} 2015, \apjs, 216, 27

\bibitem[{{Brunner} {et~al.}(2021){Brunner}, {Liu}, {Lamer}, {Georgakakis},
  {Merloni}, {Brusa}, {Bulbul}, {Dennerl}, {Friedrich}, {Liu}, {Maitra},
  {Nandra}, {Ramos-Ceja}, {Sanders}, {Stewart}, {Boller}, {Buchner}, {Clerc},
  {Comparat}, {Dwelly}, {Eckert}, {Finoguenov}, {Freyberg}, {Ghirardini},
  {Gueguen}, {Haberl}, {Kreykenbohm}, {Krumpe}, {Osterhage}, {Pacaud},
  {Predehl}, {Reiprich}, {Robrade}, {Salvato}, {Santangelo}, {Schrabback},
  {Schwope}, \& {Wilms}}]{2021arXiv210614517B}
{Brunner}, H., {Liu}, T., {Lamer}, G., {et~al.} 2021, arXiv e-prints,
  arXiv:2106.14517

\bibitem[{{Canalias} \& {Masdemont}(2004)}]{canalias04}
{Canalias}, E. \& {Masdemont}, J.~J. 2004, International Astronautical
  Federation - 55th International Astronautical Congress, 1, 536

\bibitem[{{Castro-Tirado} {et~al.}(1992){Castro-Tirado}, {Brandt}, \&
  {Lund}}]{Castro_1992}
{Castro-Tirado}, A.~J., {Brandt}, S., \& {Lund}, N. 1992, \iaucirc, 5590, 2

\bibitem[{{Churazov} {et~al.}(1994){Churazov}, {Gilfanov}, {Sunyaev},
  {Khavenson}, {Novikov}, {Dyachkov}, {Kremnev}, {Sukhanov}, {Cordier}, {Paul},
  {Laurent}, {Claret}, {Bouchet}, {Roques}, {Mandrou}, \&
  {Vedrenne}}]{Churazov_1994}
{Churazov}, E., {Gilfanov}, M., {Sunyaev}, R., {et~al.} 1994, \apjs, 92, 381

\bibitem[{{Churazov} {et~al.}(2021){Churazov}, {Khabibullin}, {Lyskova},
  {Sunyaev}, \& {Bykov}}]{churazov21}
{Churazov}, E., {Khabibullin}, I., {Lyskova}, N., {Sunyaev}, R., \& {Bykov},
  A.~M. 2021, \aap, 651, A41

\bibitem[{{Churazov} {et~al.}(2014){Churazov}, {Sunyaev}, {Isern},
  {Kn{\"o}dlseder}, {Jean}, {Lebrun}, {Chugai}, {Grebenev}, {Bravo}, {Sazonov},
  \& {Renaud}}]{churazov14}
{Churazov}, E., {Sunyaev}, R., {Isern}, J., {et~al.} 2014, \nat, 512, 406

\bibitem[{{Churazov} {et~al.}(2005){Churazov}, {Sunyaev}, {Sazonov},
  {Revnivtsev}, \& {Varshalovich}}]{churazov05}
{Churazov}, E., {Sunyaev}, R., {Sazonov}, S., {Revnivtsev}, M., \&
  {Varshalovich}, D. 2005, \mnras, 357, 1377

\bibitem[{{Coutinho} {et~al.}(2018){Coutinho}, {Bornemann}, {Budau}, {Burwitz},
  {F{\"u}rmetz}, {Gaida}, {Hartner}, {Kink}, {Meidinger}, {M{\"u}ller}, \&
  {Predehl}}]{Coutinho2018}
{Coutinho}, D., {Bornemann}, W., {Budau}, B., {et~al.} 2018, in Society of
  Photo-Optical Instrumentation Engineers (SPIE) Conference Series, Vol. 10699,
  \procspie, 106995F

\bibitem[{{Dodin} {et~al.}(2020){Dodin}, {Potanin}, {Shatsky}, {Belinski},
  {Atapin}, {Burlak}, {Egorov}, {Tatarnikov}, {Postnov}, {Belvedersky},
  {Burenin}, {Gilfanov}, {Medvedev}, {Meshcheryakov}, {Sazonov}, {Khorunzhev},
  \& {Sunyaev}}]{dodin20}
{Dodin}, A.~V., {Potanin}, S.~A., {Shatsky}, N.~I., {et~al.} 2020, Astronomy
  Letters, 46, 429

\bibitem[{{Doroshenko} {et~al.}(2021){Doroshenko}, {Santangelo}, {Tsygankov},
  \& {Ji}}]{doroshenko21}
{Doroshenko}, V., {Santangelo}, A., {Tsygankov}, S.~S., \& {Ji}, L. 2021, \aap,
  647, A165

\bibitem[{{Eismont} {et~al.}(2020){Eismont}, {Kovalenko}, {Nazarov}, {Nazirov},
  {Korotkov}, {Pogodin}, {Mzhelskii}, {Mikhailov}, {Ditrikh}, \&
  {Tregubov}}]{Eismont_2020}
{Eismont}, N.~A., {Kovalenko}, I.~D., {Nazarov}, V.~N., {et~al.} 2020,
  Astronomy Letters, 46, 263

\bibitem[{{Freyberg} {et~al.}(2020){Freyberg}, {Perinati}, {Pacaud}, {Eraerds},
  {Churazov}, {Dennerl}, {Predehl}, {Merloni}, {Meidinger}, {Bulbul},
  {Friedrich}, {Gilfanov}, {Tenzer}, {Pommranz}, {Eckert}, {Schmitt}, {Brusa},
  \& {Santangelo}}]{Freyberg2020}
{Freyberg}, M., {Perinati}, E., {Pacaud}, F., {et~al.} 2020, in Society of
  Photo-Optical Instrumentation Engineers (SPIE) Conference Series, Vol. 11444,
  Society of Photo-Optical Instrumentation Engineers (SPIE) Conference Series,
  114441O

\bibitem[{{Friedrich} {et~al.}(2008){Friedrich}, {Br{\"a}uninger}, {Budau},
  {Burkert}, {Eder}, {Freyberg}, {Hartner}, {M{\"u}hlegger}, {Predehl},
  {Erhard}, {Gutruf}, {Jugler}, {Kampf}, {Borghi}, {Citterio}, {Rossi},
  {Valsecchi}, {Vernani}, \& {Zimmermann}}]{Friedrich2008}
{Friedrich}, P., {Br{\"a}uninger}, H., {Budau}, B., {et~al.} 2008, in Society
  of Photo-Optical Instrumentation Engineers (SPIE) Conference Series, Vol.
  7011, \procspie, 70112T

\bibitem[{{Friedrich} {et~al.}(2014){Friedrich}, {Roh{\'e}}, {Gaida},
  {Hartwig}, {Soller}, {Br{\"a}uninger}, {Budau}, {Burkert}, {Burwitz}, {Eder},
  {Hartner}, {Menz}, \& {Predehl}}]{2014SPIE.9144E..4RF}
{Friedrich}, P., {Roh{\'e}}, C., {Gaida}, R., {et~al.} 2014, in Society of
  Photo-Optical Instrumentation Engineers (SPIE) Conference Series, Vol. 9144,
  Space Telescopes and Instrumentation 2014: Ultraviolet to Gamma Ray, ed.
  T.~{Takahashi}, J.-W.~A. {den Herder}, \& M.~{Bautz}, 91444R

\bibitem[{F{\"u}rmetz {et~al.}(2008)F{\"u}rmetz, Pfeffermann, Predehl,
  Roh{\'e}, \& Tiedemann}]{Fuermetz2008}
F{\"u}rmetz, M., Pfeffermann, E., Predehl, P., Roh{\'e}, C., \& Tiedemann, L.
  2008, in Space Telescopes and Instrumentation 2008: Ultraviolet to Gamma Ray,
  Vol. 7011, International Society for Optics and Photonics, 70113Y

\bibitem[{{Ghirlanda} {et~al.}(2015){Ghirlanda}, {Salvaterra}, {Campana},
  {Vergani}, {Japelj}, {Bernardini}, {Burlon}, {D'Avanzo}, {Melandri},
  {Gomboc}, {Nappo}, {Paladini}, {Pescalli}, {Salafia}, \&
  {Tagliaferri}}]{Ghirlanda2015}
{Ghirlanda}, G., {Salvaterra}, R., {Campana}, S., {et~al.} 2015, \aap, 578, A71

\bibitem[{{Gilfanov} {et~al.}(1993){Gilfanov}, {Churazov}, {Sunyaev},
  {Khavenson}, {Novikov}, {Dyachkov}, {Kremnev}, {Sukhanov}, {Bouchet},
  {Mandrou}, {Roques}, {Vedrenne}, {Cordier}, {Goldwurm}, {Laurent}, \&
  {Paul}}]{1993ApJ...418..844G}
{Gilfanov}, M., {Churazov}, E., {Sunyaev}, R., {et~al.} 1993, \apj, 418, 844

\bibitem[{{Gilfanov} {et~al.}(2021){Gilfanov}, {Sazonov}, {Medvedev},
  {Khorunzhev}, {Sunyaev}, {Yao}, {Kulkarni}, \& {Sharma}}]{Gilfanov2021}
{Gilfanov}, M., {Sazonov}, S., {Medvedev}, P., {et~al.} 2021, The Astronomer's
  Telegram, 14800, 1

\bibitem[{{Gilfanov} {et~al.}(2020){Gilfanov}, {Sazonov}, {Sunyaev},
  {Medvedev}, {Khorunzhev}, {Semena}, {Yao}, {Kulkarni}, {Gezari}, \& {van
  Velzen}}]{gilfanov20}
{Gilfanov}, M., {Sazonov}, S., {Sunyaev}, R., {et~al.} 2020, The Astronomer's
  Telegram, 14246, 1

\bibitem[{{Gilfanov} {et~al.}(1989){Gilfanov}, {Sunyaev}, {Churazov},
  {Loznikov}, {Efremov}, {Kaniovskjii}, {Kuznetsov}, {Yamburenko},
  {Melioranskii}, {Skinner}, {Al-Emam}, {Patterson}, {Willmore}, {Brinkman},
  {Heise}, {Intzand}, {Jager}, {Voges}, {Pietsch}, {Doebereiner}, {Englhauser},
  {Truemper}, {Reppin}, {Oegelman}, {Kendizorra}, {Mony}, {Maisack},
  {Staubert}, {Parmar}, \& {Smith}}]{gilfanov89}
{Gilfanov}, M., {Sunyaev}, R., {Churazov}, E., {et~al.} 1989, Soviet Astronomy
  Letters, 15, 291

\bibitem[{{Gruber} {et~al.}(1999){Gruber}, {Matteson}, {Peterson}, \&
  {Jung}}]{1999ApJ...520..124G}
{Gruber}, D.~E., {Matteson}, J.~L., {Peterson}, L.~E., \& {Jung}, G.~V. 1999,
  \apj, 520, 124

\bibitem[{{Gubarev} {et~al.}(2014){Gubarev}, {Ramsey}, {Kolodziejczak},
  {O'Dell}, {Elsner}, {Zavlin}, {Swartz}, {Pavlinsky}, {Tkachenko}, \&
  {Lapshov}}]{Gubarev_2014}
{Gubarev}, M., {Ramsey}, B., {Kolodziejczak}, J.~J., {et~al.} 2014, Society of
  Photo-Optical Instrumentation Engineers (SPIE) Conference Series, Vol. 9144,
  {The calibration of flight mirror modules for the ART-XC instrument on board
  the SRG mission}, 91444U

\bibitem[{{Gubarev} {et~al.}(2012){Gubarev}, {Ramsey}, {O'Dell}, {Elsner},
  {Kilaru}, {McCracken}, {Pavlinsky}, {Tkachenko}, \& {Lapshov}}]{Gubarev_2012}
{Gubarev}, M., {Ramsey}, B., {O'Dell}, S.~L., {et~al.} 2012, Society of
  Photo-Optical Instrumentation Engineers (SPIE) Conference Series, Vol. 8443,
  {The Marshall Space Flight Center development of mirror modules for the
  ART-XC instrument aboard the Spectrum-Roentgen-Gamma mission}, 84431U

\bibitem[{{Haiman} {et~al.}(2005){Haiman}, {Allen}, {Bahcall}, {Bautz},
  {Boehringer}, {Borgani}, {Bryan}, {Cabrera}, {Canizares}, {Citterio},
  {Evrard}, {Finoguenov}, {Griffiths}, {Hasinger}, {Henry}, {Jahoda},
  {Jernigan}, {Kahn}, {Lamb}, {Majumdar}, {Mohr}, {Molendi}, {Mushotzky},
  {Pareschi}, {Peterson}, {Petre}, {Predehl}, {Rasmussen}, {Ricker}, {Ricker},
  {Rosati}, {Sanderson}, {Stanford}, {Voit}, {Wang}, {White}, \&
  {White}}]{Haiman_2005}
{Haiman}, Z., {Allen}, S., {Bahcall}, N., {et~al.} 2005, arXiv e-prints, astro

\bibitem[{{Hilton} {et~al.}(2021){Hilton}, {Sif{\'o}n}, {Naess},
  {Madhavacheril}, {Oguri}, {Rozo}, {Rykoff}, {Abbott}, {Adhikari}, {Aguena},
  {Aiola}, {Allam}, {Amodeo}, {Amon}, {Annis}, {Ansarinejad}, {Aros-Bunster},
  {Austermann}, {Avila}, {Bacon}, {Battaglia}, {Beall}, {Becker}, {Bernstein},
  {Bertin}, {Bhandarkar}, {Bhargava}, {Bond}, {Brooks}, {Burke}, {Calabrese},
  {Carrasco Kind}, {Carretero}, {Choi}, {Choi}, {Conselice}, {da Costa},
  {Costanzi}, {Crichton}, {Crowley}, {D{\"u}nner}, {Denison}, {Devlin},
  {Dicker}, {Diehl}, {Dietrich}, {Doel}, {Duff}, {Duivenvoorden}, {Dunkley},
  {Everett}, {Ferraro}, {Ferrero}, {Fert{\'e}}, {Flaugher}, {Frieman},
  {Gallardo}, {Garc{\'\i}a-Bellido}, {Gaztanaga}, {Gerdes}, {Giles}, {Golec},
  {Gralla}, {Grandis}, {Gruen}, {Gruendl}, {Gschwend}, {Gutierrez}, {Han},
  {Hartley}, {Hasselfield}, {Hill}, {Hilton}, {Hincks}, {Hinton}, {Ho},
  {Honscheid}, {Hoyle}, {Hubmayr}, {Huffenberger}, {Hughes}, {Jaelani}, {Jain},
  {James}, {Jeltema}, {Kent}, {Knowles}, {Koopman}, {Kuehn}, {Lahav}, {Lima},
  {Lin}, {Lokken}, {Loubser}, {MacCrann}, {Maia}, {Marriage}, {Martin},
  {McMahon}, {Melchior}, {Menanteau}, {Miquel}, {Miyatake}, {Moodley},
  {Morgan}, {Mroczkowski}, {Nati}, {Newburgh}, {Niemack}, {Nishizawa},
  {Ogando}, {Orlowski-Scherer}, {Page}, {Palmese}, {Partridge},
  {Paz-Chinch{\'o}n}, {Phakathi}, {Plazas}, {Robertson}, {Romer}, {Carnero
  Rosell}, {Salatino}, {Sanchez}, {Schaan}, {Schillaci}, {Sehgal}, {Serrano},
  {Shin}, {Simon}, {Smith}, {Soares-Santos}, {Spergel}, {Staggs}, {Storer},
  {Suchyta}, {Swanson}, {Tarle}, {Thomas}, {To}, {Trac}, {Ullom}, {Vale}, {Van
  Lanen}, {Vavagiakis}, {De Vicente}, {Wilkinson}, {Wollack}, {Xu}, \&
  {Zhang}}]{Hilton_2020}
{Hilton}, M., {Sif{\'o}n}, C., {Naess}, S., {et~al.} 2021, \apjs, 253, 3

\bibitem[{{H{\"u}tsi}(2006)}]{Hutsi_2006}
{H{\"u}tsi}, G. 2006, \aap, 446, 43

\bibitem[{{Jonker} {et~al.}(2020){Jonker}, {Stone}, {Generozov}, {van Velzen},
  \& {Metzger}}]{Jonker2020}
{Jonker}, P.~G., {Stone}, N.~C., {Generozov}, A., {van Velzen}, S., \&
  {Metzger}, B. 2020, \apj, 889, 166

\bibitem[{{Kardashev} {et~al.}(2013){Kardashev}, {Khartov}, {Abramov},
  {Avdeev}, {Alakoz}, {Aleksandrov}, {Ananthakrishnan}, {Andreyanov},
  {Andrianov}, {Antonov}, {Artyukhov}, {Arkhipov}, {Baan}, {Babakin},
  {Babyshkin}, {Bartel'}, {Belousov}, {Belyaev}, {Berulis}, {Burke},
  {Biryukov}, {Bubnov}, {Burgin}, {Busca}, {Bykadorov}, {Bychkova},
  {Vasil'kov}, {Wellington}, {Vinogradov}, {Wietfeldt}, {Voitsik},
  {Gvamichava}, {Girin}, {Gurvits}, {Dagkesamanskii}, {D'Addario},
  {Giovannini}, {Jauncey}, {Dewdney}, {D'yakov}, {Zharov}, {Zhuravlev},
  {Zaslavskii}, {Zakhvatkin}, {Zinov'ev}, {Ilinen}, {Ipatov}, {Kanevskii},
  {Knorin}, {Casse}, {Kellermann}, {Kovalev}, {Kovalev}, {Kovalenko}, {Kogan},
  {Komaev}, {Konovalenko}, {Kopelyanskii}, {Korneev}, {Kostenko}, {Kotik},
  {Kreisman}, {Kukushkin}, {Kulishenko}, {Cooper}, {Kut'kin}, {Cannon},
  {Larionov}, {Lisakov}, {Litvinenko}, {Likhachev}, {Likhacheva}, {Lobanov},
  {Logvinenko}, {Langston}, {McCracken}, {Medvedev}, {Melekhin}, {Menderov},
  {Murphy}, {Mizyakina}, {Mozgovoi}, {Nikolaev}, {Novikov}, {Novikov},
  {Oreshko}, {Pavlenko}, {Pashchenko}, {Ponomarev}, {Popov}, {Pravin-Kumar},
  {Preston}, {Pyshnov}, {Rakhimov}, {Rozhkov}, {Romney}, {Rocha}, {Rudakov},
  {R{\"a}is{\"a}nen}, {Sazankov}, {Sakharov}, {Semenov}, {Serebrennikov},
  {Schilizzi}, {Skulachev}, {Slysh}, {Smirnov}, {Smith}, {Soglasnov},
  {Sokolovskii}, {Sondaar}, {Stepan'yants}, {Turygin}, {Turygin}, {Tuchin},
  {Urpo}, {Fedorchuk}, {Finkel'shtein}, {Fomalont}, {Fejes}, {Fomina},
  {Khapin}, {Tsarevskii}, {Zensus}, {Chuprikov}, {Shatskaya}, {Shapirovskaya},
  {Sheikhet}, {Shirshakov}, {Schmidt}, {Shnyreva}, {Shpilevskii}, {Ekers}, \&
  {Yakimov}}]{Kardashev2013}
{Kardashev}, N.~S., {Khartov}, V.~V., {Abramov}, V.~V., {et~al.} 2013,
  Astronomy Reports, 57, 153

\bibitem[{{Kawamuro} {et~al.}(2018){Kawamuro}, {Ueda}, {Shidatsu}, {Hori},
  {Morii}, {Nakahira}, {Isobe}, {Kawai}, {Mihara}, {Matsuoka}, {Morita},
  {Nakajima}, {Negoro}, {Oda}, {Sakamoto}, {Serino}, {Sugizaki}, {Tanimoto},
  {Tomida}, {Tsuboi}, {Tsunemi}, {Ueno}, {Yamaoka}, {Yamada}, {Yoshida},
  {Iwakiri}, {Kawakubo}, {Sugawara}, {Sugita}, {Tachibana}, \&
  {Yoshii}}]{Kawamuro_2018}
{Kawamuro}, T., {Ueda}, Y., {Shidatsu}, M., {et~al.} 2018, \apjs, 238, 32

\bibitem[{{Khabibullin} {et~al.}(2020{\natexlab{a}}){Khabibullin}, {Medvedev},
  {Churazov}, {Gilfanov}, {Sazonov}, {Sunyaev}, {Burenin}, {Pavlinsky},
  {Russian Srg/Erosita Consortium}, \& {Srg/Art-Xc Team}}]{khabibullin20b}
{Khabibullin}, I., {Medvedev}, P., {Churazov}, E., {et~al.} 2020{\natexlab{a}},
  The Astronomer's Telegram, 13499, 1

\bibitem[{{Khabibullin} {et~al.}(2016){Khabibullin}, {Medvedev}, \&
  {Sazonov}}]{2016MNRAS.455.1414K}
{Khabibullin}, I., {Medvedev}, P., \& {Sazonov}, S. 2016, \mnras, 455, 1414

\bibitem[{{Khabibullin} {et~al.}(2012){Khabibullin}, {Sazonov}, \&
  {Sunyaev}}]{Khabibullin_2012}
{Khabibullin}, I., {Sazonov}, S., \& {Sunyaev}, R. 2012, \mnras, 426, 1819

\bibitem[{{Khabibullin} {et~al.}(2014){Khabibullin}, {Sazonov}, \&
  {Sunyaev}}]{Khabibullin_2014}
{Khabibullin}, I., {Sazonov}, S., \& {Sunyaev}, R. 2014, \mnras, 437, 327

\bibitem[{{Khabibullin} {et~al.}(2020{\natexlab{b}}){Khabibullin}, {Sunyaev},
  {Churazov}, {Gilfanov}, {Medvedev}, \& {Sazonov}}]{khabibullin20}
{Khabibullin}, I., {Sunyaev}, R., {Churazov}, E., {et~al.} 2020{\natexlab{b}},
  The Astronomer's Telegram, 13494, 1

\bibitem[{{Khorunzhev} {et~al.}(2020){Khorunzhev}, {Meshcheryakov}, {Burenin},
  {Lyapin}, {Medvedev}, {Sazonov}, {Eselevich}, {Sunyaev}, \&
  {Gilfanov}}]{khorunzhev20}
{Khorunzhev}, G.~A., {Meshcheryakov}, A.~V., {Burenin}, R.~A., {et~al.} 2020,
  Astronomy Letters, 46, 149

\bibitem[{{Khorunzhev} {et~al.}(2021){Khorunzhev}, {Meshcheryakov}, {Medvedev},
  {Borisov}, {Burenin}, {Krivonos}, {Uklein}, {Shablovinskaya}, {Afanasiev},
  {Dodonov}, {Sunyaev}, {Sazonov}, \& {Gilfanov}}]{khor21}
{Khorunzhev}, G.~A., {Meshcheryakov}, A.~V., {Medvedev}, P.~S., {et~al.} 2021,
  Astronomy Letters, 47, 123

\bibitem[{{Krivonos} {et~al.}(2017){Krivonos}, {Tkachenko}, {Burenin},
  {Filippova}, {Lapshov}, {Mereminskiy}, {Molkov}, {Pavlinsky}, {Sazonov},
  {Gubarev}, {Kolodziejczak}, {O'Dell}, {Swartz}, {Zavlin}, \&
  {Ramsey}}]{Krivonos_2017}
{Krivonos}, R., {Tkachenko}, A., {Burenin}, R., {et~al.} 2017, Experimental
  Astronomy, 44, 147

\bibitem[{{Lamer} {et~al.}(2021){Lamer}, {Schwope}, {Predehl}, {Traulsen},
  {Wilms}, \& {Freyberg}}]{lamer21}
{Lamer}, G., {Schwope}, A.~D., {Predehl}, P., {et~al.} 2021, \aap, 647, A7

\bibitem[{{Levin} {et~al.}(2014){Levin}, {Pavlinsky}, {Akimov}, {Kuznetsova},
  {Rotin}, {Krivchenko}, {Lapshov}, \& {Oleinikov}}]{Levin_2014}
{Levin}, V., {Pavlinsky}, M., {Akimov}, V., {et~al.} 2014, Society of
  Photo-Optical Instrumentation Engineers (SPIE) Conference Series, Vol. 9144,
  {ART-XC/SRG: status of the x-ray focal plane detector development}, 914413

\bibitem[{{Levin} {et~al.}(2016){Levin}, {Pavlinsky}, {Akimov}, {Kuznetsova},
  {Rotin}, {Krivchenko}, {Lapshov}, \& {Oleynikov}}]{Levin_2016}
{Levin}, V., {Pavlinsky}, M., {Akimov}, V., {et~al.} 2016, Society of
  Photo-Optical Instrumentation Engineers (SPIE) Conference Series, Vol. 9905,
  {Results of ground tests and calibration of x-ray focal plane detectors for
  ART-XC/SRG instrument}, 990551

\bibitem[{{Liu} {et~al.}(2021){Liu}, {Rau}, {Malyali}, {Merloni}, \&
  {Grotova}}]{Liu2021}
{Liu}, Z., {Rau}, A., {Malyali}, A., {Merloni}, A., \& {Grotova}, I. 2021, The
  Astronomer's Telegram, 14407, 1

\bibitem[{{Lumb} {et~al.}(2002){Lumb}, {Warwick}, {Page}, \& {De
  Luca}}]{2002A&A...389...93L}
{Lumb}, D.~H., {Warwick}, R.~S., {Page}, M., \& {De Luca}, A. 2002, \aap, 389,
  93

\bibitem[{{Lutovinov} {et~al.}(2021){Lutovinov}, {Tsygankov}, {Mereminskiy},
  {Molkov}, {Semena}, {Arefiev}, {Bikmaev}, {Djupvik}, {Gilfanov}, {Karasev},
  {Lapshov}, {Medvedev}, {Shtykovsky}, {Sunyaev}, {Tkachenko}, {Anand},
  {Ashley}, {De}, {Kasliwal}, {Kulkarni}, {van Roestel}, \&
  {Yao}}]{Lutovinov2021}
{Lutovinov}, A.~A., {Tsygankov}, S.~S., {Mereminskiy}, I.~A., {et~al.} 2021,
  arXiv e-prints, arXiv:2107.05587

\bibitem[{{Malyali} {et~al.}(2019){Malyali}, {Rau}, \& {Nandra}}]{Malyali2019}
{Malyali}, A., {Rau}, A., \& {Nandra}, K. 2019, \mnras, 489, 5413

\bibitem[{{Medvedev} {et~al.}(2021){Medvedev}, {Gilfanov}, {Sazonov},
  {Schartel}, \& {Sunyaev}}]{medvedev21}
{Medvedev}, P., {Gilfanov}, M., {Sazonov}, S., {Schartel}, N., \& {Sunyaev}, R.
  2021, \mnras, 504, 576

\bibitem[{{Medvedev} {et~al.}(2020){Medvedev}, {Sazonov}, {Gilfanov},
  {Burenin}, {Khorunzhev}, {Meshcheryakov}, {Sunyaev}, {Bikmaev}, \&
  {Irtuganov}}]{Medvedev_2020}
{Medvedev}, P., {Sazonov}, S., {Gilfanov}, M., {et~al.} 2020, \mnras, 497, 1842

\bibitem[{{Meidinger} {et~al.}(2014){Meidinger}, {Andritschke}, {Bornemann},
  {Coutinho}, {Emberger}, {H{\"a}lker}, {Kink}, {Mican}, {M{\"u}ller},
  {Pietschner}, {Predehl}, \& {Reiffers}}]{Meidinger2014}
{Meidinger}, N., {Andritschke}, R., {Bornemann}, W., {et~al.} 2014, in Society
  of Photo-Optical Instrumentation Engineers (SPIE) Conference Series, Vol.
  9144, \procspie, 91441W

\bibitem[{{Mereminskiy} {et~al.}(2020{\natexlab{a}}){Mereminskiy}, {Lutovinov},
  {Semena}, {Molkov}, {Filippova}, {Tkachenko}, \& {Lapshov}}]{atel14051}
{Mereminskiy}, I., {Lutovinov}, A., {Semena}, A., {et~al.} 2020{\natexlab{a}},
  The Astronomer's Telegram, 14051, 1

\bibitem[{{Mereminskiy} {et~al.}(2020{\natexlab{b}}){Mereminskiy}, {Medvedev},
  {Lutovinov}, {Gilfanov}, {Semena}, {Sunyaev}, {Molkov}, \&
  {Arefiev}}]{atel14206}
{Mereminskiy}, I., {Medvedev}, P., {Lutovinov}, A., {et~al.}
  2020{\natexlab{b}}, The Astronomer's Telegram, 14206, 1

\bibitem[{{Mereminskiy} {et~al.}(2020{\natexlab{c}}){Mereminskiy}, {Medvedev},
  {Semena}, {Pavlinsky}, {Molkov}, {Lutovinov}, {Burenin}, {Sazonov},
  {Sunyaev}, \& {Gilfanov}}]{atel13571}
{Mereminskiy}, I., {Medvedev}, P., {Semena}, A., {et~al.} 2020{\natexlab{c}},
  The Astronomer's Telegram, 13571, 1

\bibitem[{{Mereminskiy} {et~al.}(2020{\natexlab{d}}){Mereminskiy}, {Semena},
  {Pavlinsky}, {Lutovinov}, {Molkov}, {Lapshov}, \& {Tkachenk}}]{atel13606}
{Mereminskiy}, I., {Semena}, A., {Pavlinsky}, M., {et~al.} 2020{\natexlab{d}},
  The Astronomer's Telegram, 13606, 1

\bibitem[{{Mereminskiy} {et~al.}(2021){Mereminskiy}, {Dodin}, {Lutovinov},
  {Semena}, {Arefiev}, {Atapin}, {Belinski}, {Burenin}, {Burlak}, {Eselevich},
  {Fedotieva}, {Gilfanov}, {Ikonnikova}, {Krivonos}, {Lapshov}, {Lyapin},
  {Medvedev}, {Molkov}, {Postnov}, {Pshirkov}, {Sazonov}, {Shakura},
  {Shtykovsky}, {Sunyaev}, {Tatarnikov}, {Tkachenko}, \&
  {Zheltoukhov}}]{Mereminskiy2021}
{Mereminskiy}, I.~A., {Dodin}, A.~V., {Lutovinov}, A.~A., {et~al.} 2021, arXiv
  e-prints, arXiv:2107.05588

\bibitem[{{Merloni} {et~al.}(2012){Merloni}, {Predehl}, {Becker},
  {B{\"o}hringer}, {Boller}, {Brunner}, {Brusa}, {Dennerl}, {Freyberg},
  {Friedrich}, {Georgakakis}, {Haberl}, {Hasinger}, {Meidinger}, {Mohr},
  {Nandra}, {Rau}, {Reiprich}, {Robrade}, {Salvato}, {Santangelo}, {Sasaki},
  {Schwope}, {Wilms}, \& {German eROSITA Consortium}}]{Merloni2012}
{Merloni}, A., {Predehl}, P., {Becker}, W., {et~al.} 2012, arXiv e-prints,
  arXiv:1209.3114

\bibitem[{{Pavlinsky} {et~al.}(2021{\natexlab{a}}){Pavlinsky}, {Sazonov},
  {Burenin}, {Filippova}, {Krivonos}, {Arefiev}, {Buntov}, {Chen}, {Ehlert},
  {Lapshov}, {Levin}, {Lutovinov}, {Lyapin}, {Mereminskiy}, {Molkov}, {Ramsey},
  {Semena}, {Semena}, {Shtykovsky}, {Sunyaev}, {Tkachenko}, {Swartz}, \&
  {Vikhlinin}}]{Pavlinsky2021}
{Pavlinsky}, M., {Sazonov}, S., {Burenin}, R., {et~al.} 2021{\natexlab{a}},
  arXiv e-prints, arXiv:2107.05879

\bibitem[{{Pavlinsky} {et~al.}(2021{\natexlab{b}}){Pavlinsky}, {Tkachenko},
  {Levin}, {Alexandrovich}, {Arefiev}, {Babyshkin}, {Batanov}, {Bodnar},
  {Bogomolov}, {Bubnov}, {Buntov}, {Burenin}, {Chelovekov}, {Chen}, {Drozdova},
  {Ehlert}, {Filippova}, {Frolov}, {Gamkov}, {Garanin}, {Garin}, {Glushenko},
  {Gorelov}, {Grebenev}, {Grigorovich}, {Gureev}, {Gurova}, {Ilkaev},
  {Katasonov}, {Krivchenko}, {Krivonos}, {Korotkov}, {Kudelin}, {Kuznetsova},
  {Lazarchuk}, {Lomakin}, {Lapshov}, {Lipilin}, {Lutovinov}, {Mereminskiy},
  {Molkov}, {Nazarov}, {Oleinikov}, {Pikalov}, {Ramsey}, {Roiz}, {Rotin},
  {Ryadov}, {Sankin}, {Sazonov}, {Sedov}, {Semena}, {Semena}, {Serbinov},
  {Shirshakov}, {Shtykovsky}, {Shvetsov}, {Sunyaev}, {Swartz}, {Tambov},
  {Voron}, \& {Yaskovich}}]{pavlinsky21}
{Pavlinsky}, M., {Tkachenko}, A., {Levin}, V., {et~al.} 2021{\natexlab{b}},
  \aap, 650, A42

\bibitem[{{Pavlinsky} {et~al.}(2019{\natexlab{a}}){Pavlinsky}, {Tkachenko},
  {Levin}, {Krivchenko}, {Rotin}, {Kuznetsova}, {Lapshov}, {Krivonos},
  {Semena}, {Semena}, {Serbinov}, {Shtykovsky}, {Yaskovich}, {Oleinikov},
  {Glushenko}, {Mereminskiy}, {Molkov}, {Sazonov}, \&
  {Arefiev}}]{Pavlinsky_2019b}
{Pavlinsky}, M., {Tkachenko}, A., {Levin}, V., {et~al.} 2019{\natexlab{a}},
  Experimental Astronomy, 48, 233

\bibitem[{{Pavlinsky} {et~al.}(2018){Pavlinsky}, {Tkachenko}, {Levin},
  {Krivchenko}, {Rotin}, {Kuznetsova}, {Lapshov}, {Krivonos}, {Semena},
  {Semena}, {Serbinov}, {Shtykovsky}, {Yaskovich}, {Oleinikov}, {Glushenko},
  {Mereminskiy}, {Molkov}, {Sazonov}, \& {Arefiev}}]{Pavlinsky_2018}
{Pavlinsky}, M., {Tkachenko}, A., {Levin}, V., {et~al.} 2018, Experimental
  Astronomy, 45, 315

\bibitem[{{Pavlinsky} {et~al.}(2019{\natexlab{b}}){Pavlinsky}, {Tkachenko},
  {Levin}, {Krivchenko}, {Rotin}, {Kuznetsova}, {Lapshov}, {Krivonos},
  {Semena}, {Semena}, {Serbinov}, {Shtykovsky}, {Yaskovich}, {Oleinikov},
  {Glushenko}, {Mereminskiy}, {Molkov}, {Sazonov}, \&
  {Arefiev}}]{Pavlinsky_2019a}
{Pavlinsky}, M., {Tkachenko}, A., {Levin}, V., {et~al.} 2019{\natexlab{b}},
  Experimental Astronomy, 47, 1

\bibitem[{{Pavlinsky} {et~al.}(1994){Pavlinsky}, {Grebenev}, \&
  {Sunyaev}}]{Pavlinsky_1994}
{Pavlinsky}, M.~N., {Grebenev}, S.~A., \& {Sunyaev}, R.~A. 1994, \apj, 425, 110

\bibitem[{{Pavlinsky} {et~al.}(2020){Pavlinsky}, {Tkachenko}, {Levin}, V.,
  {Rotin}, {Kuznetsova}, {Lapshov}, {Semena}, P., {Serbinov}, {Krivonos},
  {Shtykovsky}, {Yaskovich}, {Oleinikov}, {Mereminskiy}, {Glushenko},
  {Molkova}, {Sazonov}, \& {Arefiev}}]{Pavlinsky2020}
{Pavlinsky}, M.~N., {Tkachenko}, A.~Y., {Levin}, V.~V., {et~al.} 2020,
  Instruments and Experimental Techniques, 63, 243

\bibitem[{{Peebles} \& {Yu}(1970)}]{Peebles_1970}
{Peebles}, P.~J.~E. \& {Yu}, J.~T. 1970, \apj, 162, 815

\bibitem[{{Planck Collaboration} {et~al.}(2013){Planck Collaboration}, {Ade},
  {Aghanim}, {Arnaud}, {Ashdown}, {Atrio-Barandela}, {Aumont}, {Baccigalupi},
  {Balbi}, {Banday}, {Barreiro}, {Bartlett}, {Battaner}, {Benabed},
  {Beno{\^\i}t}, {Bernard}, {Bersanelli}, {Bikmaev}, {B{\"o}hringer},
  {Bonaldi}, {Bond}, {Borrill}, {Bouchet}, {Bourdin}, {Brown}, {Brown},
  {Burenin}, {Burigana}, {Cabella}, {Cardoso}, {Carvalho}, {Catalano},
  {Cay{\'o}n}, {Chiang}, {Chon}, {Christensen}, {Churazov}, {Clements},
  {Colafrancesco}, {Colombo}, {Coulais}, {Crill}, {Cuttaia}, {Da Silva},
  {Dahle}, {Danese}, {Davis}, {de Bernardis}, {de Gasperis}, {de Rosa}, {de
  Zotti}, {Delabrouille}, {D{\'e}mocl{\`e}s}, {D{\'e}sert}, {Dickinson},
  {Diego}, {Dolag}, {Dole}, {Donzelli}, {Dor{\'e}}, {D{\"o}rl}, {Douspis},
  {Dupac}, {En{\ss}lin}, {Eriksen}, {Finelli}, {Flores-Cacho}, {Forni},
  {Frailis}, {Franceschi}, {Frommert}, {Galeotta}, {Ganga},
  {G{\'e}nova-Santos}, {Giard}, {Gilfanov}, {Gonz{\'a}lez-Nuevo}, {G{\'o}rski},
  {Gregorio}, {Gruppuso}, {Hansen}, {Harrison}, {Henrot-Versill{\'e}},
  {Hern{\'a}ndez-Monteagudo}, {Hildebrandt}, {Hivon}, {Hobson}, {Holmes},
  {Hornstrup}, {Hovest}, {Huffenberger}, {Hurier}, {Jaffe}, {Jagemann},
  {Jones}, {Juvela}, {Keih{\"a}nen}, {Khamitov}, {Kneissl}, {Knoche}, {Knox},
  {Kunz}, {Kurki-Suonio}, {Lagache}, {L{\"a}hteenm{\"a}ki}, {Lamarre},
  {Lasenby}, {Lawrence}, {Le Jeune}, {Leonardi}, {Lilje}, {Linden-V{\o}rnle},
  {L{\'o}pez-Caniego}, {Lubin}, {Mac{\'\i}as-P{\'e}rez}, {Maffei}, {Maino},
  {Mand olesi}, {Maris}, {Marleau}, {Mart{\'\i}nez-Gonz{\'a}lez}, {Masi},
  {Massardi}, {Matarrese}, {Matthai}, {Mazzotta}, {Mei}, {Melchiorri}, {Melin},
  {Mendes}, {Mennella}, {Mitra}, {Miville-Desch{\^e}nes}, {Moneti}, {Montier},
  {Morgante}, {Munshi}, {Murphy}, {Naselsky}, {Natoli}, {N{\o}rgaard-Nielsen},
  {Noviello}, {Novikov}, {Novikov}, {Osborne}, {Pajot}, {Paoletti},
  {Perdereau}, {Perrotta}, {Piacentini}, {Piat}, {Pierpaoli}, {Piffaretti},
  {Plaszczynski}, {Pointecouteau}, {Polenta}, {Ponthieu}, {Popa}, {Poutanen},
  {Pratt}, {Prunet}, {Puget}, {Rachen}, {Rebolo}, {Reinecke}, {Remazeilles},
  {Renault}, {Ricciardi}, {Riller}, {Ristorcelli}, {Rocha}, {Roman}, {Rosset},
  {Rossetti}, {Rubi{\~n}o-Mart{\'\i}n}, {Rudnick}, {Rusholme}, {Sandri},
  {Savini}, {Schaefer}, {Scott}, {Smoot}, {Stivoli}, {Sudiwala}, {Sunyaev},
  {Sutton}, {Suur-Uski}, {Sygnet}, {Tauber}, {Terenzi}, {Toffolatti}, {Tomasi},
  {Tristram}, {Tuovinen}, {T{\"u}rler}, {Umana}, {Valenziano}, {Van Tent},
  {Varis}, {Vielva}, {Villa}, {Vittorio}, {Wade}, {Wandelt}, {Welikala},
  {White}, {Yvon}, {Zacchei}, {Zaroubi}, \& {Zonca}}]{Planck_2013}
{Planck Collaboration}, {Ade}, P.~A.~R., {Aghanim}, N., {et~al.} 2013, \aap,
  554, A140

\bibitem[{{Planck Collaboration} {et~al.}(2016){Planck Collaboration}, {Ade},
  {Aghanim}, {Arnaud}, {Ashdown}, {Aumont}, {Baccigalupi}, {Banday},
  {Barreiro}, {Barrena}, {Bartlett}, {Bartolo}, {Battaner}, {Battye},
  {Benabed}, {Beno{\^\i}t}, {Benoit-L{\'e}vy}, {Bernard}, {Bersanelli},
  {Bielewicz}, {Bikmaev}, {B{\"o}hringer}, {Bonaldi}, {Bonavera}, {Bond},
  {Borrill}, {Bouchet}, {Bucher}, {Burenin}, {Burigana}, {Butler}, {Calabrese},
  {Cardoso}, {Carvalho}, {Catalano}, {Challinor}, {Chamballu}, {Chary},
  {Chiang}, {Chon}, {Christensen}, {Clements}, {Colombi}, {Colombo}, {Combet},
  {Comis}, {Couchot}, {Coulais}, {Crill}, {Curto}, {Cuttaia}, {Dahle},
  {Danese}, {Davies}, {Davis}, {de Bernardis}, {de Rosa}, {de Zotti},
  {Delabrouille}, {D{\'e}sert}, {Dickinson}, {Diego}, {Dolag}, {Dole},
  {Donzelli}, {Dor{\'e}}, {Douspis}, {Ducout}, {Dupac}, {Efstathiou},
  {Eisenhardt}, {Elsner}, {En{\ss}lin}, {Eriksen}, {Falgarone}, {Fergusson},
  {Feroz}, {Ferragamo}, {Finelli}, {Forni}, {Frailis}, {Fraisse}, {Franceschi},
  {Frejsel}, {Galeotta}, {Galli}, {Ganga}, {G{\'e}nova-Santos}, {Giard},
  {Giraud-H{\'e}raud}, {Gjerl{\o}w}, {Gonz{\'a}lez-Nuevo}, {G{\'o}rski},
  {Grainge}, {Gratton}, {Gregorio}, {Gruppuso}, {Gudmundsson}, {Hansen},
  {Hanson}, {Harrison}, {Hempel}, {Henrot-Versill{\'e}},
  {Hern{\'a}ndez-Monteagudo}, {Herranz}, {Hildebrandt}, {Hivon}, {Hobson},
  {Holmes}, {Hornstrup}, {Hovest}, {Huffenberger}, {Hurier}, {Jaffe}, {Jaffe},
  {Jin}, {Jones}, {Juvela}, {Keih{\"a}nen}, {Keskitalo}, {Khamitov}, {Kisner},
  {Kneissl}, {Knoche}, {Kunz}, {Kurki-Suonio}, {Lagache}, {Lamarre}, {Lasenby},
  {Lattanzi}, {Lawrence}, {Leonardi}, {Lesgourgues}, {Levrier}, {Liguori},
  {Lilje}, {Linden-V{\o}rnle}, {L{\'o}pez-Caniego}, {Lubin},
  {Mac{\'\i}as-P{\'e}rez}, {Maggio}, {Maino}, {Mak}, {Mandolesi}, {Mangilli},
  {Martin}, {Mart{\'\i}nez-Gonz{\'a}lez}, {Masi}, {Matarrese}, {Mazzotta},
  {McGehee}, {Mei}, {Melchiorri}, {Melin}, {Mendes}, {Mennella}, {Migliaccio},
  {Mitra}, {Miville-Desch{\^e}nes}, {Moneti}, {Montier}, {Morgante},
  {Mortlock}, {Moss}, {Munshi}, {Murphy}, {Naselsky}, {Nastasi}, {Nati},
  {Natoli}, {Netterfield}, {N{\o}rgaard-Nielsen}, {Noviello}, {Novikov},
  {Novikov}, {Olamaie}, {Oxborrow}, {Paci}, {Pagano}, {Pajot}, {Paoletti},
  {Pasian}, {Patanchon}, {Pearson}, {Perdereau}, {Perotto}, {Perrott},
  {Perrotta}, {Pettorino}, {Piacentini}, {Piat}, {Pierpaoli}, {Pietrobon},
  {Plaszczynski}, {Pointecouteau}, {Polenta}, {Pratt}, {Pr{\'e}zeau}, {Prunet},
  {Puget}, {Rachen}, {Reach}, {Rebolo}, {Reinecke}, {Remazeilles}, {Renault},
  {Renzi}, {Ristorcelli}, {Rocha}, {Rosset}, {Rossetti}, {Roudier}, {Rozo},
  {Rubi{\~n}o-Mart{\'\i}n}, {Rumsey}, {Rusholme}, {Rykoff}, {Sandri}, {Santos},
  {Saunders}, {Savelainen}, {Savini}, {Schammel}, {Scott}, {Seiffert},
  {Shellard}, {Shimwell}, {Spencer}, {Stanford}, {Stern}, {Stolyarov},
  {Stompor}, {Streblyanska}, {Sudiwala}, {Sunyaev}, {Sutton}, {Suur-Uski},
  {Sygnet}, {Tauber}, {Terenzi}, {Toffolatti}, {Tomasi}, {Tramonte},
  {Tristram}, {Tucci}, {Tuovinen}, {Umana}, {Valenziano}, {Valiviita}, {Van
  Tent}, {Vielva}, {Villa}, {Wade}, {Wandelt}, {Wehus}, {White}, {Wright},
  {Yvon}, {Zacchei}, \& {Zonca}}]{Planck_2016}
{Planck Collaboration}, {Ade}, P.~A.~R., {Aghanim}, N., {et~al.} 2016, \aap,
  594, A27

\bibitem[{{Predehl} {et~al.}(2021){Predehl}, {Andritschke}, {Arefiev},
  {Babyshkin}, {Batanov}, {Becker}, {B{\"o}hringer}, {Bogomolov}, {Boller},
  {Borm}, {Bornemann}, {Br{\"a}uninger}, {Br{\"u}ggen}, {Brunner}, {Brusa},
  {Bulbul}, {Buntov}, {Burwitz}, {Burkert}, {Clerc}, {Churazov}, {Coutinho},
  {Dauser}, {Dennerl}, {Doroshenko}, {Eder}, {Emberger}, {Eraerds},
  {Finoguenov}, {Freyberg}, {Friedrich}, {Friedrich}, {F{\"u}rmetz},
  {Georgakakis}, {Gilfanov}, {Granato}, {Grossberger}, {Gueguen}, {Gureev},
  {Haberl}, {H{\"a}lker}, {Hartner}, {Hasinger}, {Huber}, {Ji}, {Kienlin},
  {Kink}, {Korotkov}, {Kreykenbohm}, {Lamer}, {Lomakin}, {Lapshov}, {Liu},
  {Maitra}, {Meidinger}, {Menz}, {Merloni}, {Mernik}, {Mican}, {Mohr},
  {M{\"u}ller}, {Nandra}, {Nazarov}, {Pacaud}, {Pavlinsky}, {Perinati},
  {Pfeffermann}, {Pietschner}, {Ramos-Ceja}, {Rau}, {Reiffers}, {Reiprich},
  {Robrade}, {Salvato}, {Sanders}, {Santangelo}, {Sasaki}, {Scheuerle},
  {Schmid}, {Schmitt}, {Schwope}, {Shirshakov}, {Steinmetz}, {Stewart},
  {Str{\"u}der}, {Sunyaev}, {Tenzer}, {Tiedemann}, {Tr{\"u}mper}, {Voron},
  {Weber}, {Wilms}, \& {Yaroshenko}}]{Predehl_2020}
{Predehl}, P., {Andritschke}, R., {Arefiev}, V., {et~al.} 2021, \aap, 647, A1

\bibitem[{{Predehl} {et~al.}(2020){Predehl}, {Sunyaev}, {Becker}, {Brunner},
  {Burenin}, {Bykov}, {Cherepashchuk}, {Chugai}, {Churazov}, {Doroshenko},
  {Eismont}, {Freyberg}, {Gilfanov}, {Haberl}, {Khabibullin}, {Krivonos},
  {Maitra}, {Medvedev}, {Merloni}, {Nandra}, {Nazarov}, {Pavlinsky}, {Ponti},
  {Sanders}, {Sasaki}, {Sazonov}, {Strong}, \& {Wilms}}]{predehl20}
{Predehl}, P., {Sunyaev}, R.~A., {Becker}, W., {et~al.} 2020, \nat, 588, 227

\bibitem[{{Reiprich} {et~al.}(2021){Reiprich}, {Veronica}, {Pacaud},
  {Ramos-Ceja}, {Ota}, {Sanders}, {Kara}, {Erben}, {Klein}, {Erler}, {Kerp},
  {Hoang}, {Br{\"u}ggen}, {Marvil}, {Rudnick}, {Biffi}, {Dolag},
  {Aschersleben}, {Basu}, {Brunner}, {Bulbul}, {Dennerl}, {Eckert}, {Freyberg},
  {Gatuzz}, {Ghirardini}, {K{\"a}fer}, {Merloni}, {Migkas}, {Nandra},
  {Predehl}, {Robrade}, {Salvato}, {Whelan}, {Diaz-Ocampo}, {Hernandez-Lang},
  {Zenteno}, {Brown}, {Collier}, {Diego}, {Hopkins}, {Kapinska}, {Koribalski},
  {Mroczkowski}, {Norris}, {O'Brien}, \& {Vardoulaki}}]{reiprich21}
{Reiprich}, T.~H., {Veronica}, A., {Pacaud}, F., {et~al.} 2021, \aap, 647, A2

\bibitem[{{Revnivtsev} {et~al.}(2009){Revnivtsev}, {Sazonov}, {Churazov},
  {Forman}, {Vikhlinin}, \& {Sunyaev}}]{Revnivtsev2009}
{Revnivtsev}, M., {Sazonov}, S., {Churazov}, E., {et~al.} 2009, \nat, 458, 1142

\bibitem[{{Saxton} {et~al.}(2008){Saxton}, {Read}, {Esquej}, {Freyberg},
  {Altieri}, \& {Bermejo}}]{Saxton_2008}
{Saxton}, R.~D., {Read}, A.~M., {Esquej}, P., {et~al.} 2008, \aap, 480, 611

\bibitem[{{Sazonov} {et~al.}(2020){Sazonov}, {Burenin}, {Khorunzhev}, {Lyapin},
  {Medvedev}, {Uskov}, {Zaznobin}, {Afanasiev}, {Dodonov}, {Uklein},
  {Belinsky}, {Tatarnikov}, {Vozyakova}, {Zheltoukhov}, {Chugai}, {Sunyaev}, \&
  {Gilfanov}}]{sazonov20}
{Sazonov}, S., {Burenin}, R., {Khorunzhev}, G., {et~al.} 2020, The Astronomer's
  Telegram, 13987, 1

\bibitem[{{Sazonov} {et~al.}(2021){Sazonov}, {Gilfanov}, {Medvedev}, {Yao},
  {Khorunzhev}, {Semena}, {Sunyaev}, {Burenin}, {Lyapin}, {Mescheryakov},
  {Uskov}, {Zaznobin}, {Postnov}, {Dodin}, {Belinski}, {Cherepashchuk},
  {Eselevich}, {Dodonov}, {Grokhovskaya}, {Kotov}, {Bikmaev}, {Zhuchkov},
  {Gumerov}, {van Velzen}, \& {Kulkarni}}]{Sazonov2021}
{Sazonov}, S., {Gilfanov}, M., {Medvedev}, P., {et~al.} 2021, arXiv e-prints,
  arXiv:2108.02449

\bibitem[{{Schuecker} {et~al.}(2003){Schuecker}, {B{\"o}hringer}, {Collins}, \&
  {Guzzo}}]{2003Schuecker}
{Schuecker}, P., {B{\"o}hringer}, H., {Collins}, C.~A., \& {Guzzo}, L. 2003,
  \aap, 398, 867

\bibitem[{{Schwope} {et~al.}(2021){Schwope}, {Buckley}, {Kawka}, {K{\"o}nig},
  {Lutovinov}, {Maitra}, {Mereminskiy}, {Miller-Jones}, {Pichardo Marcano},
  {Rau}, {Semena}, {Townsend}, \& {Wilms}}]{Schwope2021}
{Schwope}, A., {Buckley}, D. A.~H., {Kawka}, A., {et~al.} 2021, arXiv e-prints,
  arXiv:2106.14538

\bibitem[{{Schwope} {et~al.}(2020){Schwope}, {Semena}, {Maitra}, {Buckley},
  {Kawka}, {K{\"o}nig}, {Lutovinov}, {Mereminskiy}, {Miller-Jones}, {Rau},
  {Townsend}, \& {Wilms}}]{atel14219}
{Schwope}, A., {Semena}, A., {Maitra}, C., {et~al.} 2020, The Astronomer's
  Telegram, 14219, 1

\bibitem[{{Snowden} {et~al.}(1997){Snowden}, {Egger}, {Freyberg}, {McCammon},
  {Plucinsky}, {Sanders}, {Schmitt}, {Tr{\"u}mper}, \& {Voges}}]{Snowden_1997}
{Snowden}, S.~L., {Egger}, R., {Freyberg}, M.~J., {et~al.} 1997, \apj, 485, 125

\bibitem[{{Str{\"u}der} {et~al.}(2001){Str{\"u}der}, {Briel}, {Dennerl},
  {Hartmann}, {Kendziorra}, {Meidinger}, {Pfeffermann}, {Reppin}, {Aschenbach},
  {Bornemann}, {Br{\"a}uninger}, {Burkert}, {Elender}, {Freyberg}, {Haberl},
  {Hartner}, {Heuschmann}, {Hippmann}, {Kastelic}, {Kemmer}, {Kettenring},
  {Kink}, {Krause}, {M{\"u}ller}, {Oppitz}, {Pietsch}, {Popp}, {Predehl},
  {Read}, {Stephan}, {St{\"o}tter}, {Tr{\"u}mper}, {Holl}, {Kemmer}, {Soltau},
  {St{\"o}tter}, {Weber}, {Weichert}, {von Zanthier}, {Carathanassis}, {Lutz},
  {Richter}, {Solc}, {B{\"o}ttcher}, {Kuster}, {Staubert}, {Abbey}, {Holland},
  {Turner}, {Balasini}, {Bignami}, {La Palombara}, {Villa}, {Buttler},
  {Gianini}, {Lain{\'e}}, {Lumb}, \& {Dhez}}]{Strueder2001}
{Str{\"u}der}, L., {Briel}, U., {Dennerl}, K., {et~al.} 2001, \aap, 365, L18

\bibitem[{{Sunyaev} {et~al.}(1987){Sunyaev}, {Kaniovsky}, {Efremov},
  {Gilfanov}, {Churazov}, {Grebenev}, {Kuznetsov}, {Melioranskiy},
  {Yamburenko}, {Yunin}, {Stepanov}, {Chulkov}, {Pappe}, {Boyarskiy},
  {Gavrilova}, {Loznikov}, {Prudkoglyad}, {Rodin}, {Reppin}, {Pietsch},
  {Engelhauser}, {Truemper}, {Voges}, {Kendziorra}, {Bezler}, {Staubert},
  {Brinkman}, {Heise}, {Mels}, {Jager}, {Skinner}, {Al-Emam}, {Patterson},
  {Willmore}, {Gilfanov}, \& {Churazov}}]{Sunyaev_1987a}
{Sunyaev}, R., {Kaniovsky}, A., {Efremov}, V., {et~al.} 1987, \nat, 330, 227

\bibitem[{{Sunyaev} {et~al.}(1991){Sunyaev}, {Aref'ev}, {Borozdin}, {Churazov},
  {Efremov}, {Gilfanov}, {Grebenev}, {Kaniovsky}, {Kuznetsov}, {Lapshov},
  {Yamburenko}, {Englhauser}, {Doebereiner}, {Pietsch}, {Reppin}, {Truemper},
  {Boer}, {Kendziorra}, {Maisack}, {Mony}, {Staubert}, {Skinner}, {Patterson},
  {Willmore}, {Eman}, {Pan}, {Nottingham}, {Brinkman}, {Heise}, {in't Zand}, \&
  {Jager}}]{sunyaev91a}
{Sunyaev}, R.~A., {Aref'ev}, V., {Borozdin}, K., {et~al.} 1991, Advances in
  Space Research, 11, 5

\bibitem[{{Sunyaev} {et~al.}(1993){Sunyaev}, {Markevitch}, \&
  {Pavlinsky}}]{1993ApJ...407..606S}
{Sunyaev}, R.~A., {Markevitch}, M., \& {Pavlinsky}, M. 1993, \apj, 407, 606

\bibitem[{{Sunyaev} \& {Zeldovich}(1980)}]{sunyaev80}
{Sunyaev}, R.~A. \& {Zeldovich}, I.~B. 1980, \araa, 18, 537

\bibitem[{{Sunyaev} \& {Zeldovich}(1970)}]{Sunyaev_1970}
{Sunyaev}, R.~A. \& {Zeldovich}, Y.~B. 1970, \apss, 7, 3

\bibitem[{{Syunyaev} {et~al.}(1991){Syunyaev}, {Arefev}, {Borozdin},
  {Gilfanov}, {Efremov}, {Kaniovskii}, {Churazov}, {Kendziorra}, {Mony},
  {Kretschmar}, {Maisack}, {Staubert}, {Dobereiner}, {Englhauser}, {Pietsch},
  {Reppin}, {Trumper}, {Skinner}, {Nottingham}, {Pan}, \&
  {Willmore}}]{sunyaev91b}
{Syunyaev}, R.~A., {Arefev}, V.~A., {Borozdin}, K.~N., {et~al.} 1991, Soviet
  Astronomy Letters, 17, 409

\bibitem[{{Syunyaev} {et~al.}(1994){Syunyaev}, {Borozdin}, {Aleksandrovich},
  {Arefev}, {Kaniovskii}, {Efremov}, {Maisack}, {Reppin}, \&
  {Skinner}}]{sunyaev94}
{Syunyaev}, R.~A., {Borozdin}, K.~N., {Aleksandrovich}, N.~L., {et~al.} 1994,
  Astronomy Letters, 20, 777

\bibitem[{{Syunyaev} {et~al.}(1987){Syunyaev}, {Kaniovskii}, {Efremov},
  {Gilfanov}, {Dhurazov}, {Grebenev}, {Kuznetsov}, {Melioranskii},
  {Yamburenko}, {Yunin}, {Stepanov}, {Chulkov}, {Pappe}, {Boyarskii},
  {Gavrilova}, {Loznikov}, {Prudkoglyad}, {Rodin}, {Reppin}, {Pietsch},
  {Engelhauser}, {Trumper}, {Voges}, {Kendziorra}, {Bezler}, {Staubert},
  {Brinkman}, {Heise}, {Mels}, {Jager}, {Skinner}, {Al-Emam}, {Patterson}, \&
  {Willmore}}]{Sunyaev_1987b}
{Syunyaev}, R.~A., {Kaniovskii}, A., {Efremov}, V., {et~al.} 1987, Soviet
  Astronomy Letters, 13, 431

\bibitem[{{Tkachenko} {et~al.}(2002){Tkachenko}, {Terekhov}, {Sunyaev},
  {Kuznetsov}, {Barat}, {Dezalay}, \& {Vedrenne}}]{Tkachenko_2002}
{Tkachenko}, A.~Y., {Terekhov}, O.~V., {Sunyaev}, R.~A., {et~al.} 2002,
  Astronomy Letters, 28, 353

\bibitem[{{Tr{\"u}mper}(1982)}]{Truemper1982}
{Tr{\"u}mper}, J. 1982, Advances in Space Research, 2, 241

\bibitem[{{Vainshtein} \& {Syunyaev}(1980)}]{1980SvAL....6..353V}
{Vainshtein}, L.~A. \& {Syunyaev}, R.~A. 1980, Soviet Astronomy Letters, 6, 353

\bibitem[{{Voges} {et~al.}(1999){Voges}, {Aschenbach}, {Boller},
  {Br{\"a}uninger}, {Briel}, {Burkert}, {Dennerl}, {Englhauser}, {Gruber},
  {Haberl}, {Hartner}, {Hasinger}, {K{\"u}rster}, {Pfeffermann}, {Pietsch},
  {Predehl}, {Rosso}, {Schmitt}, {Tr{\"u}mper}, \& {Zimmermann}}]{Voges_1999}
{Voges}, W., {Aschenbach}, B., {Boller}, T., {et~al.} 1999, \aap, 349, 389

\bibitem[{{Wilms} {et~al.}(2020){Wilms}, {Kreykenbohm}, {Weber}, {Falkner},
  {Dauser}, {Knies}, {Koenig}, {Malyali}, {Rau}, {Merloni}, {Bogensberger},
  {Brunner}, {Buchner}, {Carpano}, {Freyberg}, {Haberl}, {Maitra}, {Salvato},
  {Doroshenko}, {Ducci}, {Ji}, {Schmitt}, \& {Schwope}}]{Wilms_2020}
{Wilms}, J., {Kreykenbohm}, I., {Weber}, P., {et~al.} 2020, The Astronomer's
  Telegram, 13416, 1

\bibitem[{{Wolf} {et~al.}(2021){Wolf}, {Nandra}, {Salvato}, {Liu}, {Buchner},
  {Brusa}, {Hoang}, {Moss}, {Arcodia}, {Br{\"u}ggen}, {Comparat}, {de
  Gasperin}, {Georgakakis}, {Hotan}, {Lamer}, {Merloni}, {Rau}, {Rottgering},
  {Shimwell}, {Urrutia}, {Whiting}, \& {Williams}}]{wolf20}
{Wolf}, J., {Nandra}, K., {Salvato}, M., {et~al.} 2021, \aap, 647, A5

\bibitem[{{Wolter}(1952{\natexlab{a}})}]{Wolter1952a}
{Wolter}, H. 1952{\natexlab{a}}, Annalen der Physik, 445, 94

\bibitem[{{Wolter}(1952{\natexlab{b}})}]{Wolter1952b}
{Wolter}, H. 1952{\natexlab{b}}, Annalen der Physik, 445, 286

\bibitem[{{Yao} {et~al.}(2020{\natexlab{a}}){Yao}, {Kulkarni}, {Burdge},
  {Caiazzo}, {De}, {Dong}, {Kasliwal}, {Kupfer}, {van Roestel}, {Sollerman},
  {Bagdasaryan}, {Bellm}, {Cenko}, {Drake}, {Duev}, {Fremling}, {Graham},
  {Kaye}, {Masci}, {Miranda}, {Prince}, {Riddle}, {Rusholme}, \&
  {Soumagnac}}]{yao21b}
{Yao}, Y., {Kulkarni}, S.~R., {Burdge}, K.~B., {et~al.} 2020{\natexlab{a}},
  arXiv e-prints, arXiv:2012.00169

\bibitem[{{Yao} {et~al.}(2020{\natexlab{b}}){Yao}, {Kulkarni}, {Gendreau},
  {Jaisawal}, {Enoto}, {Grefenstette}, {Marshall}, {Garc{\'\i}a}, {Ludlam},
  {Pike}, {Ng}, {Zhang}, {Altamirano}, {Jaodand}, {Cenko}, {Remillard},
  {Steiner}, {Negoro}, {Brightman}, {Lien}, {Wolff}, {Ray}, {Mukai},
  {Wadiasingh}, {Arzoumanian}, {Kawai}, {Mihara}, \& {Strohmayer}}]{yao21a}
{Yao}, Y., {Kulkarni}, S.~R., {Gendreau}, K.~C., {et~al.} 2020{\natexlab{b}},
  arXiv e-prints, arXiv:2012.00160

\bibitem[{{Zaznobin} {et~al.}(2021{\natexlab{a}}){Zaznobin}, {Sazonov},
  {Burenin}, {Uskov}, {Semena}, {Gilfanov}, {Medvedev}, {Sunyaev}, \&
  {Eselevich}}]{Zaznobin2021b}
{Zaznobin}, I., {Sazonov}, S., {Burenin}, R., {et~al.} 2021{\natexlab{a}},
  arXiv e-prints, arXiv:2107.05611

\bibitem[{{Zaznobin} {et~al.}(2021{\natexlab{b}}){Zaznobin}, {Uskov},
  {Sazonov}, {Burenin}, {Medvedev}, {Khorunzhev}, {Lyapin}, {Krivonos},
  {Filippova}, {Gilfanov}, {Sunyaev}, {Eselevich}, {Bikmaev}, {Irtuganov}, \&
  {Nikolaeva}}]{zaznobin21}
{Zaznobin}, I.~A., {Uskov}, G.~S., {Sazonov}, S.~Y., {et~al.}
  2021{\natexlab{b}}, Astronomy Letters, 47, 89

\end{thebibliography}

\end{document}